\documentclass[12pt,a4paper]{article}
\usepackage[T1]{fontenc}
\usepackage[utf8]{inputenc}
\usepackage{lmodern}
\usepackage[margin=1in]{geometry}
\usepackage{needspace}
\usepackage{microtype}
\usepackage{amsmath,amssymb,amsthm,mathtools}
\allowdisplaybreaks[1]
\usepackage{booktabs,tabularx,array}
\usepackage{enumitem}
\usepackage[authoryear,round]{natbib}
\usepackage{xurl}
\usepackage[unicode,bookmarksnumbered,hidelinks]{hyperref}
\hypersetup{hypertexnames=false,pdftitle={Truthful compromise and insurance on a sphere},pdfauthor={Siwei Chen and Pengbo Wang},pdfsubject={Increasing-concave incentives, efficiency at profiles with few distinct reports, and compromise insurance}}

\newcommand{\R}{\mathbb R}

\newcommand{\Prob}{\mathcal P}

\newtheorem{theorem}{Theorem}
\newtheorem{proposition}{Proposition}
\newtheorem{corollary}{Corollary}
\newtheorem{lemma}{Lemma}

\theoremstyle{definition}

\theoremstyle{remark}

\numberwithin{equation}{section}
\newcommand{\doi}[1]{doi: \begingroup\def\UrlBreaks{\do\/\do-}\href{https://doi.org/#1}{\nolinkurl{#1}}\endgroup}
\title{Truthful compromise and insurance on a sphere}
\author{Siwei Chen\thanks{\linespread{1}\selectfont\raggedright Corresponding author.\\
\textit{Affiliation (both authors):} Lingnan College, Sun Yat-sen University, Guangzhou, China.\\
\textit{E-mail addresses:} \href{mailto:chensw6@mail.sysu.edu.cn}{\nolinkurl{chensw6@mail.sysu.edu.cn}} (S. Chen), \href{mailto:illusory0101@gmail.com}{\nolinkurl{illusory0101@gmail.com}} (P. Wang).\\
\textit{ORCID:} 0000-0002-8486-2251 (S. Chen).\\
\textit{Funding:} Chen's research is supported by the National Natural Science Foundation of China (Grant No. 72373170).}
\quad Pengbo Wang}
\date{}
\begin{document}
\flushbottom
\maketitle
\begin{abstract}
Can public choice without transfers insure conflicting interests when action intensity is fixed? We require truthful reporting for every increasing-concave evaluation of alignment with private ideals. On a sphere, efficiency at profiles with at most three ideal directions forces fixed-weight random dictatorship; weaker tests suffice in two and three dimensions. From four dimensions, binary efficiency permits committees to insure members by placing outcome risk orthogonally to their reports. We identify restrictions shared by all mechanisms with these incentives and binary efficiency: binary winning probabilities determine mean outcomes at every profile and individual alignment distributions at planar three-camp profiles. When these probabilities depend only on camp size, complete insurance at balanced equilateral disagreement requires dimension proportional to population. In contrast, committees reduce each participant's welfare loss relative to certain zero alignment to any prescribed positive fraction of the equal-weight random-dictatorship loss in a population-independent dimension.
\end{abstract}
\section{Introduction}\label{sec:intro}
Without transfers, a public decision can insure conflicting interests only through the outcome itself. When action intensity is fixed, averaging conflicting private ideal directions is infeasible. Randomization preserves intensity, but its insurance value depends on how the outcome varies relative to each participant's ideal.

For example, an institution may require a public evaluation index to have unit variance. With standardized, uncorrelated indicators, index weights lie on the unit sphere, and alignment equals correlation with a participant's ideal unit-variance index. The mechanism randomizes over index weights, so the risk studied here comes from the public choice of weights. More generally, a fixed positive definite quadratic intensity index becomes a sphere after a change of coordinates.

We take alignment as a fixed cardinal performance index and allow every increasing-concave evaluation of it. Increasing-concave strategy-proofness (ICV-SP) requires the truthful alignment distribution to second-order stochastically dominate every misreport distribution. A public outcome can remain random while every participant's alignment is certain. We ask when incentives and efficiency permit this insurance, and how the available dimension limits it.

The first finding is an efficiency boundary. Under full-domain ICV-SP, unanimity alone forces fixed-weight random dictatorship in two dimensions. Efficiency at orthogonal two-position profiles does so in three dimensions; efficiency at all three-position profiles suffices in every dimension (Theorem~\ref{thm:icv}). The institution can then allocate only fixed chances of selecting individual ideals. We use ex-post and first-order stochastic-dominance (SD) efficiency, ordinal safeguards that leave room for insurance gains under concave evaluations.

From four dimensions onward, binary efficiency permits truthful compromise for populations with at least three participants. An odd committee uses majority votes on directional projections to choose a mean in the unit ball. Risk orthogonal to its reports restores unit intensity while making each selected member's alignment certain conditional on the committee. Truth maximizes mean alignment, and concavity makes deviation risk unattractive (Proposition~\ref{thm:icv-e2-counterexample}). Random selection of the committee can still leave individual risk.

The second finding concerns restrictions shared by all binary-efficient ICV-SP mechanisms in dimension at least four. A camp's binary control is its probability of selecting its ideal when the camp and its complement report two distinct directions. This probability remains a guarantee at every profile. Binary controls identify every mean outcome and, at planar three-camp profiles, every participant's alignment distribution (Theorem~\ref{thm:control-identification}). The result holds without symmetry or regularity across reports. The planar qualification has content: for a particular control vector, two full-domain rules share all means but give different individual variances at every linearly independent triple.

The control restrictions separate complete insurance from a proportional reduction in welfare loss. When binary winning probabilities depend only on camp size, sufficiently large separated camps face a positive risk floor at fixed dimension, even with unequal camp sizes and arbitrary outsiders (Proposition~\ref{prop:fixed-d-risk}). At balanced equilateral disagreement, complete insurance consequently needs dimension proportional to population. Yet committees achieve any prescribed positive fraction of the equal-weight random-dictatorship welfare loss in a population-independent dimension. Both losses are measured relative to certain zero alignment, separately for each participant and each permitted evaluation.

Two design comparisons refine this distinction. With binary control fixed, guaranteed individual chances of control create unavoidable atoms. The remaining probability has a common welfare bound. When that probability is positive, one rule attains the bound at every profile exactly when residual binary contests have deterministic winners and dimension exceeds population (Proposition~\ref{prop:all-profile-insurance}).

We also compare anonymous, report-independent delegations to committees whose component rules insure all selected members at every profile. This class permits arbitrary component voting rules and mixtures of committee sizes. Its optimum is attained by the largest feasible odd majority committee, which minimizes alignment variance and maximizes every permitted evaluation at balanced equilateral disagreement (Proposition~\ref{prop:committee-optimality}).

Finally, relaxing incentives provides another route to insurance in a fixed dimension of at least four. For sufficiently large balanced three-camp populations, the smallest worst-case incentive error allowing complete insurance that preserves equal-weight random-dictatorship means has inverse-population order (Proposition~\ref{prop:insurance-error}). Utilities are normalized at the endpoints, and insurance must give everyone zero alignment at every labeling of the reference disagreement. Thus additional dimension and a vanishing incentive error offer distinct ways to insure the same benchmark.

\subsection{Related literature}\label{sec:literature}
On the sphere in three-dimensional space, \citet[Corollaries~1--2]{cps2016} obtain deterministic dictatorship from Pareto optimality and strict or coalitional strategy-proofness. \citet[Theorem~1]{chenwang2026} obtain it from ordinary strategy-proofness and diagonal unanimity on compact full-dimensional strictly convex frontiers. We use their body-valued menu decomposition to identify binary lottery laws and residual risk. Their boundary-valued dictatorship theorem does not apply to interior lottery means.

\citet{zhou1991} establishes multidimensional public-good dictatorship on a rich domain of continuous convex preferences. For probabilistic public-good choice, \citet[Theorem~2.1]{dps2002} obtain random dictatorship from SD-strategy-proofness and unanimity on convex Euclidean outcome sets of dimension at least two, with all continuous strictly convex single-peaked preferences. Distance rankings extend to the ball; extending a directional-report rule while preserving incentives on this richer domain is not automatic. Our lotteries remain sphere-valued and evaluations are increasing-concave transforms of alignment.

On unrestricted finite strict-order domains, \citet{gibbard1977} characterizes ex-post efficient SD-strategy-proof rules as random dictatorships; \citet{csz2014} study domains retaining this conclusion. The decomposition of unanimous SD-strategy-proof rules into truthful deterministic rules on minimally rich single-peaked finite ordered domains \citep{prss2014} does not apply to spherical ICV-SP.

\citet[Theorem~1]{cz2019} characterize minimally rich, connected$^+$ finite multidimensional domains admitting unanimous SD-strategy-proof rules with a common-second-best compromise property. We fix a continuous spherical domain and study the insurance supplied by compromise lotteries. \citet{lederer2026} restricts cardinal utility sets and studies incentives, decisiveness and efficiency. Our ICV-SP restriction has a risk-aversion interpretation; spherical geometry determines the efficiency boundary and insurance possibilities.

\citet{maskinriley1984} study insurance versus screening in auctions with risk-averse buyers. \citet[Theorem~3.1]{dughmiperes2012} use outcome-dependent transfers to insure truthful net payoffs, making truthful-in-expectation mechanisms robust to unknown increasing-concave evaluations. Our committee construction likewise combines truthful alignment that is certain conditional on the committee with Jensen's inequality. Without transfers, insurance must come from the sphere-valued public outcome, creating control and dimension restrictions.

\section{Model}\label{sec:model}
Fix a population $N=\{1,\ldots,n\}$ with $n\geq1$ and an integer $d\geq2$. Outcomes and types are $S=\mathbb S^{d-1}\subset\R^d$. Type $p$ ranks outcome $x$ by alignment $p\cdot x$, equivalently by negative squared distance since $\|p-x\|^2=2-2p\cdot x$. There are no transfers.

The institution commits to $P:S^N\to\Prob_B(S)$ and, after reports $r$, draws $X\sim P(r)$. Each $P(r)$ is a countably additive Borel probability; no measurability or continuity across reports is assumed. A jointly Borel rule is measurable into the weak topology on probabilities, equivalently $r\mapsto\int f\,dP(r)$ is Borel for every continuous $f:S\to\R$.

A rule is increasing-concave strategy-proof (ICV-SP) if
\[
 \int\phi(p_i\cdot x)\,dP(p_i,r_{-i})(x)
 \geq\int\phi(p_i\cdot x)\,dP(s_i,r_{-i})(x)
\]
for every participant $i$, true direction, alternative report, opponents' profile and bounded nondecreasing concave $\phi:[-1,1]\to\R$. Write $\Phi$ for this class of evaluations. Such functions may have a downward jump at $-1$. Alignment is the fixed cardinal index; positive affine rescaling preserves the class. Participants may have different evaluations, with no interpersonal utility comparison.

Complete insurance at a profile makes every participant's reported alignment almost surely constant, even if the public outcome remains random. A camp consists of participants reporting the same direction. A balanced equilateral profile has three camps of equal size whose directions satisfy $p_1+p_2+p_3=0$.

Write $\mu\succeq_p^{\rm SD}\nu$ for first-order stochastic dominance if $\mu\{x:p\cdot x\geq t\}\geq\nu\{x:p\cdot x\geq t\}$ for every $t$, strictly if some inequality is strict. An outcome is Pareto efficient at $r$ if no other sphere point weakly raises all reported alignments and strictly raises one. A lottery is ex-post efficient if it gives such outcomes probability one; it is SD-Pareto efficient if no lottery weakly SD-improves every participant and strictly SD-improves one. Define
\begin{equation}\label{eq:ek}
 \mathcal E_k=\{r:|\{r_i:i\in N\}|\leq k\},
\end{equation}
\begin{equation}\label{eq:orthogonal-tests}
 \mathcal E_2^\perp=\mathcal E_1\cup
 \{r\in\mathcal E_2:\{r_i:i\in N\}=\{a,b\},\ a\cdot b=0\}.
\end{equation}
Efficiency on these sets covers every division into camps and orientation. Incentives always cover the full domain. Under ICV-SP, the orthogonal tests are equivalent to efficiency on all of $\mathcal E_2$ (Lemma~\ref{lem:binary}), so we refer to either as binary efficiency. Either notion of efficiency on $\mathcal E_1$ implies unanimity $P(p,\ldots,p)=\delta_p$: the common ideal dominates every other point, and its lottery dominates every other law, strictly at the top cutoff.

The efficiency commitment is ordinal and does not require Pareto optimality within the concave class. At orthogonal reports $a,b$, the lottery $(\delta_a+\delta_b)/2$ satisfies both efficiencies: its endpoints are efficient, and the top cutoffs force any SD dominator to retain their masses. Yet $(a+b)/\sqrt2$ gives each camp certain alignment $1/\sqrt2$, and
$[\phi(0)+\phi(1)]/2\leq\phi(1/2)\leq\phi(1/\sqrt2)$,
strictly for linear evaluation.

Fixed-weight random dictatorship selects $r_i$ with report-independent probability $\alpha_i\geq0$, $\sum_i\alpha_i=1$. Anonymity means invariance under permutations of reports across labels. Orthogonal neutrality (orthogonal equivariance) means $P(Or)=O_\#P(r)$ for every orthogonal map, including reflections.

\section{Efficiency, control, and insurance}\label{sec:majority}
\subsection{Efficiency and control}\label{subsec:binary}
We first ask when truthful choice can allocate only fixed chances of selecting individual ideals.
\begin{theorem}\label{thm:icv}
Fix finite $d\geq2$ and a finite nonempty population $N$. For a rule $P:S^N\to\Prob_B(S)$, the following are equivalent:
\begin{enumerate}
\item $P$ is ICV-SP on the full report domain and ex-post Pareto efficient on $\mathcal E_3$.
\item $P$ is ICV-SP on the full report domain and SD-Pareto efficient on $\mathcal E_3$.
\item There are constants $\alpha_i\geq0$, $\sum_i\alpha_i=1$, such that
\begin{equation}\label{eq:sphere-e2-rd}
 P(r)=\sum_{i\in N}\alpha_i\delta_{r_i}\qquad(r\in S^N).
\end{equation}
\end{enumerate}
If $d=2,3$, $\mathcal E_3$ in statements 1 and 2 can be replaced by $\mathcal E_2$, or by the smaller family $\mathcal E_2^\perp$.
If $d=2$, either efficiency requirement can be replaced by unanimity alone.
\end{theorem}

Anonymity specializes the characterization to equal weights. Its key step is to identify a camp's binary chance of control and show that this chance remains guaranteed at every profile.

The next two lemmas use full-domain ICV-SP and binary efficiency, so they also apply to the compromise rules studied below. For $\varnothing\ne A\subsetneq N$, write $P_A(a,b)$ for the law when camp $A$ reports $a$ and its complement reports $b$.
\begin{lemma}\label{lem:binary}
Full-domain ICV-SP and either efficiency on $\mathcal E_2^\perp$ imply a constant $w_A\in[0,1]$ such that
\[
 P_A(a,b)=w_A\delta_a+(1-w_A)\delta_b\quad(a,b\in S).
\]
\end{lemma}
Individual incentive comparisons telescope when a camp's members share a true direction. The planar menu representation and orthogonal efficiency fix one weight across planes and angles; finite concave comparisons propagate the endpoint law to all pairs, including antipodes (Appendices~\ref{app:plane}--\ref{app:binary}).

Set $w_\varnothing=0$, $w_N=1$; exchanging the two camps gives $w_{A^c}=1-w_A$. These binary chances are guarantees at richer profiles.
\begin{lemma}\label{lem:peak-mass}
Under the hypotheses of Lemma~\ref{lem:binary}, $P(r)\{a\}\geq w_A$ whenever every label in $A$ reports $a$, regardless of outsiders. For disjoint $A,B$, $w_{A\cup B}\geq w_A+w_B$.
\end{lemma}
\begin{proof}
Start with $A$ at $a$ and all outsiders at $-a$. The binary law gives mass $w_A$ at $a$. Restore outsiders one by one. Each departing type is $-a$, whose unique worst outcome is $a$. For $U=-a\cdot X$, ICV-SP with $\phi(u)=\min\{u,-1+\varepsilon\}$ says that $\mathbb E(-1+\varepsilon-U)_+$ cannot decrease after this departure. Dividing by $\varepsilon$ and using bounded convergence gives
\[
 \lim_{\varepsilon\downarrow0}\varepsilon^{-1}
 \mathbb E(-1+\varepsilon-U)_+=\Pr(U=-1).
\]
The atom at $a$ cannot decrease. For three nonempty disjoint blocks $A,B,C=N\setminus(A\cup B)$, assign three distinct peaks. The guarantees give $w_A+w_B+w_C\leq1$, and complementarity gives $w_{A\cup B}=1-w_C$. Empty or complementary blocks use $w_\varnothing=0,w_N=1$ directly.
\end{proof}

Random dictatorship requires additive control: each camp's chance equals the sum of its members' individual chances. The proof of Theorem~\ref{thm:icv} subtracts the guaranteed peak masses and identifies the mean of the remaining lottery at three-camp profiles. Geometric separation rules out positive remaining mass in dimension three; three-position efficiency confines higher-dimensional outcomes to the same span. A circle deviation handles dimension two. Additive weights and peak guarantees then exhaust probability (Appendix~\ref{app:efficiency-proof}). If efficiency is required only at binary profiles, compromise becomes possible from dimension four. We next construct such rules.

The three-position requirement also has a local version. For a specified nested chain of $n-2$ camp divisions, efficiency along two-sided sequences of small out-of-plane perturbations of equilateral disagreement suffices to restore random dictatorship. The sequences can depend on the rule; the count concerns divisions, not individual profiles. Appendix~\ref{app:local-tests} states and proves this refinement.

\subsection{Truthful insurance}\label{subsec:compromise}
For three ideal directions summing to zero in a plane $L\subset\R^4$, equal-weight random dictatorship gives each participant alignments $1,-1/2$ with probabilities $1/3,2/3$. Uniform outcome risk on the unit circle in $L^\perp$ instead gives everyone certain zero alignment. Jensen's inequality gives insurance gains at the same mean. Orthogonal risk could also insure this particular profile in three dimensions, but this construction needs an orthogonal direction even at linearly independent reports to define one full-domain rule. Four dimensions provide it.

Fix odd $3\leq k\leq\min\{n,d-1\}$ and select a $k$-member committee uniformly, independently of reports. For its reports $t=(t_1,\ldots,t_k)$, we construct a mean in their span and add orthogonal risk to restore unit intensity. Let $\sigma_d$ be uniform spherical probability and $a_d=\int|e_1\cdot v|\,d\sigma_d(v)$. For the majority function $M_k$ on $k$ signs, with $\operatorname{sgn}(0)=1$, define
\begin{equation}\label{eq:majority-mean}
 g_k(t)=\frac1{a_d}\int_S v\,
 M_k(\operatorname{sgn}(t_1\cdot v),\ldots,\operatorname{sgn}(t_k\cdot v))\,d\sigma_d(v).
\end{equation}
The vector $g_k(t)$ lies in the unit ball and in $L_t=\operatorname{span}\{t_1,\ldots,t_k\}$. Let $Z_t$ be uniform on the unit sphere of $L_t^\perp$, which is nonzero since $k<d$, and set
\begin{equation}\label{eq:majority-lift}
 T_k(t)=\operatorname{Law}\bigl[g_k(t)+\sqrt{1-\|g_k(t)\|^2}\,Z_t\bigr].
\end{equation}
The outcome has unit norm and mean $g_k(t)$.

Averaging over the sampled committee gives
\begin{equation}\label{eq:triplet-population}
 P_{n,k}(r)=\binom nk^{-1}\sum_{J\subseteq N:\,|J|=k}T_k(r_J).
\end{equation}
A participant is pivotal in a directional vote only when the other $k-1$ signs tie. Truth then chooses the preferred sign, giving $p\cdot g_k(p,u)\geq p\cdot g_k(s,u)$. For a selected member, truthful alignment is certain conditional on the committee, whereas a deviation may expose the participant to alignment risk. Hence
\begin{equation}\label{eq:body-committee-ic}
 \mathbb E_{T_k(s,u)}\phi(p\cdot X)
 \leq\phi(p\cdot g_k(s,u))
 \leq\phi(p\cdot g_k(p,u))
 =\mathbb E_{T_k(p,u)}\phi(p\cdot X).
\end{equation}
Mixing over committees with report-independent probabilities preserves ICV-SP, although committee sampling can retain alignment risk. Appendix~\ref{app:compromise-proof} verifies feasibility and the endpoint case of Jensen's inequality.
\begin{proposition}\label{thm:icv-e2-counterexample}
For every finite $n\geq3$, $d\geq4$ and odd $3\leq k\leq\min\{n,d-1\}$, the rule $P_{n,k}$ is jointly Borel, anonymous and orthogonally neutral. It is ICV-SP on the full report domain and both ex-post and SD-Pareto efficient on $\mathcal E_2$, but is not fixed-weight random dictatorship.

At a balanced equilateral profile with $m$ participants per camp (so $n=3m$), let $C_i$ count camp-$i$ members in the sampled committee. Every participant's alignment has law
\begin{equation}\label{eq:population-insurance}
 (1-\beta_{m,k})\left(\tfrac13\delta_1+\tfrac23\delta_{-1/2}\right)
 +\beta_{m,k}\delta_0,\qquad
 \beta_{m,k}=\Pr\!\left(\max_i C_i\leq(k-1)/2\right)>0.
\end{equation}
Every participant weakly prefers this law to equal-weight random dictatorship for every permitted evaluation, strictly for strictly concave evaluations. Within this construction, complete insurance holds exactly when $k\geq2m+1$.
\end{proposition}

For approximate insurance, choose the largest odd $k\leq\min\{n,d-1\}$ and write $\delta_{m,k}=1-\beta_{m,k}$. The counting argument in Appendix~\ref{app:compromise-proof} gives, uniformly in $m$,
\begin{equation}\label{eq:uniform-near-insurance}
 \delta_{m,k}\leq\min\left\{1,\frac3{\sqrt2}\left(\frac89\right)^{(d-2)/2}\right\},
 \qquad \operatorname{Var}(p_i\cdot X)=\delta_{m,k}/2.
\end{equation}
Let $L_\phi=\phi(0)-[\phi(1)+2\phi(-1/2)]/3\geq0$ be the welfare gap between certain zero alignment and equal-weight random dictatorship. The scalar law gives
\begin{equation}\label{eq:near-insurance-welfare}
 \phi(0)-\mathbb E\phi(p_i\cdot X)=\delta_{m,k}L_\phi.
\end{equation}
Consequently dimension of order $\log(1/\varepsilon)$ suffices to keep every participant's welfare loss at most $\varepsilon L_\phi$, uniformly over populations $3m$ and permitted evaluations, with exact incentives.

\subsection{Control determines means and planar risk}\label{subsec:control-identification}
The committee construction establishes that truthful compromise is possible. Which restrictions on its means and risks are shared by every ICV-SP rule satisfying binary efficiency?
\Needspace*{3\baselineskip}
\begin{theorem}\label{thm:control-identification}
Let $d\geq4$ and $P$ be full-domain ICV-SP and efficient in either sense on $\mathcal E_2^\perp$. At every profile, its mean outcome is
\begin{equation}\label{eq:control-mean}
 m_P(r)=G_w(r):=\frac1{a_d}\int_S v
 \bigl[2w_{\{i:r_i\cdot v>0\}}-1\bigr]\,d\sigma_d(v).
\end{equation}
For three nonempty camps $A_j$ partitioning $N$ with reports $p=(p_1,p_2,p_3)$, let $g_3(p)$ be the mean in \eqref{eq:majority-mean}. If $\dim\operatorname{span}\{p_1,p_2,p_3\}\leq2$, each camp-$i$ participant's alignment law is
\begin{equation}\label{eq:planar-control-statement}
 \sum_{j=1}^3w_{A_j}\delta_{p_i\cdot p_j}
 +\left(1-\sum_jw_{A_j}\right)\delta_{p_i\cdot g_3(p)}.
\end{equation}
\end{theorem}

The mean formula aggregates directional contests: the positive-projection camp $A_v(r)=\{i:r_i\cdot v>0\}$ contributes $2w_{A_v(r)}-1$. Thus fixed control fixes every expected alignment, and at planar three-camp reports every alignment law. The identified objects are expected outcomes and, at these planar profiles, individual alignment distributions.

This distributional restriction is sharp for a natural control vector. For every $d\geq4$ and $n\geq3$, two anonymous, orthogonally neutral, Borel ICV-SP rules can share all binary weights and all-profile means while giving every participant different alignment variance at every three-camp profile with linearly independent directions. At planar triples, their reported alignment laws coincide. Appendix~\ref{app:rank-boundary} constructs one such pair of full-domain rules and proves its finite incentive inequalities. This separates the identification of means from that of individual risk.

At the following planar disagreements, unanimous weak preference over a given random dictatorship, for every permitted evaluation, is equivalent to preserving each participant's mean alignment. Fix three distinct planar directions whose triangle contains the origin in its relative interior, and positive $\lambda_i$ with $\sum_i\lambda_i=1$ and $\sum_i\lambda_i p_i=0$. Put $\rho_i=1-2\lambda_i$, $\kappa=1-\sum_jw_{A_j}$. Consider a fixed-weight random dictatorship with camp weights $q_i=\sum_{j\in A_i}\alpha_j$. All participants weakly prefer $P$ to this benchmark, for every permitted evaluation, exactly when
\begin{equation}\label{eq:camp-mean-match}
 q_i=w_{A_i}+\kappa\rho_i\quad(i=1,2,3).
\end{equation}
These equalities make each participant's alignment distribution a mean-preserving contraction of the benchmark distribution \citep{rothschildstiglitz1970}. If $\kappa>0$, every strictly concave evaluation gives a strict gain.
\begin{proof}
The positive relation and distinct unit directions give $0<\lambda_i<1/2$. Thus $\rho$ is a positive probability vector and $g:=p_1+p_2+p_3=\sum_i\rho_i p_i$ has norm below one. Outside null great spheres the three signs cannot all agree, so their majority is their sum and $g_3=g$.

Unanimous weak preference requires nonnegative mean gains. Their $\lambda$-weighted sum is zero, so every gain vanishes. Both means lie in the report plane; their projections identify the vector and affine independence identifies the weights, giving \eqref{eq:camp-mean-match}. The scalar law then gives welfare gain
\begin{equation}\label{eq:insurance-gain}
 \kappa\left[\phi(p_i\cdot g)-\sum_j\rho_j\phi(p_i\cdot p_j)\right]\geq0
\end{equation}
by finite Jensen, strictly when $\kappa>0$ and $\phi$ is strictly concave. If $\kappa=0$ the matched laws coincide.
\end{proof}

The proof of Theorem~\ref{thm:control-identification} starts from each participant's attainable mean menu: the means obtainable by changing that participant's report. Linear incentives make truthful means supporting points, and the menus must be compatible across participants. Binary weights identify boundary values when opponents report signed copies of one direction; concave comparisons control their approach. Appendix~\ref{app:general-mean} combines these traces with even and odd decompositions under report reversal to identify the mean almost everywhere. Finite incentive comparisons recover every actual profile. Appendix~\ref{app:planar-control} then eliminates residual alignment variation at planar triples. These arguments derive the needed regularity from incentives. The same control structure yields two diagnostics of random dictatorship: pairwise additivity in dimension four and differentiability of specified second-moment paths in every dimension at least four. Appendix~\ref{app:pair-control-proof} states and proves them.

\subsection{Dimension and unavoidable risk}\label{subsec:population-risk}
Suppose equally sized camps have equal binary winning probabilities: $w_A=\omega_{|A|}$, with $\omega_0=0$, $\omega_n=1$. Anonymity implies this condition, but the risk bounds require only these equalities. At balanced equilateral profiles, Theorem~\ref{thm:control-identification} gives alignment law $\omega_m\delta_1+2\omega_m\delta_{-1/2}+(1-3\omega_m)\delta_0$, with mean zero and variance $3\omega_m/2$. Complete insurance therefore asks whether a one-third camp's winning probability can vanish. The next proposition rules this out at fixed dimension when camps are sufficiently large. Define
\[
 b_d(n,s)=\tfrac12\exp\{-1000d^2[1+\log(n/s)]\},\qquad d-2\leq s\leq n.
\]
\Needspace*{3\baselineskip}
\begin{proposition}\label{prop:fixed-d-risk}
Fix $d\geq4$ and a finite population $n$. Let $P$ be full-domain ICV-SP and efficient in either sense on $\mathcal E_2^\perp$, with binary weights $w_A=\omega_{|A|}$. Then
\begin{equation}\label{eq:control-share-bound}
 \omega_s\geq b_d(n,s)\qquad(d-2\leq s\leq n).
\end{equation}
At any profile with disjoint camps of sizes $a,b\geq d-2$ reporting distinct directions $u,v$, every participant with report $p$ satisfies
\begin{equation}\label{eq:asymmetric-risk-bound}
 \operatorname{Var}(p\cdot X)\geq
 \frac{b_d(n,a)b_d(n,b)}{b_d(n,a)+b_d(n,b)}[p\cdot(u-v)]^2.
\end{equation}
The other participants may report arbitrarily. At balanced equilateral profiles with $n=3m$ and $m\geq d-2$, each participant satisfies
\begin{equation}\label{eq:fixed-d-risk}
 \operatorname{Var}(p_i\cdot X)\geq
 \max\{c_d,e^{-1000d^2}\},\qquad c_d=\tfrac34\eta_d^{2(d-1)}>0.
\end{equation}
Here $\eta_d$ is the geometry-dependent factor in \eqref{eq:capacity-recurrence}.
For such a balanced profile $r^*$ and any $0<c<e^{-1000d^2}$, arbitrary individual report perturbations satisfy
\begin{equation}\label{eq:dispersed-risk}
 \max_i\|r_i-r_i^*\|\leq
 \frac{(1-ce^{1000d^2})^{3/2}}{128m}
 \quad\Longrightarrow\quad
 \operatorname{Var}_{P(r)}(r_i\cdot X)>c\quad(i\in N).
\end{equation}
\end{proposition}

The two camp guarantees give simultaneous masses at least $\omega_a,\omega_b$ at $u,v$. For $\mu=\mathbb E[p\cdot X]$,
\[
 \operatorname{Var}(p\cdot X)\geq
 \omega_a(p\cdot u-\mu)^2+\omega_b(p\cdot v-\mu)^2
 \geq\frac{\omega_a\omega_b}{\omega_a+\omega_b}[p\cdot(u-v)]^2.
\]
The coefficient increases in both weights. To bound them, Appendix~\ref{app:fixed-d-risk} surrounds a central camp with regular-simplex groups: if control falls too far with camp size, the residual mean and support become incompatible. Iterating the resulting recurrence gives the bounds. For \eqref{eq:dispersed-risk}, incentives retain probability near each camp's ideal as reports disperse, without continuity of the rule. The sufficient per-person radius shrinks with population. These conservative bounds establish a positive risk floor; they do not quantify the smallest attainable risk.

If two camps each have at least $d-2$ members and population share at least $\alpha>0$, every member of either camp has variance at least $b(d,\alpha)(1-u\cdot v)^2/2$, where $b(d,\alpha)=\tfrac12e^{-1000d^2[1+\log(1/\alpha)]}$. The bound is uniform over camp directions whose separation is bounded away from zero, with arbitrary outsiders and no equilateral or planar restriction.

These bounds imply that the dimension needed for complete insurance grows linearly with population. Let $D_{\min}(m)$ be the least dimension allowing a full-domain ICV-SP rule with either efficiency on $\mathcal E_2^\perp$ and $w_A=\omega_{|A|}$ to completely insure a reference balanced equilateral profile of $3m$ participants. Theorem~\ref{thm:icv} and equal singleton weights force equal-weight random dictatorship in dimensions two and three. Proposition~\ref{prop:fixed-d-risk} excludes $4\leq d\leq m+2$, and the committee construction gives
\begin{equation}\label{eq:insurance-dimension-bracket}
 m+3\leq D_{\min}(m)\leq2m+2\quad(m\geq1).
\end{equation}
Hence $D_{\min}(1)=4$. By contrast, \eqref{eq:near-insurance-welfare} achieves any fixed proportional welfare tolerance at these profiles in a population-independent dimension.

\section{Insurance at every profile under fixed control}\label{app:all-profile-insurance}

The preceding bounds concern risk at particular disagreements. We now hold binary control fixed and ask whether one rule can offer the best residual insurance at every profile.

Fix $n\geq3$, $d\geq4$ and a feasible binary control vector $w=(w_A)_{A\subseteq N}$. Let $\mathcal F(w)$ contain all full-domain ICV-SP rules efficient in either sense on $\mathcal E_2^\perp$ with binary weights $w$. Feasibility means $\mathcal F(w)\ne\varnothing$. Competitors need not be anonymous, neutral, measurable or continuous across reports. A \emph{self-dual simple game} is a monotone map $v:2^N\to\{0,1\}$ with $v_\varnothing=0$ and $v_{A^c}=1-v_A$. Thus each two-camp contest has a determined winning camp. A dictatorial simple game has $v_A=\mathbf1_{\{j\in A\}}$ for one fixed label $j$.

Write $s_i=w_{\{i\}}$ and $\kappa=1-\sum_i s_i\geq0$. The $s_i$ are guaranteed individual chances of control. If $\kappa>0$, define the residual control vector
\begin{equation}\label{eq:residual-control}
 \zeta_A=\frac{w_A-\sum_{i\in A}s_i}{\kappa}.
\end{equation}
Use \eqref{eq:control-mean} to define $G_v$ for any normalized monotone complementary vector $v$, including $\zeta$ as verified below, whether or not it is separately feasible.

\Needspace*{3\baselineskip}
\begin{proposition}\label{prop:all-profile-insurance}
Suppose $\kappa>0$. At every actual profile $r$, for every participant $i$, every $\phi\in\Phi$ and every $P\in\mathcal F(w)$,
\begin{equation}\label{eq:global-jensen-bound}
 \mathbb E_{P(r)}\phi(r_i\cdot X)
 \leq\sum_j s_j\phi(r_i\cdot r_j)
       +\kappa\phi(r_i\cdot G_\zeta(r)).
\end{equation}
There exists a single $F_w^*\in\mathcal F(w)$ attaining these bounds simultaneously for every $r,i,\phi$ if and only if $\zeta$ is $\{0,1\}$-valued and $d\geq n+1$.
If $\kappa=0$, every rule in $\mathcal F(w)$ is the fixed-weight random dictatorship with weights $s_i$.
\end{proposition}

The bound separates the risk attached to guaranteed individual control from the remaining alignment risk. Attaining it everywhere requires a certain winner in each residual binary contest and a direction orthogonal to all reports, even when the reports are independent.

\subsection{A common welfare bound}
\begin{proof}[Proof of the bound in Proposition~\ref{prop:all-profile-insurance}]

Superadditivity and complementarity imply
\[
 \sum_{i\in A}s_i\leq w_A\leq1-\sum_{i\notin A}s_i.
\]
Consequently $\zeta$ in \eqref{eq:residual-control} is valued in $[0,1]$, normalized, complementary and superadditive on disjoint sets. It is monotone, and $\zeta_{\{i\}}=0$ for every $i$.

At a profile with colliding reports, group labels into maximal camps $A_q=\{j:r_j=q\}$. The peak guarantee gives
\[
 P(r)\{q\}\geq w_{A_q}\geq\sum_{j\in A_q}s_j.
\]
Hence $R_r=P(r)-\sum_j s_j\delta_{r_j}$ is an actual nonnegative measure of mass $\kappa$. Linearity of the sign integral and
$a_d^{-1}\int_S v\operatorname{sgn}(p\cdot v)\,d\sigma_d(v)=p$
give
\begin{equation}\label{eq:residual-mean-identity}
 G_w(r)=\sum_j s_jr_j+\kappa G_\zeta(r),
 \qquad \int_S x\,dR_r(x)=\kappa G_\zeta(r).
\end{equation}
Jensen's inequality for $R_r/\kappa$ proves \eqref{eq:global-jensen-bound}. For an interior scalar mean, a finite concave function has a supporting affine upper bound, even if its lower endpoint has a downward jump. At an endpoint mean the scalar is constant. Thus Jensen covers the entire permitted class. For strictly concave $\phi$, equality holds precisely when that participant's residual alignment is constant. No incentive property of $R_r/\kappa$ has been assumed. If $\kappa=0$, the guaranteed atoms already exhaust probability at every profile, proving the final assertion.
\end{proof}

\subsection{Attainment for self-dual simple games}
\begin{proof}[Proof of sufficiency in Proposition~\ref{prop:all-profile-insurance}]

Let $v$ be any self-dual simple game and assume $d\geq n+1$. Its sign-integral mean $G_v$ is continuous: outside the finitely many great spheres $r_i\cdot z=0$, the signs stabilize along every converging report sequence, so dominated convergence applies. For every unit $u$,
\[
 |u\cdot G_v(r)|\leq a_d^{-1}\int_S|u\cdot z|\,d\sigma_d(z)=1.
\]
Reflection through the report span $L_r=\operatorname{span}\{r_1,\ldots,r_n\}$ leaves the signs unchanged and reverses the normal component of the integral. Thus $G_v(r)\in L_r$ and $\|G_v(r)\|\leq1$.

Let $Z_r$ be uniform on the unit sphere of $L_r^\perp$ and define the actual lottery
\begin{equation}\label{eq:boolean-lift}
 L_v(r)=\operatorname{Law}\left[
 G_v(r)+\sqrt{1-\|G_v(r)\|^2}\,Z_r\right].
\end{equation}
The normal space is nonzero at every profile since $\dim L_r\leq n<d$. To obtain one Borel construction across changes of rank, form the $d\times n$ report matrix $A(r)$ and the projector $\Pi_r=I-A(r)A(r)^\dagger$, where the Moore--Penrose inverse is Borel (continuous on each fixed-rank stratum). If $H$ is a standard Gaussian, $\Pi_rH/\|\Pi_rH\|$ has the required uniform law. On the null event $\Pi_rH=0$, use the normalized first nonzero vector among $\Pi_re_1,\ldots,\Pi_re_d$. This defines a jointly Borel realization. Orthogonality makes its norm exactly one, its mean $G_v(r)$, and each truthful alignment equal to $r_i\cdot G_v(r)$ almost surely. All participants' reports enter $L_r$, even those of labels whose votes do not affect $v$.

Monotonicity gives the finite incentive comparisons. Fix the opponents and an integration direction $z$ away from their null sign boundaries. Put $B=\{j\ne i:r_j\cdot z>0\}$. Label $i$ can choose only between $v_B$ and $v_{B\cup\{i\}}$, with $v_B\leq v_{B\cup\{i\}}$. If its true direction $p$ has $p\cdot z>0$, truthful reporting chooses the larger value; if $p\cdot z<0$, it chooses the smaller. Multiplying by $p\cdot z$ and integrating yields
\[
 p\cdot G_v(p,r_{-i})\geq p\cdot G_v(s,r_{-i})
 \quad(s\in S).
\]
Truthful alignment under \eqref{eq:boolean-lift} is constant. At a deviating profile Jensen and monotonicity therefore imply
\[
 \mathbb E_{L_v(s,r_{-i})}\phi(p\cdot X)
 \leq\phi(p\cdot G_v(s,r_{-i}))
 \leq\phi(p\cdot G_v(p,r_{-i})).
\]
The endpoint argument above again applies. This proves full-domain ICV-SP, including changes that alter the report span.

At a binary profile $(a^A,b^{A^c})$, if $v_A=1$, the sign integrand is $\operatorname{sgn}(a\cdot z)$: when the two signs disagree the winning camp is $A$, and when they agree normalization determines the value. If $v_A=0$ it is $\operatorname{sgn}(b\cdot z)$. Thus $G_v$ is the winning ideal and $L_v$ is its Dirac law, including collisions and antipodes. It satisfies both efficiencies at every binary profile.

Apply this construction to $v=\zeta$. The fixed mixture
\begin{equation}\label{eq:attaining-mixture}
 F_w^*(r)=\sum_j s_j\delta_{r_j}+\kappa L_\zeta(r)
\end{equation}
is ICV-SP, has binary weights $w$, and is efficient at binary profiles. At distinct binary reports, any SD-dominating lottery must put at least the original masses on both ideals, by the upper-tail tests at alignment one. Those masses already exhaust probability, proving SD efficiency; unanimity is immediate. Every residual alignment is constant, so it attains \eqref{eq:global-jensen-bound} simultaneously. This constructs one rule attaining every bound in the full comparison class.
\end{proof}

\subsection{Necessity of the control and dimension conditions}
\begin{proof}[Proof of necessity in Proposition~\ref{prop:all-profile-insurance}]

First, if some $\zeta_A\in(0,1)$, then $A$ is nonempty and proper. Take a binary profile with distinct reports $a,b$. Its normalized residual is exactly $\zeta_A\delta_a+(1-\zeta_A)\delta_b$. An $a$-reporter's residual alignment takes distinct values $1$ and $a\cdot b$ with positive probabilities. Strict Jensen for the permitted function $\phi(u)=u-u^2/2$ rules out equality in \eqref{eq:global-jensen-bound}. Simultaneous attainment therefore requires $\zeta$ to be $\{0,1\}$-valued.

We next establish the dimension obstruction for any nondictatorial self-dual simple game $v$. Such a rule has no winning singleton: a winning singleton and complementarity would make that label a dictator. Choose a minimal winning coalition $A$. It has at least two members and is proper, since each $(n-1)$-member coalition wins. Split $A$ into two nonempty proper subsets $B,C$, and put $D=A^c$. All three camps lose separately; their pairwise unions win by complementarity. The induced three-input rule is majority.

Give these camps orthonormal reports $e_1,e_2,e_3$. For their signs $t_j$, majority is $(t_1+t_2+t_3-t_1t_2t_3)/2$. For each vector component, reflection in a different one of the first three coordinates makes the sign-product integral zero, while $a_d^{-1}\int z\operatorname{sgn}(e_j\cdot z)\,d\sigma_d=e_j$. Hence
\[
 G_v(r^0)=\tfrac12(e_1+e_2+e_3),\qquad \|G_v(r^0)\|^2=3/4.
\]
If $d\leq n$, keep one original agent in each camp fixed and choose $d-3$ other labels. Perturb their reports toward the independent directions $e_4,\ldots,e_d$, normalizing to unit length. For every nonzero sufficiently small perturbation, the reports span $\R^d$, whereas continuity of $G_v$ keeps its norm below one.

At any such profile, certainty of every participant's alignment would imply $r_i\cdot(X-G_v(r))=0$ almost surely for all $i$. Full rank then forces $X=G_v(r)$ almost surely, contradicting $X\in S$. This uses continuity of the explicit mean only, not regularity of any lottery law.

For simultaneous attainment in Proposition~\ref{prop:all-profile-insurance}, equality for the strictly concave evaluation forces every residual alignment to be constant at every profile. Apply the same argument to the actual residual probability, whose mean is $G_\zeta$. Its self-dual simple game has zero singleton weights and cannot be a dictator. Consequently $d\geq n+1$ is necessary, completing the proof.
\end{proof}

Complete insurance at every profile follows as a consequence. There exists one $P\in\mathcal F(w)$ making every participant's alignment constant at every profile if and only if $w$ is $\{0,1\}$-valued and either $d\geq n+1$ or $w$ is dictatorial. A fractional binary weight immediately precludes certainty at a binary disagreement. For a nondictatorial self-dual simple game $w$, the rank argument gives necessity of $d\geq n+1$, and $L_w$ gives sufficiency. A dictator is deterministic in every dimension covered here. The same mechanism insures all participants, including those whose reports do not affect its control rule.

\section{Optimal insurance by committees}\label{sec:committee-optimality}
The majority construction uses report-independent delegation to committees whose selected members are fully insured. We now optimize within this class: can another voting rule or a mixture of committee sizes reduce risk at balanced equilateral disagreement?

A report-independent delegation rule first draws $\theta$ from a fixed probability space $(T,\mu)$, independently of reports. Each draw specifies a nonempty set $J_\theta\subseteq N$ and a component rule $Q_\theta$ on all reports of those participants. Each component is full-domain ICV-SP and efficient in either sense on all of $\mathcal E_2(J_\theta)$, including unanimity. Every member of $J_\theta$ has constant alignment at every profile under that component; reports outside $J_\theta$ are ignored. The mixture
\[
 P(r)=\int_T Q_\theta(r_{J_\theta})\,d\mu(\theta)
\]
is assumed well-defined, with measurable probability evaluations in the mixing variable, and anonymous. Components need not be anonymous or neutral. Thus component insurance is required at every profile, while the risk comparison concerns balanced equilateral disagreement.

For $n=3m$, let $C_k$ have the hypergeometric distribution obtained by drawing $k$ labels without replacement from $m$ marked and $2m$ unmarked labels. Define
\begin{equation}\label{eq:committee-control-tail}
 p_k=\Pr(C_k>k/2)+\tfrac12\Pr(C_k=k/2),\qquad
 k_* =\max\{k\leq\min(n,d-1):k\text{ is odd}\}.
\end{equation}
The tie term is zero for odd $k$.

\begin{proposition}\label{prop:committee-optimality}
Fix $d\geq4$ and $n=3m$. Among the anonymous report-independent delegation rules just defined, the smallest variance of each participant's alignment at balanced equilateral disagreement is
\[
 \frac32p_{k_*}.
\]
The uniform $k_*$-member majority rule $P_{n,k_*}$ attains this minimum and maximizes each participant's expected evaluation for every permitted increasing-concave function at these profiles. Complete insurance is attainable within this class if and only if $d\geq2m+2$.
\end{proposition}

The optimum compares arbitrary component voting rules and mixtures of different committee sizes. It is not the minimum over all ICV-SP mechanisms. In particular, an improvement on the dimension $2m+2$ must leave this class: it cannot arise solely by replacing majority or changing the report-independent mixture of fully insured committees.

\begin{proof}
Binary identification and constant component alignments make each component's control a self-dual simple game $v$ on its participating set $J$. A fractional weight at two distinct reports would give a participant two different alignments. A nondictatorial component with $k=|J|\geq3$ must have $k\leq d-1$, by the all-profile insurance characterization in Section~\ref{app:all-profile-insurance}. If its control is dictatorial, mean identification gives the dictator's unit ideal at every profile and hence a deterministic dictatorship. It can be represented on one label. Components with one or two labels are also dictators, by unanimity and the binary law.

Average a component's winning probability over the physical $m$-subsets $A\subset N$. A particular $B\subseteq J$ occurs as $A\cap J$ in $\binom{n-k}{m-|B|}$ such subsets. The average control is therefore
\[
 p(v,J)=\sum_{B\subseteq J}v(B)q_{|B|},\qquad
 q_s=\frac{\binom{n-k}{m-s}}{\binom nm},
\]
where binomial coefficients outside their integer ranges are zero. Pair $B$ with $J\setminus B$. Complementarity makes the pair contribute either $q_s$ or $q_{k-s}$. For $s\leq k/2$, one has $q_s\geq q_{k-s}$. If $q_{k-s}>0$, put $M=n-k$, $a=m-k+s$, and $b=m-s$. Then $0\leq a\leq b$, $a+b=2m-k\leq M$, and $b\leq M$. Symmetry and unimodality of binomial coefficients give $\binom Mb\geq\binom Ma$. If $q_{k-s}=0$ the comparison is immediate.

Each complementary pair thus costs least when its larger coalition wins. Equal-sized complements have equal costs. Summing gives $p(v,J)\geq p_k$, with equality for odd majority, or for any complementary tie-breaking rule when $k$ is even. Such tie-breaking rules are monotone.

The sequence $p_k$ is nonincreasing. First, $p_{2l}=p_{2l-1}$: deleting a uniformly chosen label from a uniform $2l$-subset preserves its majority off a tie, and converts a tie into a marked majority with probability $1/2$. Next let $k=2l+1$ and $k+2\leq n$, and draw two additional labels. A marked majority is gained only from $C_k=l$ followed by two marked labels, and lost only from $C_k=l+1$ followed by two unmarked labels. For $l<m$, the hypergeometric ratio is
\[
 \frac{\Pr(C_k=l+1)}{\Pr(C_k=l)}=\frac{m-l}{2m-l},
\]
and hence
\begin{equation}\label{eq:committee-tail-descent}
 p_{k+2}-p_k
 =-\frac{m(m-l)\Pr(C_k=l)}{(n-k)(n-k-1)}\leq0.
\end{equation}
If $l\geq m$, both majority probabilities are zero. At $l=m$, $\Pr(C_k=m)$ need not vanish, but neither the gain nor the loss transition can occur. These cases establish the monotonicity including the zero plateau.

Every nondictatorial component has $k\leq\min(n,d-1)$ and contributes at least $p_{k_*}$; an even maximum size has the same tail as the preceding odd size. A dictator contributes $p_1=1/3\geq p_{k_*}$. Anonymity of the mixture identifies its common $m$-camp control with its finite label average, so
\[
 \omega_m=\int_T p(v_\theta,J_\theta)\,d\mu(\theta)\geq p_{k_*}.
\]
Interchanging the finite average and the well-defined mixture is valid. Theorem~\ref{thm:control-identification} gives the actual alignment law
\[
 \omega_m\delta_1+2\omega_m\delta_{-1/2}+(1-3\omega_m)\delta_0.
\]
Its variance is $3\omega_m/2$. Concavity implies $\phi(1)+2\phi(-1/2)\leq3\phi(0)$, so lowering $\omega_m$ also weakly raises every permitted expected evaluation. The majority construction attains $\omega_m=p_{k_*}$.

For odd $k$, $p_k=0$ exactly when $k\geq2m+1$: below that threshold an $m$-camp can supply a strict majority, with positive sampling probability. Thus zero risk is possible exactly when $d\geq2m+2$. At that dimension, every selected $(2m+1)$-committee contains all three camps, since any two camps together have only $2m$ labels. None can have a majority, its mean is zero, and its orthogonal outcome risk also insures the unselected participants. This proves attainability for the entire population.
\end{proof}

\section{Mean-preserving insurance and incentives}\label{app:insurance-error}

The preceding sections maintain exact incentives and vary dimension or control. We now fix dimension and allow an incentive error, while requiring complete insurance that preserves each participant's mean under equal-weight random dictatorship. At balanced equilateral disagreement these means are zero.

Our error criterion is uniform over all opponents' reports and normalized increasing-concave evaluations. It differs from the interim criterion of \citet{azevedobudish2019}, which evaluates approximate incentives under every full-support i.i.d. distribution of opponents' reports.

Let $\mathcal U$ be the nondecreasing concave functions $\phi:[-1,1]\to[0,1]$ with $\phi(-1)=0$ and $\phi(1)=1$, including the permitted jump at $-1$. Every nonconstant evaluation in $\Phi$ can be normalized this way by a positive affine transformation. For a full-domain assignment $P$ of actual Borel sphere probabilities, define its incentive error, the worst-case gain from misreporting, by
\[
 I(P)=\sup_{i,p,s,r_{-i},\phi\in\mathcal U}
 \left\{\int\phi(p\cdot x)\,dP(s,r_{-i})(x)
       -\int\phi(p\cdot x)\,dP(p,r_{-i})(x)\right\}.
\]
A rule is $\varepsilon$-strategy-proof for this normalized utility class if $I(P)\leq\varepsilon$. The option $s=p$ gives $I(P)\geq0$. Without normalization, any profitable gain could be scaled arbitrarily.

For $r\in\mathcal E_2^\perp$, let $C(r)$ be its efficient singleton at unanimity or its closed short arc at orthogonal binary disagreement, and put
\[
 \Delta(P)=\sup_{r\in\mathcal E_2^\perp}
       \int\operatorname{dist}(x,C(r))\,dP(r)(x).
\]
Distance is Euclidean. Thus $\Delta=0$ is exact ex-post efficiency on the binary tests. For $n=3m$, fix unit directions $(p_1,p_2,p_3)$ with $p_1+p_2+p_3=0$ and write $r_m=(p_1^m,p_2^m,p_3^m)$. Define
\[
 R(P,r)=\frac1n\sum_i\int(r_i\cdot x)^2\,dP(r)(x),\qquad
 \mathcal L(P,r)=\frac1{n!}\sum_{\pi\in S_n}R(P,r_\pi).
\]
Here $(r_\pi)_i=r_{\pi(i)}$. Zero $\mathcal L(P,r_m)$ means zero alignment almost surely for every participant at every label assignment of this fixed triple. Anonymity reduces the all-labelings check to $r_m$; the required constant alignment is still zero. Without this restriction, choosing a fixed participant's report deterministically gives complete insurance and $I=\Delta=0$, although $\mathcal L(P,r_m)=1/2$.

Let $\mathcal H_{n,d}$ contain all full-domain assignments of Borel probabilities on $S$, without report measurability, continuity or symmetry, and set
\[
 e_{n,d}(\delta)=\inf\{I(P):P\in\mathcal H_{n,d},\ 
          \Delta(P)\leq\delta,\ \mathcal L(P,r_m)=0\}.
\]

\Needspace*{3\baselineskip}
\begin{proposition}\label{prop:insurance-error}
For every fixed $d\geq4$ there are constants $b_d>0$ and $\bar\delta_d>0$, independent of population, such that
\begin{equation}\label{eq:insurance-error-order}
 \frac{b_d}{n}\leq e_{n,d}(\delta)\leq\min\{1,100/n\},
 \qquad n=3m,\quad m\geq d-1,\quad 0\leq\delta\leq\bar\delta_d.
\end{equation}
The upper bound uses a weakly continuous, anonymous, orthogonally neutral rule satisfying both efficiencies on all of $\mathcal E_2$ and giving every participant zero alignment almost surely at every balanced equilateral profile. Thus the same order holds if these additional restrictions are imposed.
\end{proposition}

The lower constant and efficiency allowance depend on dimension and are nonexplicit. The proof combines finite incentive comparisons with compactness of sphere probabilities. Approximate binary weights must remain close to the controls of exact small-population rules; grouping physical participants then makes the relevant error scale with $nI(P)$. The upper bound blends equal-weight random dictatorship with an orthogonal insurance rule through a continuous spectral weight. Appendix~\ref{app:insurance-error-proof} gives both arguments, including the endpoint behavior of the normalized evaluations.

\Needspace*{3\baselineskip}
\section{Conclusion}
Fixed intensity prevents the institution from insuring participants by shrinking the public action. Incentives nevertheless permit insurance when outcome risk can be placed orthogonally to reported ideals. Binary control governs both this opportunity and its limits: it fixes every mean and, at planar three-camp disagreement, every individual alignment law.

The design implications depend on the insurance target. Under size-dependent binary control, eliminating risk at balanced equilateral disagreement requires dimension to grow with population; reducing every participant's welfare loss to a fixed proportion of the random-dictatorship benchmark does not. A uniform guarantee across all profiles imposes an additional restriction on residual control. In fixed dimension, a vanishing incentive error provides a further route to mean-preserving insurance at balanced disagreement. These distinctions connect the institution's control rights, available directions and incentive commitment to the insurance it can provide.

\appendix
\section*{Appendix}
\addcontentsline{toc}{section}{Appendix}
Appendix~\ref{app:tools} collects the incentive, menu and analytic tools. Appendix~\ref{app:efficiency-proof} develops the three-camp geometry and proves Theorem~\ref{thm:icv} and its local efficiency refinement. Appendix~\ref{app:compromise-proof} verifies the committee construction. Appendix~\ref{app:uniform-risk-gap} identifies means and planar risk, derives the risk bounds, and gives the control diagnostics. Appendix~\ref{app:insurance-error-proof} proves Proposition~\ref{prop:insurance-error}, and Appendix~\ref{app:rank-boundary} constructs the nonplanar counterexample. Throughout, capped-linear comparisons use $\phi_t(u)=\min(u,t)=t-(t-u)_+$, so expected shortfalls enter with reversed inequalities.
\section{Incentive and analytic tools}\label{app:tools}
\subsection{Incentive-compatible means}\label{app:plane}
Let $B$ be the closed unit ball. A binary map $F:S\times S\to B$ is linearly incentive compatible (IC) in both coordinates if
\[
 p\cdot F(p,q)\geq p\cdot F(s,q),\qquad
 q\cdot F(p,q)\geq q\cdot F(p,s)\quad(p,q,s\in S).
\]
Its diagonal is $F(p,p)=p$. Neither measurability nor continuity is assumed. The next lemma is the unit-disk specialization of \citet[Lemma~3 and Appendix~B]{chenwang2026}. Its proof supplies the continuity and menu decomposition used in the binary-law argument below.

A mean menu is the closed convex hull of the values obtained by varying one report. For a compact convex set $C$, write $h_C(u)=\max_{x\in C}u\cdot x$ for its support function and $E_C(u)=\{x\in C:u\cdot x=h_C(u)\}$ for its exposed face. A Minkowski summand of $B$ is a convex set $C$ for which $C+D=B$ for some convex set $D$, where $C+D=\{x+y:x\in C,y\in D\}$.

\begin{lemma}\label{lem:planar-body}
For $d=2$, $F$ is jointly continuous and one-Lipschitz in either angular coordinate. On each lifted strip $0<\pm(a-b)<\pi$,
\begin{equation}\label{eq:disk-strip}
 F(e(a),e(b))=A_\pm(a)+B_\pm(b),\quad
 A'_\pm=t\ell_\pm,\quad B'_\pm=t(1-\ell_\pm),\quad0\leq\ell_\pm\leq1,
\end{equation}
where $e(a)=(\cos a,\sin a)$ and $t=e'$. Every convexified row or column menu is a Minkowski summand of the disk.
\end{lemma}
\begin{proof}
For a one-coordinate increment $0<u<\pi$, IC puts its output change $\Delta$ in the cone
\[
 t(a+u/2)\cdot\Delta\geq0,\qquad
 |e(a+u/2)\cdot\Delta|\leq\tan(u/2)t(a+u/2)\cdot\Delta.
\]
At a nonparallel pair choose two independent linear functionals with the same strict signs on both input tangents. Orient each coordinate by those signs. On a small box the cone makes both scalarizations coordinatewise nondecreasing. A bounded coordinatewise monotone function is Riemann integrable: on an $m\times m$ grid its upper-minus-lower Darboux sum is at most $2|R|\operatorname{osc}(f)/m$. Thus $F$ is jointly locally integrable on the two strips in the completed Lebesgue sense.

For fixed $b$, $c_b(a)=e(a)\cdot F(e(a),e(b))$ is the support function of the closed convex hull of its column. The support inequalities give $c'_b=t\cdot F$ almost everywhere. Fubini and integration by parts imply $e(a)\cdot\partial_aF=e(b)\cdot\partial_bF=0$ distributionally. Differentiating transversely and inverting the two nonparallel directions yields $\partial_a\partial_bF=0$. Hence $F=A(a)+B(b)$ jointly almost everywhere on each strip. To see the global decomposition, on a box a distribution with zero derivative in the second coordinate depends only on the first coordinate: subtract a transverse mass-one test and take a compactly supported transverse primitive. Overlapping boxes glue the derivatives along the interval fibers of a strip; take their one-dimensional primitives and apply the same argument to the remainder.

In distributional derivatives, orthogonality gives $DA=t\alpha$, $DB=t\beta$. These scalar distributions are positive. For fixed $u$, almost every $a$ admits a $b$ for which both additive values at $(a,b),(a+u,b)$ are valid, since the common transverse interval has length $\pi-u$. The cone inequality gives $t(a+u/2)\cdot[A(a+u)-A(a)]\geq0$. Multiply by a nonnegative test, integrate, divide by $u$, and send $u$ to zero. The smooth test quotient converges uniformly on a compact set, proving $\alpha\geq0$. The second coordinate gives $\beta\geq0$. Positive distributions are locally finite measures (positivity bounds a test by its sup norm times a fixed dominating cutoff), so $A,B$ are locally BV.

Reporting the other type gives
\[
 0\leq1-e(a)\cdot F(e(a),e(b))\leq1-e(a)\cdot e(b).
\]
Together with $F\in B$ this implies uniform convergence to $e(a)$ at the diagonal. Pull back the joint additive null set by $(r,\varepsilon)\mapsto(r+\varepsilon,r)$, of Jacobian one. Fubini supplies a sequence of valid offsets tending to zero. Translation continuity in $L^1_{\rm loc}$ then gives $A+B=e$. Consequently $\alpha+\beta=da$ and both have densities in $[0,1]$. Their primitives are continuous.

On the same small oriented box, each strictly lower or upper rectangle has positive area and contains a point outside the joint exceptional set. Both independent scalarizations bracket the actual target by the values at these points. Continuity of the additive representatives identifies both scalarizations in the limit, hence the vector value. Thus the additive representation holds at every such value.

At the antipodal seam write $L_+(b)=A_+(b+\pi)+B_+(b)$ and $L_-(b)=A_-(b-\pi)+B_-(b)$. Both limits expose the same column menu in direction $-e(b)$, so $e(b)\cdot J(b)=0$, where $J=L_+-L_-$. Their derivatives are tangential: $DJ=t\gamma$ for a signed measure. The product rule gives $0=D(e\cdot J)=(t\cdot J)db$ and hence $J=0$ by continuity. The resulting continuous column selection $g_b$ has support $c_b$, with $c'_b=t\cdot g_b$ everywhere by two-sided support inequalities. Any exposed point has these same normal and tangent coordinates, so is $g_b$. The actual seam selection therefore equals its trace.

Columns have derivative $t\lambda_b\,da$, $0\leq\lambda_b\leq1$, including across the seams, and are one-Lipschitz. Rows obey the same argument. Coordinatewise telescoping gives joint continuity.

Finally $d_b=1-c_b$ satisfies $d_b''+d_b=1-\lambda_b\geq0$. The continuous curve $z_b=d_be+d_b't$ has derivative $t(1-\lambda_b)$. Integrating along the shorter oriented arc gives $e(\beta)\cdot z_b(\alpha)\leq d_b(\beta)$ for all angles, with equality at $\alpha=\beta$. Its convex hull has support $d_b$, furnishing a complement to the column menu. The row assertion follows with labels exchanged.
\end{proof}

\begin{lemma}\label{lem:spherical-mean}
For every $d\geq2$, a binary map with the stated IC and diagonal properties is jointly continuous. Every row and column satisfies
\[
 \|F(p,q)-F(p',q)\|\leq\frac\pi2\|p-p'\|.
\]
\end{lemma}
\begin{proof}
The disk lemma proves $d=2$. For $d\geq3$, we specialize the joint-envelope argument of \citet[Appendices~C--D]{chenwang2026}. The planar lemma first controls projections of the mean menus; support-function bounds then recover the ambient mean and its continuity. Extend $F$ with degree zero in each positive radial variable to $\Omega=(\R^d\setminus\{0\})^2$. Put
\[
 U(P,Q)=P\cdot F(P,Q)=h_{C_Q}(P),\qquad
 V(P,Q)=Q\cdot F(P,Q)=h_{R_P}(Q),
\]
where $C_Q,R_P$ are the closed convexified menus. Restrict both reports to any two-plane containing $Q$ and project outcomes into its disk. The disk lemma applies; the projected column menu equals $\Pi_L C_Q$ because truthful reports in $L$ maximize the whole ambient menu. Its exposed faces are singletons. For independent $P,Q$, differences within $E_{C_Q}(P)$ are therefore orthogonal to $Q$. For parallel $P,Q$, use every plane through their axis to obtain a singleton ambient face. Rows have the exchanged property.

If $Q_j\to Q$, each cluster point of $F(P,Q_j)$ lies in $E_{R_P}(Q)$ by the row support inequalities. The preceding property makes its $P$-inner product constant. Thus $U(P,Q_j)\to U(P,Q)$. Its uniform Lipschitz bound in $P$ gives joint continuity of $U$; similarly for $V$. Convex differentiability in $P$ has a jointly Borel exception set, describable by rational difference quotients, with null sections. Fubini makes it jointly null. Off this set $F=\nabla_P U$. Hence the actual $F$ is completed-Lebesgue measurable and locally bounded.

In fixed Euclidean coordinates set $A=D_PF=D^2_{PP}U$ and $D=D_QF=D^2_{QQ}V$. They are symmetric PSD matrix measures: each constant-vector scalarization of a convex Hessian is a positive distribution; polarization gives the entries. Their traces are locally finite by integration by parts against cutoffs and boundedness of $F$. Mixed derivative symmetry makes $\partial_{Q_k}\partial_{P_j}F_i$ symmetric in all three indices. Differentiating radial homogeneity consequently gives
\[
 (P\cdot\nabla_Q)A=0.
\]
On the independent-pair set $\mathcal O$, its flow is the shear $(P,Q)\mapsto(P,Q+tP)$, of Jacobian one. Scalar componentwise distribution pullback preserves $A$ along that flow. Homogeneity also preserves $A$ under $Q\mapsto\varepsilon Q$. Thus $\Phi_\varepsilon^*A=A$ for $\Phi_\varepsilon(P,Q)=(P,P+\varepsilon Q)$; explicitly
\[
 \langle\Phi_\varepsilon^*M,\varphi\rangle
 =\langle M,\varepsilon^{-d}\varphi(P,(Q-P)/\varepsilon)\rangle.
\]
The inverse Jacobian is retained and positive matrix measures remain positive under this scalar pullback.

The diagonal gap, $\|F\|\leq1$, and $\widehat{P+\varepsilon Q}\to\widehat P$ imply locally uniform $F(P,P+\varepsilon Q)\to P/\|P\|$. The weak chain rule gives
\[
 D_PF(P,P+\varepsilon Q)=A+\Phi_\varepsilon^*D.
\]
Taking weak limits yields $0\preceq A\preceq H(P)\,dP\,dQ$ on $\mathcal O$, where $H=D^2\|P\|=(I-\widehat P\widehat P^\top)/\|P\|$.

The matrix measure assigns zero mass to parallel pairs. On a fixed compact neighborhood write $a=(P\cdot Q)/\|P\|^2$, $W=Q-aP$. A smooth cutoff $\chi_\delta=\psi\theta(\|W\|^2/\delta^2)$ equal to one near that set has support volume $O(\delta^{d-1})$ and $P$-gradient $O(\delta^{-1})$, since $\nabla_P\|W\|^2=-2aW$. Thus its positive trace mass is at most $|\int F\cdot\nabla_P\chi_\delta|=O(\delta^{d-2})\to0$. Trace domination removes every matrix entry there. The bound therefore holds on all $\Omega$.

For each fixed vector $v$ and nonnegative $P$-test $\varphi$, the function
$Q\mapsto\int(\|P\|-U(P,Q))\partial_v^2\varphi(P)dP$
is continuous and its joint distribution is positive. A negative value at any actual $Q$ would contradict positivity on a neighborhood. Hence $D^2_{PP}U(\cdot,Q)\preceq H$ for every $Q$, as well as $D^2_{PP}U\succeq0$.

Mollification on balls avoiding zero now gives a continuous gradient with local Lipschitz bounds inherited from $H$; the mollified functions converge locally uniformly. Their Hessian bounds make the gradients locally bounded and equi-Lipschitz. Arzel\`a--Ascoli gives subsequential locally uniform gradient limits; each is the distributional gradient of $U$, hence all limits coincide. Thus $U$ is $C^1$ and its gradients satisfy those bounds. Every actual $F(P,Q)$, being a subgradient, equals it.

To obtain the bound along every fixed great-circle arc, mollify $U(\cdot,Q)$ on an annulus about the unit sphere with kernel radius $\varepsilon$. On the sphere, $0\preceq D^2U_\varepsilon\preceq H*\rho_\varepsilon$ and $\|H*\rho_\varepsilon\|_{\rm op}\leq(1-\varepsilon)^{-1}$. Integrating the smooth gradient along the chosen arc, then using its locally uniform convergence and letting $\varepsilon\downarrow0$, bounds the output difference by arc length. The shorter arc has length at most $(\pi/2)\|p-p'\|$. Rows have the same bound, proving joint continuity by telescoping. The resulting continuity bound applies to the mean map.
\end{proof}

\subsection{Binary lottery laws}\label{app:binary}

\begin{lemma}\label{lem:sd-efficiency}
For finitely many continuous utility indices on a compact metric outcome space, SD-Pareto efficiency gives probability one to the set of Pareto-efficient outcomes.
\end{lemma}
\begin{proof}
Let $u:X\to\R^m$ be the utility map. For $a\in u(X)$, the set $D(a)=\{y:u(y)\geq a\}$ is nonempty compact and has closed graph. The maximum $v(a)=\max_{D(a)}\sum_j u_j(y)$ is upper semicontinuous: a subsequence of maximizers along a converging sequence identifies the limsup. Its maximizing correspondence has Borel graph and compact nonempty sections, so the Arsenin--Kunugui Borel selection theorem supplies a Borel maximizer $s(a)$ \citep{kechris1995}. All these objects concern one fixed profile.

The Borel map $T(x)=s(u(x))$ weakly increases every utility, and strictly increases at least one whenever $x$ is inefficient. That inefficient set is Borel, since it is $\{x:v(u(x))>\sum_j u_j(x)\}$. If a lottery gives it positive mass, one utility strictly increases on a set of positive mass. Its expectation strictly rises, and the layer-cake identity then gives a strict upper-tail improvement at some cutoff. The pushforward by $T$ is an SD-Pareto improvement, a contradiction.
\end{proof}

\begin{lemma}\label{lem:scalarization}
If directions $p_1,\ldots,p_m$ have a common strict positive direction, their efficient outcomes are exactly the unit directions of their nonzero positive cone.
\end{lemma}
\begin{proof}
An interior dominator in the ball can move along the common strict positive direction until reaching the sphere. Efficient points on the ball and sphere therefore agree. Separate the ball's convex utility image from the open positive orthant above an efficient utility vector. The resulting nonzero nonnegative weights give a nonzero direction $q$ in the stated cone. Its unique maximizer on the ball is $q/\|q\|$. Conversely any dominator of that point would also maximize $q\cdot x$, contradicting uniqueness, including when some weights vanish.
\end{proof}

\begin{proof}[Proof of Lemma~\ref{lem:binary}]
Lemma~\ref{lem:sd-efficiency} reduces either efficiency hypothesis to probability one on efficient outcomes. We first identify the binary mean and then recover the lottery using concave incentive comparisons. Change reports inside a block one at a time: each departing member has the same true direction and evaluation, so individual inequalities telescope to a block inequality. Thus $P_A$ is ICV-SP in both coordinates. Write $m(a,b)=\int x\,dP_A(a,b)$. Its mean is bilaterally linearly IC, lies in the unit ball, and satisfies $m(a,a)=a$.

\textit{Step 1: orthogonal efficiency fixes the mean.}
Restrict reports to a two-plane $V$ and project $m$ onto $V$. Lemma~\ref{lem:planar-body} applies. With $e(s)=(\cos s,\sin s)$ and $t=e'$, its pointwise strip representation gives
\begin{equation}\label{eq:icv-strip}
 F_V(e(\alpha),e(\beta))=e(\beta)+\int_\beta^\alpha t(s)\ell_+(s)\,ds,
 \qquad 0<\alpha-\beta<\pi,
\end{equation}
where $0\leq\ell_+\leq1$ almost everywhere. On the negative strip $0<\beta-\alpha<\pi$, use the analogous density $\ell_-$. Both densities are $2\pi$-periodic: the additive primitives $A(\alpha)+B(\beta)$ have increments under simultaneous $2\pi$ shifts summing to zero throughout the connected strip. Overlapping coordinate intervals make each increment constant. Differentiation makes $A'=t\ell_\pm$ periodic. The periodicity comes from the angular parametrization of identical reports.

Set $h_0=\pi/4$. At $a=e(q+h_0)$, $b=e(q-h_0)$, the reports are orthogonal. Lemma~\ref{lem:scalarization} confines efficient outcomes to their short arc in $V$. Every such outcome satisfies $e(q)\cdot x\geq\cos h_0$, with equality only at the endpoints. Substitution into \eqref{eq:icv-strip} gives
\[
 D_{h_0}(q)=\int_0^{h_0}[\ell_+(q-x)-\ell_+(q+x)]\sin x\,dx\geq0.
\]
Its circular average is zero by periodicity. This convolution is continuous, by translation continuity of its integrable kernel, so it vanishes everywhere. Each nonzero Fourier mode satisfies $\widehat\ell_+(k)I_k=0$, where
\[
 I_k=\int_0^{\pi/4}\sin(kx)\sin x\,dx.
\]
Here $I_1=\pi/8-1/4>0$, and for $k\geq2$,
\[
 I_k=\frac{\sin((k-1)\pi/4)}{2(k-1)}
      -\frac{\sin((k+1)\pi/4)}{2(k+1)}\ne0.
\]
For odd $k\geq3$, exactly one numerator vanishes and the other is $\pm1$. For even $k$, the numerators are $\pm\sqrt2/2$: opposite signs reinforce, and equal signs cannot cancel with unequal denominators. Negative modes use $I_{-k}=-I_k$. Fourier uniqueness makes $\ell_+$ a constant $\lambda_{V,+}$. On the negative strip the midpoint excess has the opposite sign; its zero circular average yields the same argument and a constant $\lambda_{V,-}$.

The planar lemma recovers actual values continuously at antipodal seams. Thus
\[
 a\cdot m(a,-a)=2\lambda_{V,+}-1=2\lambda_{V,-}-1
 \quad(a\in V\cap S).
\]
Write their common value as $\lambda_V$. The function
$L(a)=[1+a\cdot m(a,-a)]/2$ equals $\lambda_V$ on every report plane. Nonparallel directions share a plane; parallel ones are joined by a third direction since $d\geq2$. Hence all planes have one constant $\lambda\in[0,1]$.

The support function now recovers the ambient mean from its planar projections. For fixed $b$, let $C_b$ be the closed convex hull of $\{m(s,b):s\in S\}$. Linear IC and the projected identity give its support function, for every nonzero $z\in\R^d$,
\[
 h_{C_b}(z)=\lambda\|z\|+(1-\lambda)z\cdot b.
\]
The function is differentiable away from zero; its exposed face is therefore a singleton. Every actual truthful selection satisfies
\begin{equation}\label{eq:binary-mean}
 m(a,b)=\lambda a+(1-\lambda)b\qquad(a,b\in S).
\end{equation}
At orthogonal pairs, the expected midpoint projection equals the arc's endpoint value. Efficiency therefore puts all probability at the endpoints, giving the Bernoulli law with weight $\lambda$. If $\lambda=0$ or $1$, the mean in \eqref{eq:binary-mean} is a unit vector everywhere and already identifies every lottery. Assume $0<\lambda<1$ below.

\textit{Step 2: incentives propagate the actual binary law.}
At a pair with angle $\theta\in(0,\pi)$, put $c=\cos\theta$, $A=a\cdot X$, and $B=b\cdot X$. Comparing truth to the opponent's report with the cap $\min(u,c)$ and using unanimity gives $A,B\in[c,1]$ almost surely. Define
\begin{align*}
 S_1&=\mathbb E(1-A)^2,& S_2&=\mathbb E(1-B)^2,\\
 F_1&=(1-\lambda)(1-c)^2,& F_2&=\lambda(1-c)^2,\\
 D_1&=F_1-S_1=\mathbb E(A-c)(1-A),&
 D_2&=F_2-S_2=\mathbb E(B-c)(1-B),\\
 D&=\mathbb E(1-A)(1-B)=\mathbb E(A-c)(B-c).
\end{align*}
These quantities are nonnegative; the equalities use \eqref{eq:binary-mean}, in particular $\mathbb E(A+B)=1+c$. Set $z=(A-c)/(1-c)$ and $w=(B-c)/(1-c)$. They lie in $[0,1]$, and spherical feasibility $A+B\leq2\cos(\theta/2)$ implies
$2-z-w\geq[1+\cos(\theta/2)]^{-1}$. The identity
\[
 z(1-z)+w(1-w)-(2-z-w)zw=(z+w)(1-z)(1-w)\geq0
\]
therefore gives
\begin{equation}\label{eq:binary-deficit-bound}
 0\leq D\leq[1+\cos(\theta/2)](D_1+D_2)\leq2(D_1+D_2).
\end{equation}

Work on a fixed compact angular interval inside $(0,\pi)$. All constants below are uniform on it, for steps remaining in the interval. Let $v=(ca-b)/\sin\theta$ point away from $b$ in the tangent space at $a$, and put $a_\pm=\cos h\,a\pm\sin h\,v$. Their angles with $b$ are $\theta\pm h$. Direct calculation gives
\[
 T_1:=\mathbb E[(1-A)v\cdot X]=\frac{-F_1-cS_1+D}{\sin\theta}.
\]
The admissible concave index $\phi(u)=-(1-u)^2$ and truth $a_\pm$ versus report $a$ imply
\[
 S_1(a_-,b)\leq S_1+2hT_1+O(h^2),\qquad
 S_1(a_+,b)\leq S_1-2hT_1+O(h^2).
\]
The remainders are uniform because sphere coordinates are bounded; no derivative of the lottery map is used. Also
\[
 F_1(\theta)-F_1(\theta-h)
 =F_1(\theta+h)-F_1(\theta)+O(h^2)
 =\frac{2hF_1(1+c)}{\sin\theta}+O(h^2).
\]
Subtracting and substituting $S_1=F_1-D_1$ yields both directions:
\begin{align}
 D_1(a,b)&\leq D_1(a_-,b)
       +\frac{2h(cD_1+D)}{\sin\theta}+Ch^2,\label{eq:binary-forward}\\
 D_1(a,b)&\leq D_1(a_+,b)
       -\frac{2h(cD_1+D)}{\sin\theta}+Ch^2.\label{eq:binary-backward}
\end{align}
Moving the second report gives identical estimates for $D_2$. Either signed first-order term is bounded above by $Ch(D_1+D_2)$, using \eqref{eq:binary-deficit-bound} and the positive lower bound on $\sin\theta$.

Let $M(\theta)$ be the supremum of $\max(D_1,D_2)$ over all ordered pairs of angle $\theta$. Take the pairwise maximum first and then the supremum; no attainment or measurability is needed. For small $h>0$ the two estimates give
\[
 M(\theta)\leq\frac{M(\theta-h)+Ch^2}{1-2Ch},\qquad
 M(\theta)\leq\frac{M(\theta+h)+Ch^2}{1-2Ch}.
\]
The orthogonal law gives $M(\pi/2)=0$. For any target $\theta_*\in(0,\pi)$, partition the interval between $\pi/2$ and $\theta_*$ into $k$ equal steps of length $h$. Use the first estimate above $\pi/2$ and the second below it. With $q_h=(1-2Ch)^{-1}$ and $C>0$, iteration gives
\[
 M(\theta_*)\leq Ch^2\sum_{j=1}^k q_h^j
 =\frac h2(q_h^k-1)\longrightarrow0.
\]
Indeed $kh$ is fixed and $q_h^k$ stays bounded. Thus every deficit vanishes. From $D_1=0$, $A\in\{c,1\}$ almost surely, and its mean puts mass $\lambda$ at $1$. The fiber $a\cdot X=1$ is $\{a\}$. Removing this atom leaves mean $b$, whose unit norm forces the remaining law to be $\delta_b$.

\textit{Step 3: antipodal reports.}
Fix $(a,-a)$ and take nonantipodal $a_j\to a$. Compare truth $a$ to report $a_j$, and truth $a_j$ to report $a$, with every capped-linear utility $\phi_t(u)=\min(u,t)$. Since these functions are one-Lipschitz,
$|\phi_t(a_j\cdot x)-\phi_t(a\cdot x)|\leq\|a_j-a\|$ uniformly. The two inequalities and the established nonantipodal laws give
\[
 \int\phi_t(a\cdot x)\,dP_A(a,-a)
 =\lambda\phi_t(1)+(1-\lambda)\phi_t(-1).
\]
Cap expectations determine the scalar distribution through one-sided derivatives. The fibers at $1$ and $-1$ are $\{a\}$ and $\{-a\}$. This identifies the actual antipodal law. Unanimity covers coincident reports. Writing $w_A=\lambda$ proves
\begin{equation}\label{eq:icv-binary}
 P_A(a,b)=w_A\delta_a+(1-w_A)\delta_b\quad(a,b\in S),
\end{equation}
and exchanging the blocks gives $w_{A^c}=1-w_A$. This completes the binary-law identification.
\end{proof}

\subsection{Three-camp residual probabilities}\label{app:residual-tools}
Retain the full-domain ICV-SP and $\mathcal E_2^\perp$-efficiency assumptions of Lemma~\ref{lem:binary} throughout this subsection.

The next two lemmas concern a population of exactly three labels, with weights $w_i=w_{\{i\}}$. For a larger population, restrict reports to three nonempty camps $A_1,A_2,A_3$. Unilateral ICV-SP inequalities telescope within each camp because all its members share the same true direction and evaluation. The induced three-camp rule inherits ICV-SP and binary efficiency. Apply the same three-label formulas to that rule, using $w_i=w_{A_i}$ and its three camp reports. In either case put $\kappa=1-\sum_{i=1}^3w_i\geq0$. If $\kappa>0$, peak guarantees and superadditivity make
\begin{equation}\label{eq:residual}
 Q(r)=\kappa^{-1}\left[P(r)-\sum_iw_i\delta_{r_i}\right]
\end{equation}
a probability at every profile, including report collisions. Its residual comparisons will always be obtained from $P$.

The residual can still put mass on reported ideals; merge deviations nevertheless restrict its support.

\begin{lemma}\label{lem:residual-support}
At every three-label profile,
\begin{equation}\label{eq:icv-J}
 Q(r)(J(r))=1,\qquad J(r)=S\cap\bigcap_i
 \{x:r_i\cdot x\geq t_i\},\quad t_i=\max_{j\ne i}r_i\cdot r_j.
\end{equation}
\end{lemma}
\begin{proof}
Merge label $i$ into an opponent attaining $t_i$. The binary formula gives the same fixed opponents' atoms plus the merged active atom and residual mass at that peak. At shortfall threshold $t_i$, the active atom has zero shortfall before and after the deviation, as does the merged residual. Opponents' atoms cancel. Thus $\kappa\int(t_i-r_i\cdot x)_+dQ(r)\leq0$. Intersect the three closed probability-one sets. Ties change nothing.
\end{proof}

\begin{lemma}\label{lem:mean-continuity}
On the branch $\kappa>0$, both $m(r)=\int x\,dP(r)$ and $\bar q(r)=\int x\,dQ(r)$ are jointly continuous.
\end{lemma}

\begin{proof}[Proof of Lemma~\ref{lem:mean-continuity}]
Freeze label 3 at $c$. The guaranteed fixed atom makes
\[
 R_c(a,b)=\frac{P(a,b,c)-w_3\delta_c}{1-w_3}
\]
a probability. Cancelling that fixed term preserves ICV-SP in both active coordinates, hence linear IC of its ball-valued mean. The binary law gives $P(a,a,c)=(1-w_3)\delta_a+w_3\delta_c$, so the mean of $R_c$ equals $a$ on its diagonal. Lemma~\ref{lem:spherical-mean} supplies a uniform coordinate modulus, independent of $c$. Freeze another label for each other coordinate and telescope:
\[
 \|m(r)-m(r')\|\leq\frac\pi2\sum_i\|r_i-r_i'\|.
\]
Subtract the continuous peak means and divide by $\kappa$ to obtain continuity of $\bar q$. The joint continuity will be used to identify residual means at limiting report profiles.
\end{proof}

\subsection{Zero-trace uniqueness}\label{app:zero-trace}
The following analytic lemma supplies the uniqueness argument for three-camp means in Appendix~\ref{app:icv-three} and for general means in Appendix~\ref{app:general-mean}. It gives almost-everywhere equality; each application separately recovers every actual report profile.

\begin{lemma}[Zero-trace uniqueness]\label{lem:zero-trace}
Let $d\geq3$, $n\geq2$, and $S=\mathbb S^{d-1}$. Let $F:S^n\to\R^d$ be bounded and measurable for completed product surface measure. Its separately zero-homogeneous extension $\bar F(x)=F(x_1/\|x_1\|,\ldots,x_n/\|x_n\|)$ satisfies
\[
 \partial_{x_{i,a}}\bar F_b=\partial_{x_{i,b}}\bar F_a
 \quad(i=1,\ldots,n;\ a,b=1,\ldots,d)
\]
in distributions on $(\R^d\setminus\{0\})^n$. Suppose that for every $i$, $h\in S$ and choice of signs $\tau_j\in\{-1,1\}$, $j\ne i$,
\begin{equation}\label{eq:zero-trace-condition}
 \lim_{\delta\downarrow0}\ 
 \sup_{\substack{r\in S^n\\
 \max_{j\ne i}\|r_j-\tau_jh\|\leq\delta}}
 \|F(r)\|=0.
\end{equation}
Then $F=0$ almost everywhere on $S^n$.
\end{lemma}
\begin{proof}
Extend $\bar F$ by zero on zero blocks and write $F$ for this extension in the proof; put $N=\{1,\ldots,n\}$. Boundedness makes the cutoff errors at these blocks $O(\epsilon^{d-1})$, so the curl identities hold on the whole Euclidean product.

\textit{Step 1: A Fourier representation.}
Take any bounded, separately zero-homogeneous, blockwise curl-free field $f$ on $(\R^d)^k$. Fourier transform its first block at $\xi$. On a patch with $\xi\ne0$, curl gives $\widehat f=\xi U$ and, for every other block,
$(\xi_b\partial_{x_{j,a}}-\xi_a\partial_{x_{j,b}})U=0$.
In a smooth local frame write $x_j=t_j\xi/\|\xi\|+y_j$, $y_j\perp\xi$. Then $U$ is distributionally constant in $y_j$: subtract a test's $y_j$-integral times a fixed unit-integral test; its remainder is a sum of compactly supported transverse derivatives, also with distribution parameters. Radial homogeneity becomes $t_j\partial_{t_j}U=0$. For $W=\partial_{t_j}U$, removing a test's value at zero and dividing by $t_j$ gives $W=D\otimes\delta_0$; subtracting $(D/2)\otimes\operatorname{sgn}t_j$ leaves zero derivative. Iteration yields
\begin{equation}\label{eq:rg-normal-form}
 \widetilde U=\sum_{\varepsilon\in\{0,1\}^{k-1}}
 D_\varepsilon(\xi)\otimes
 \bigotimes_{j=2}^k\operatorname{sgn}(t_j)^{\varepsilon_j}
 \otimes\bigotimes_{j=2}^k1_{y_j}.
\end{equation}
Tensor variables are independent before smooth pullback, so no discontinuous multiplier acts on an arbitrary frequency distribution. If $f$ is even in an active block $j\ne1$, its sign coefficient vanishes, so the partial Fourier transform of $\partial_{x_j}f$ is supported at $\xi=0$. Pairing the other variables with compact tests gives a bounded $x_1$-function with point Fourier support; point-supported distributions are finite sums of Dirac derivatives, whose inverse transforms are polynomials, so boundedness makes it constant. Thus $\partial_{x_1}\partial_{x_j}f=0$; permutation covers any even active block and any other block.

\textit{Step 2: Elimination of odd components.}
Apply this normal form to the radial extension of $F$. Let $\mathsf E_i,\mathsf O_i$ be the even and odd parts in block $i$, and write
$F=\sum_{J\subseteq N}f_J$, where
$f_J=\prod_{i\in J}\mathsf O_i\prod_{i\notin J}\mathsf E_iF$.
For nonempty $J$, \eqref{eq:rg-normal-form} makes $\partial_{x_q}f_J$ independent of every odd block $p\in J$ when $q\notin J$; its oddness in $p$ makes it zero. Hence $f_J=g_J(x_J)$ almost everywhere for bounded fully odd $g_J$. Averaging inactive variables near $h$ gives an essential zero trace with any one active input free: finitely many reflected terms are bounded by a vanishing maximum of signed trace moduli, without evaluating an almost-everywhere identity on a collision slice. Thus $g_J=0$ for $|J|=1$.

For $|J|\geq2$, Fourier transform one active block. All-oddness leaves one coefficient in \eqref{eq:rg-normal-form}. On a compact smooth patch where $\xi\cdot h\ne0$, average the other active inputs against probability densities shrinking to $h$. In the independent tensor coordinates their sign product is the nonzero constant $\operatorname{sgn}(\xi\cdot h)^{|J|-1}$; the localized transform is the unchanged coefficient times this constant and $\xi$, so the essential trace makes that coefficient zero. Taking $h=e_l$, $l=1,\ldots,d$, confines the first-block Fourier support to $\bigcap_l\{\xi:\xi\cdot e_l=0\}=\{0\}$. Pairing the remaining inputs with compact smooth tests gives a bounded function with point Fourier support, hence a constant polynomial; oddness makes that constant zero. Thus every nonempty $f_J$ vanishes.

\textit{Step 3: The even component.}
Now $F$ is even in every block, and all mixed derivatives between distinct blocks vanish. Let $\mathsf A_i$ average block $i$ against a smooth probability density supported within $\epsilon$ of $h$. Convolution, rectangular differences and integration of base points give, almost everywhere,
\[
 F=\sum_i\mathsf A_{-i}F-(n-1)\mathsf A_{\rm all}F,
 \qquad
 \mathsf A_{-i}=\prod_{j\ne i}\mathsf A_j,
 \quad \mathsf A_{\rm all}=\prod_j\mathsf A_j.
\]
Compact integration supports pass the smooth identity to local $L^1$. The traces bound every term by a modulus tending to zero; for the last use $\mathsf A_{\rm all}=\mathsf A_i\mathsf A_{-i}$. Hence $F=0$ almost everywhere. Radial invariance and polar Fubini on products of annuli give sphere-product almost-everywhere equality.
\end{proof}

\section{Efficiency and random dictatorship}\label{app:efficiency-proof}
\subsection{Three-camp means and feasibility}\label{app:icv-three}
Throughout Sections~\ref{app:icv-three}--\ref{app:three-certificate}, $d\geq3$ and $S=\mathbb S^{d-1}$. For $a\ne-b$, let $D(a,b)$ be their closed shorter spherical arc, with $D(a,a)=\{a\}$; set $D(a,-a)=S$.

For a triple $r$, define
\[
 h_i(r)=\max_{z\in D(r_j,r_k)}r_i\cdot z,\qquad
 K(r)=\{x\in S:r_i\cdot x\geq h_i(r),\ i=1,2,3\}.
\]
The residual law is the law in \eqref{eq:residual}. The following restriction uses only its binary identities and the original ICV-SP inequalities.

\begin{lemma}\label{lem:arc-support}
If $\kappa>0$, then $Q(z,a,b)=\delta_z$ for every $z\in D(a,b)$, in every ordering of the camps, and $Q(r)(K(r))=1$ at every triple.
\end{lemma}
\begin{proof}
For $z$ in the interior of a nonantipodal arc, write $z=\lambda a+\mu b$ with $\lambda,\mu>0$. The merge inequalities give $a\cdot X\geq a\cdot z$ and $b\cdot X\geq b\cdot z$ under $Q(z,a,b)$. Their positive combination gives $z\cdot X\geq1$, hence $X=z$. The endpoints follow from the repeated-report identities. For antipodal opponents and $z\ne\pm a$, write $z=\cos\theta\,a+\sin\theta\,u$, $u\perp a$, $0<\theta<\pi$. Choose $b_\varepsilon=\cos(\pi-\varepsilon)a+\sin(\pi-\varepsilon)u$ with $0<\varepsilon<\pi-\theta$. The arc contains $z$, so mean continuity gives $\bar q(z,a,-a)=z$. A sphere-valued probability with unit mean $z$ is $\delta_z$. Repeated reports cover $z=\pm a$.

At truth $p$ with opponents $a,b$, choose $z\in D(a,b)$ maximizing $p\cdot z$ and put $c=p\cdot z$. Compare the original lotteries before and after reporting $z$, using $\min(u,c)$. Opponent atoms cancel, while the active guaranteed atom contributes $w_i c$ at both reports. The residual after deviation is $\delta_z$. Thus $\int\min(p\cdot x,c)\,dQ(p,a,b)\geq c$. Its integrand is at most $c$, so $p\cdot X\geq c$ almost surely. Intersecting these three probability-one sets proves the claim. No incentive condition on $Q$ is assumed.
\end{proof}

\begin{lemma}\label{lem:three-mean}
Let $d\geq3$, $w_i\geq0$ and $\kappa=1-\sum_iw_i>0$. Suppose $q:S^3\to B$ is continuous, $m(r)=\sum_iw_ir_i+\kappa q(r)$ is linearly incentive compatible in every report, and
\[
 q(a,a,b)=a,\qquad q(a,h,-h)=a
\]
in every placement of the reports. Then $q=g_3$ at every actual triple. In particular, every induced three-camp rule has mean
\begin{equation}\label{eq:identified-three-camp-mean}
 m_P(p)=\sum_{j=1}^3w_{A_j}p_j+
 \left(1-\sum_jw_{A_j}\right)g_3(p_1,p_2,p_3).
\end{equation}
No symmetry or regularity of the original lottery map is required.
\end{lemma}
\begin{proof}
The majority mean has the stated repeated and antipodal identities and is blockwise curl-free by its sign integral. The bounded radial field $\Delta=q-g_3$ is continuous at all nonzero inputs and has zero block curls: finite mean IC gives the support gradients of $m$, and subtraction of the own norm gradients and fixed opponent means leaves the curls zero.

At every signed collision with one input free, $\Delta$ vanishes by the repeated and antipodal identities. Uniform continuity on $S^3$ gives \eqref{eq:zero-trace-condition}. Lemma~\ref{lem:zero-trace} in Appendix~\ref{app:zero-trace} yields $\Delta=0$ almost everywhere, and continuity gives equality everywhere. Its analytic proof is independent of the mechanism-specific trace argument used later for Theorem~\ref{thm:control-identification}.
\end{proof}

In dimension three, Lemma~\ref{lem:arc-support} places the actual residual on $K(r)$, while Lemma~\ref{lem:three-mean} fixes its mean. The geometric argument below produces an independent triple with $g_3(r)\notin\operatorname{conv}K(r)$, excluding positive residual mass without efficiency beyond $\mathcal E_2^\perp$.

\subsection{Alignment at three-camp disagreement}\label{app:three-camp-alignment}
We derive the three-direction majority formula used by the geometric separation in Appendix~\ref{app:three-certificate}. Write $s_{ij}=p_i\cdot p_j$. For a linearly independent triple, define
\[
 r_i=\frac{s_{jh}-s_{ij}s_{ih}}
 {\sqrt{(1-s_{ij}^2)(1-s_{ih}^2)}},\qquad
 \{i,j,h\}=\{1,2,3\}.
\]
Then
\begin{equation}\label{eq:three-camp-gram-mean}
 g_3=\sum_i a_i p_i,\qquad
 a_i=\frac12-\frac1\pi\arcsin r_i.
\end{equation}
To establish the formula, let $G$ be standard Gaussian in $\R^d$ and $q=\mathbb E|G_1|=\sqrt{2/\pi}$. Its direction is uniform on the sphere and independent of its radius. Since signs are unchanged by positive scaling and $q=a_d\mathbb E\|G\|$,
\[
 g_3=q^{-1}\mathbb E[G M_3(S_1,S_2,S_3)],\qquad
 S_i=\operatorname{sgn}(p_i\cdot G).
\]
The majority identity is $M_3=(S_1+S_2+S_3-S_1S_2S_3)/2$, and $\mathbb E[GS_i]=qp_i$. Gaussian integration by parts gives
\[
 \mathbb E[GS_1S_2S_3]
 =2\sum_i p_i\gamma(0)\,
   \mathbb E[S_jS_h\mid p_i\cdot G=0],\qquad
 \gamma(0)=(2\pi)^{-1/2}.
\]
To justify integration by parts, replace each sign by $h_\varepsilon(t)=h(t/\varepsilon)$, with $h$ smooth, odd, increasing and equal to the sign outside $[-1,1]$. Conditional on $p_i\cdot G=\varepsilon u$, the other pair has fixed positive-definite covariance and mean tending to zero. Dominated convergence in the scaled $u$-integral, with compactly supported $h'$ and $\int h'=2$, gives the derivative term; $\|G\|$ dominates the original integrand.

The conditional pair has correlation $r_i$. Two projections of an isotropic planar Gaussian onto directions separated by angle $\arccos r_i$ have opposite signs on angular proportion $\arccos(r_i)/\pi$. Their sign correlation is therefore $2\arcsin(r_i)/\pi$. Since $2\gamma(0)=q$, substitution yields \eqref{eq:three-camp-gram-mean}; the ambient radial moment cancels exactly.

\subsection{A geometric obstruction}\label{app:three-certificate}
The obstruction is that a sphere-supported probability must escape a simplex
whose vertices lie in the ball, whereas the required mean lies on its inner side.

Let $A$ be an invertible $D\times D$ matrix, $h\in\mathbb R^D$,
$x_0=A^{-1}h$ and $v_i=A^{-1}e_i$. Suppose $\|x_0\|<1$ and
there are $t_i>0$ with $\|x_0+t_iv_i\|\leq1$. Then
\begin{equation}\label{eq:simplex-support-cut}
 \ell(x):=\sum_i\frac{(Ax-h)_i}{t_i}\geq1
 \quad\text{whenever }\|x\|=1\text{ and }Ax\geq h.
\end{equation}
Indeed, if $\lambda_i=(Ax-h)_i/t_i\geq0$ sum to less than one, then
$x=(1-\sum_i\lambda_i)x_0+\sum_i\lambda_i(x_0+t_iv_i)$
is a convex combination of ball points with positive weight on the interior
point $x_0$. Its norm is strictly less than one.
Thus any probability on the indicated sphere region has $\ell(\mathbb EX)\geq1$.

Choose three unit reports with Gram matrix
\begin{equation}\label{eq:three-separation-profile}
 C=\begin{pmatrix}
 1&1/\sqrt3-\sqrt2&\sqrt3-\sqrt2\\
 1/\sqrt3-\sqrt2&1&1-2\sqrt{2/3}\\
 \sqrt3-\sqrt2&1-2\sqrt{2/3}&1
 \end{pmatrix}.
\end{equation}
Its leading principal minors are $1$, $2(\sqrt6-2)/3$ and
$4(5-2\sqrt6)/3$, all positive. Hence these reports exist and are linearly
independent; let $A$ have them as rows, so $AA^\top=C$.
The opponent projections onto each report's perpendicular plane make angles
\[
 (\alpha_1,\alpha_2,\alpha_3)=(3\pi/4,2\pi/3,5\pi/6).
\]
These values follow by substituting $C$ into
\[
 \cos\alpha_i=\frac{C_{jk}-C_{ij}C_{ik}}{\sqrt{(1-C_{ij}^2)(1-C_{ik}^2)}}.
\]
All three angles are obtuse. Along the arc from $r_j$ to $r_k$, with length $\theta$, the active alignment $f_i$ has endpoint derivatives
\[
 f_i'(0)=\frac{C_{ik}-C_{ij}C_{jk}}{\sin\theta}<0,\qquad
 f_i'(\theta)=\frac{C_{ik}C_{jk}-C_{ij}}{\sin\theta}>0.
\]
The signs follow from obtuseness of $\alpha_j$ and $\alpha_k$. Derivative zeros are $\pi$ apart and $\theta<\pi$, so the interior critical point is a minimum and each maximum is an endpoint.
Consequently the support thresholds are
\[
 h=(\sqrt3-\sqrt2,\ 1-2\sqrt{2/3},\ \sqrt3-\sqrt2)^\top,
 \qquad K=\{x\in\mathbb S^2:Ax\geq h\}.
\]
The three-direction mean formula \eqref{eq:three-camp-gram-mean}, derived in Appendix~\ref{app:three-camp-alignment},
now gives the exact vector
\[
 g_3=\tfrac34r_1+\tfrac23r_2+\tfrac56r_3.
\]

Apply \eqref{eq:simplex-support-cut} with $(t_1,t_2,t_3)=(3/5,3/5,2/3)$.
Writing $x_i=x_0+t_iv_i$, direct Gram-matrix algebra gives
\[
\begin{array}{c|l}
 &\text{squared norm}\\ \hline
 x_0&-1-3\sqrt2+2\sqrt3+\sqrt6\\
 x_1&8/25-18\sqrt2/5+7\sqrt3/5+34\sqrt6/25\\
 x_2&-43/25-3\sqrt2+13\sqrt3/5+47\sqrt6/50\\
 x_3&-5/9-7\sqrt2/3+2\sqrt3+5\sqrt6/9
\end{array}
\]
The elementary square comparisons
\[
 \sqrt2>24/17,\qquad \sqrt3<26/15,\qquad
 22/9<\sqrt6<49/20
\]
bound these four squared norms above by, respectively,
$139/204$, $12703/12750$, $43573/51000$ and $2993/3060$, all less than one.
The same comparisons yield
\[
 \ell(g_3)=\frac{13}{3}-\frac{41\sqrt2}{24}
                 +\frac{29\sqrt3}{216}-\frac{13\sqrt6}{27}
 <\frac{80747}{82620}<1.
\]
Every probability supported on $K$ has $\ell(\mathbb EX)\geq1$,
so it cannot have mean $g_3$. All inequalities are strict: the obstruction
also persists in a relative open neighborhood among positive-definite, unit-diagonal Gram matrices.

\subsection{The increasing-concave characterization}\label{app:icv}
\begin{proof}[Proof of Theorem~\ref{thm:icv}]
In dimension two, ICV-SP and unanimity imply ex-post efficiency on $\mathcal E_2^\perp$. At orthogonal reports $a,b$, change the $a$-camp to $b$ one member at a time. ICV-SP for true direction $a$ and $\phi(t)=\min\{t,0\}$ telescopes to the unanimous lottery $\delta_b$, giving $\mathbb E\min\{a\cdot X,0\}\geq0$. Hence $a\cdot X\geq0$ almost surely. Changing the $b$-camp to $a$ similarly gives $b\cdot X\geq0$. On the circle their intersection is the efficient quarter-arc: increasing either nonnegative orthogonal coordinate without reducing the other would raise squared norm above one. Unanimity covers the remaining tests. We may therefore use $\mathcal E_2^\perp$ efficiency in both low dimensions.

For $n\leq2$, use unanimity or Lemma~\ref{lem:binary} directly. Let $n\geq3$ and consider any partition into three nonempty camps $A,B,C$. Its induced three-camp rule inherits ICV-SP: individual inequalities telescope when a camp's members share a true direction. Intermediate profiles may have more than three positions, but no efficiency is invoked there. The rule also inherits efficiency on $\mathcal E_2^\perp$. Lemma~\ref{lem:binary} supplies its law at every binary pair and hence full binary efficiency. Lemmas~\ref{lem:binary}, \ref{lem:peak-mass}, \ref{lem:residual-support} and~\ref{lem:mean-continuity} therefore supply constant peak weights $w_1=w_A$, $w_2=w_B$, $w_3=w_C$, the residual law $Q$, and its continuous mean $\bar q$. If $\kappa=1-\sum_iw_i=0$, the three peak guarantees already exhaust probability. Suppose $\kappa>0$.

For $d=3$, the support and mean lemmas give the contradiction in Appendix~\ref{app:three-certificate}. For $d\geq4$, choose a triple with Gram matrix \eqref{eq:three-separation-profile} in a three-dimensional subspace $W$. Its three independent reports admit $u\in W$ with $p_i\cdot u=1$ for all $i$. For $x\in S$, put $z=\operatorname{Proj}_W x$ and
\[
 \tau(x)=\frac{-z\cdot u+\sqrt{(z\cdot u)^2+\|u\|^2(1-\|z\|^2)}}{\|u\|^2},
 \qquad y(x)=z+\tau(x)u.
\]
This continuous map takes values in $W\cap S$ and raises each reported alignment by $\tau(x)\geq0$, strictly if $x\notin W$. Ex-post efficiency therefore forces $P(W)=1$. Under SD efficiency, positive mass outside $W$ would make $y_\#P$ weakly improve all alignments and strictly raise every mean, a strict SD improvement. Thus either efficiency requirement at this three-position profile gives $P(W)=1$, and removing its ideal atoms gives $Q(W)=1$.

The arcs and thresholds within $W$ are those of the three-dimensional argument. The majority mean is also unchanged by the embedding: a uniform ambient direction projects to $RU$ in $W$, with $U$ uniform and independent of the radius $R$; reflection cancels the orthogonal component, and the common factor $\mathbb ER$ cancels against $a_d$. Consequently $g_3^d(p)=g_3^3(p)$. The same geometric separation contradicts the mean and support of $Q$. This closes $d\geq3$.

On the circle choose an equilateral profile $p=(p_1,p_2,p_3)$. The merge constraints give $J(p)=\{p_1,p_2,p_3\}$, so $Q(p)=\sum_i\beta_i\delta_{p_i}$. At least one coefficient is positive. Reindex the camp-coordinate pairs so that $\beta_3>0$. This changes notation, not the mechanism or the report profile.

Choose circle coordinates with $p_1=(1,0)$, $p_2=(-1/2,\sqrt3/2)$, $p_3=(-1/2,-\sqrt3/2)$. For $0<\theta<\pi/6$, put $s_\theta=(\cos\theta,\sin\theta)$, $t_\theta=(s_\theta,p_2,p_3)$, and $c_\theta=s_\theta\cdot p_2=\cos(2\pi/3-\theta)>-1/2$. The merge constraints imply $p_2\cdot X\geq c_\theta$, hence
\[
 q_\theta:=Q(t_\theta)\{p_2\}
 \leq\frac{p_2\cdot\bar q(t_\theta)-c_\theta}{1-c_\theta}.
\]
The right side tends to $\beta_2$ by mean continuity. Choose $\theta$ small enough that $q_\theta<\beta_2+\beta_3/2$. The identity
\[
 p_1=A_\theta s_\theta+C_\theta p_3,\qquad
 A_\theta=\frac{\sqrt3}{2\sin(\pi/3-\theta)},\quad
 C_\theta=\frac{\sin\theta}{\sin(\pi/3-\theta)}
\]
has positive coefficients. Together with $s_\theta\cdot X\geq c_\theta$ and $p_3\cdot X\geq-1/2$, it gives $p_1\cdot X\geq-1/2$. Equality forces both constraints to bind, so the planar projection is $p_2$. Unit norm then forces $X=p_2$. Consequently bounded convergence yields
\[
 \lim_{\eta\downarrow0}\frac1\eta
 \int(-1/2+\eta-p_1\cdot x)_+\,dQ(t_\theta)(x)=q_\theta.
\]
At truth the same expected shortfall equals $(\beta_2+\beta_3)\eta$ for $0<\eta<3/2$. For sufficiently small $\eta$, reporting $s_\theta$ therefore strictly lowers the residual expected shortfall. In the original law, the opponents' atoms cancel and the active own atom has zero shortfall before and after the deviation. This contradicts ICV-SP. Thus $\kappa=0$ for every three-camp partition.

For $k=1,\ldots,n-2$, put $A_k=\{1,\ldots,k\}$, $B_k=\{k+1\}$ and $C_k=\{k+2,\ldots,n\}$. Applying the three-camp conclusion gives $w_{A_k}+w_{\{k+1\}}+w_{C_k}=1$. Since $C_k=A_{k+1}^c$, complementarity gives $w_{A_{k+1}}=w_{A_k}+w_{\{k+1\}}$. Induction from $A_1=\{1\}$ and $w_{A_{n-1}}=1-w_{\{n\}}$ yields $\sum_iw_{\{i\}}=1$. Superadditivity on a coalition and its complement now implies $w_A=\sum_{i\in A}w_{\{i\}}$ for every $A$.

At any profile, group labels by their common report. Their guaranteed peak masses sum to one, so the rule is fixed-weight random dictatorship everywhere. Conversely, truthful reporting maximizes its only report-dependent component at utility one for every increasing transform. Every reported peak is Pareto efficient. For SD-Pareto efficiency, group equal reported peaks: any dominating lottery must retain at least each group's original mass at its singleton top cutoff. These group masses exhaust probability and exclude a strict improvement. This proves both versions.
\end{proof}

\subsection{Local three-camp efficiency tests}\label{app:local-tests}

Theorem~\ref{thm:icv} uses efficiency on all of $\mathcal E_3$. The following refinement gives a sharper sufficient condition and separates the number of camp divisions from the number of profiles tested. Throughout, $n\geq3$ and $d\geq4$. Choose either ex-post or SD-Pareto efficiency and use the same notion in the binary and local tests.

Order the labels as $N=\{1,\ldots,n\}$. Fix equilateral directions $p_1,p_2,p_3$ in a plane $L$ and a unit $h\perp L$. For $k=1,\ldots,n-2$, let
\[
 A_k=\{1,\ldots,k\},\qquad B_k=\{k+1\},\qquad C_k=\{k+2,\ldots,n\}.
\]
Write
\begin{equation}\label{eq:sparse-tilt-tests}
 r^k_\delta:\qquad
 A_k:\frac{p_1+\delta h}{\sqrt{1+\delta^2}},\qquad B_k:p_2,\qquad C_k:p_3.
\end{equation}
For every $k$, require efficiency along sequences $\delta^k_{m,+}>0$ and $\delta^k_{m,-}<0$ tending to zero. The sequences may depend on the rule, sign and division. No efficiency is required at their equilateral limit. The count $n-2$ is of camp divisions, each carrying two sequences of profiles.

\begin{corollary}\label{cor:local-tests}
For every finite $n\geq3,d\geq4$, full-domain ICV-SP, efficiency on $\mathcal E_2^\perp$, and the preceding local tests imply fixed-weight random dictatorship. Conversely, every fixed-weight random dictatorship satisfies these conditions.
\end{corollary}
\begin{proof}
Fix a tested chain partition. The block rule inherits full-domain ICV inequalities by telescoping unilateral deviations. Lemmas~\ref{lem:binary}, \ref{lem:peak-mass}, \ref{lem:residual-support} and~\ref{lem:mean-continuity} give the binary laws, guaranteed peak masses, residual support and continuous mean. Suppose its residual weight is positive.

At the positive and negative tilted profiles, the three directions have a common strict positive direction. Lemmas~\ref{lem:sd-efficiency} and~\ref{lem:scalarization} confine efficient support to $W=L+\R h$ and respectively the halfspaces $h\cdot x\geq0$ and $h\cdot x\leq0$. The continuous residual mean therefore has zero normal component at the common limit. Every weak cluster law is supported in $L$; the merge constraints reduce its support to $\{p_1,p_2,p_3\}$. The common mean and affine independence identify a unique cluster law.

The identified mean in Lemma~\ref{lem:three-mean} is $g_3(p)=0$ at the equilateral limit, so the cluster law is $\nu_*=(\delta_{p_1}+\delta_{p_2}+\delta_{p_3})/3$. Lemma~\ref{lem:planar-residual-insurance} gives $p_1\cdot X=0$ under the actual limiting residual. Compare a true tilted first report with the unperturbed report using $\phi(u)=\min(u,0)$. Opponent atoms cancel, and the moving own atom has evaluation zero at both profiles for sufficiently small tilts. Passing to a weakly convergent subsequence would give $\mathbb E_{\nu_*}\min(p_1\cdot X,0)\geq0$. The left side is $-1/3$, a contradiction. Thus every tested partition has zero residual weight; no additional efficiency profile or continuity of lottery laws is used.

For each $k$, its weights obey $w_{A_k}+w_{\{k+1\}}+w_{C_k}=1$. Complementarity gives $w_{A_{k+1}}=w_{A_k}+w_{\{k+1\}}$ and then $\sum_iw_{\{i\}}=1$. Superadditivity on a coalition and its complement makes all weights additive. Grouping equal reports, the guaranteed peak masses exhaust probability at every profile. The rule is therefore fixed-weight random dictatorship. The converse is immediate from Theorem~\ref{thm:icv}.
\end{proof}

\section{The compromise mechanism}\label{app:compromise-proof}
\subsection{Feasibility and incentives}\label{app:borel-realization}
\begin{proof}[Proof of the construction in Proposition~\ref{thm:icv-e2-counterexample}]
For unit $q$, $|q\cdot g_k(t)|\leq a_d^{-1}\int|q\cdot v|d\sigma_d(v)=1$; reflection through $L_t$ gives $g_k(t)\in L_t$. Hence the lift has unit norm and mean $g_k$. Signs stabilize outside the limiting great spheres along every converging report sequence, so dominated convergence makes $g_k$ continuous, including at collisions and antipodes.

For Borel realization let $A_t=[t_1\ \cdots\ t_k]$ and
\[
 \Pi_t=\lim_{j\to\infty}[I-A_t(A_t^\top A_t+j^{-1}I_k)^{-1}A_t^\top].
\]
This Borel projector has positive rank. For a standard Gaussian $G$, use $\Pi_tG/\|\Pi_tG\|$, with the normalized first nonzero $\Pi_te_j$ on the zero event. The realization is jointly Borel and its law at every profile is uniform on the unit sphere of $L_t^\perp$. The fallback changes no profile's law. Thus $T_k$, and its fixed mixture, are Borel, anonymous and orthogonally neutral.

The finite incentive argument \eqref{eq:body-committee-ic} applies to every report pair and span rank. Explicitly, with $B(v)=\sum_{j=1}^{k-1}\operatorname{sgn}(u_j\cdot v)$, the difference in truthful and deviating mean alignment is
\begin{equation}\label{eq:triple-linear-ic-supp}
 \frac1{a_d}\int_{\{B(v)=0\}}
 [|p\cdot v|-(p\cdot v)\operatorname{sgn}(s\cdot v)]\,d\sigma_d(v)\geq0.
\end{equation}
Outside this pivotal set the active report cannot change the majority. Jensen covers interior means by a supporting affine bound; endpoint means force constant alignment. Thus it also covers the permitted endpoint jump. The comparison \eqref{eq:body-committee-ic} therefore proves ICV-SP for each affected committee and hence for the mixture.

At binary profiles each committee has a strict-majority ideal. Its sign integral is that unit ideal, so the law is supported on reported peaks. This proves unanimity and ex-post efficiency. Any SD dominator must retain each peak's mass by its top cutoff; these masses exhaust probability, proving SD efficiency.

Put $\ell=(k-1)/2$. Some profile admits a positive-probability committee with three camps with directions summing to zero and counts $(\ell,\ell,1)$. No camp has a strict majority, so its sign majority equals the ordinary three-sign majority. The three centered signs cannot all agree, and their majority is their sum. Thus $g_k=p_1+p_2+p_3=0$. This component selects outcomes orthogonal to the reported directions, excluding random dictatorship.
\end{proof}

\subsection{The insurance share}\label{app:insurance-count}
\begin{proof}[Proof of insurance in Proposition~\ref{thm:icv-e2-counterexample}]
At balanced equilateral profiles, a strict-majority committee chooses its majority ideal. Otherwise every camp count is at most $\ell=(k-1)/2$, so the sign argument above gives $g_k=0$ and every camp's alignment is zero. The three majority events are disjoint and equally likely, proving \eqref{eq:population-insurance} and its variance. An evenly distributed count vector has maximum $\lceil k/3\rceil\leq\min\{m,\ell\}$ and positive probability, so $\beta_{m,k}>0$. Finite Jensen gives welfare gain
$\beta_{m,k}\{\phi(0)-[\phi(1)+2\phi(-1/2)]/3\}$,
strict under strict concavity. If $k\geq2m+1$, a majority exceeds every camp's size, giving complete insurance. If $k\leq2m-1$, a committee with $\ell+1$ members from one camp and $\ell$ outsiders has positive probability, ruling it out.

For \eqref{eq:uniform-near-insurance}, let $C$ be the number sampled from one camp. Then $\delta_{m,k}=3\Pr(C\geq(k+1)/2)$. The moment $\mathbb E2^C$ is the mean product of $k$ draws without replacement from $m$ twos and $2m$ ones. Pairwise averaging increases elementary symmetric sums: with all other entries fixed,
\[
 e_k=e_k(\text{others})+(x+y)e_{k-1}(\text{others})+xy e_{k-2}(\text{others}).
\]
Repeated averaging converges to the constant vector, so $\mathbb E2^C\leq(4/3)^k$. Markov gives
$\Pr(C\geq(k+1)/2)\leq2^{-1/2}(8/9)^{k/2}$.
For the largest feasible odd $k$, either $d\geq2m+2$ and risk is zero, or $d\leq2m+1\leq n$ and $k\geq d-2$. Combining these cases with $\delta_{m,k}\leq1$ proves \eqref{eq:uniform-near-insurance}. The scalar law gives \eqref{eq:near-insurance-welfare}, including $L_\phi=0$ and endpoint jumps. In particular
\[
 d\geq\max\left\{4,\left\lceil2+
 \frac{2\log(3/(\sqrt2\varepsilon))}{\log(9/8)}\right\rceil\right\}
\]
suffices for tolerance $\varepsilon\in(0,1)$, independently of population.
\end{proof}

\section{Control and the limits of insurance}\label{app:uniform-risk-gap}
\subsection{Means determined by binary control}\label{app:general-mean}
Lemma~\ref{lem:zero-trace} in Appendix~\ref{app:zero-trace} supplies analytic uniqueness. The proof below verifies its hypotheses from incentive compatibility and recovers every actual report profile.

\begin{proof}[Proof of the mean identity in Theorem~\ref{thm:control-identification}]
Fix $d\geq4$ and $P$ as in Theorem~\ref{thm:control-identification}; put $S=\mathbb S^{d-1}$ and $m(r)=\mathbb E_{P(r)}X$. Unanimity handles $n=1$; assume $n\geq2$. Lemmas~\ref{lem:binary} and~\ref{lem:peak-mass} give the labelled binary weights and actual camp atoms. No symmetry, neutrality or report regularity is assumed.

\noindent\textit{Step 1: Regularity and mean menus.}
Extend reports radially to $\Omega=(\R^d\setminus\{0\})^N$. Finite linear comparisons with positive radial multipliers give
$(x_i-y_i)\cdot[m(x_i,B)-m(y_i,B)]\geq0$.
For $u=e_l$, set $E=(u,\ldots,u)$ and $g=u\cdot m$. Blockwise telescoping makes $g$ nondecreasing along every $E$-line in $\Omega$, which has at most $n$ punctures. If $x\pm tE\in\Omega$, telescoping and $\|\Delta m\|\leq2$ give
\[
 g(x-tE)-\frac2t\sum_i\|y_i-x_i\|
 \leq g(y)\leq g(x+tE)+\frac2t\sum_i\|y_i-x_i\|.
\]
Each displacement is $tu$ plus a perturbation contributing at most $2/t$ times its norm, so line and joint continuity at $x$ are equivalent. Oscillation superlevel sets are closed; hence $D_g=\bigcup_{q\geq1}\{\operatorname{osc}g\geq1/q\}$ is Borel. It meets each $E$-line countably: between punctures, disjoint positive jump intervals are indexed by rationals. Tonelli makes $D_g$ null. Repeat this argument for all $d$ coordinates. The function equal to $m$ off the union of the exceptional sets and zero on that union is a Borel representative. Thus the original mean map is measurable for completed Lebesgue measure. Radial invariance and polar Fubini give completed surface measurability on $S^n$.

With $r_{-i}$ fixed,
$H(y)=\sup_s y\cdot m(s,r_{-i})$ is finite, convex and one-Lipschitz. Linear IC gives $H(y)=y\cdot m(\widehat y,r_{-i})$ and makes the selected mean a subgradient for every $y\ne0$, where $\widehat y=y/\|y\|$. Almost-everywhere differentiability identifies it with the weak gradient; completed measurability and Fubini give the distributional block curls required by Lemma~\ref{lem:zero-trace}.

\noindent\textit{Step 2: Mixed-antipodal traces.}
Fix $i$, $h\in S$, and $N\setminus\{i\}=B\sqcup C$, allowing empty groups. Set $\beta=w_B$, $\gamma=w_C$, $c=\beta-\gamma$, $k=1-\beta-\gamma$, and $\kappa_i=k-w_{\{i\}}\geq0$. If $B,C$ are nonempty and $\kappa_i>0$, Lemma~\ref{lem:arc-support} makes the induced residual $\delta_p$; if $\kappa_i=0$, the guaranteed peak atoms exhaust the law. Empty groups follow from the binary law. Thus, merging coincident atoms when necessary,
\[
 P(p,h^B,(-h)^C)=\beta\delta_h+\gamma\delta_{-h}+k\delta_p,
 \qquad m(p,h^B,(-h)^C)=ch+kp.
\]

Let $t_j$ approach these signed reports, put $\tau_j=1$ on $B$ and $-1$ on $C$, and set $\delta=\sum_{j\ne i}\|t_j-\tau_jh\|$. For each replacement of $q=\tau_jh$ by $s$, the two IC directions give, for every increasing-concave one-Lipschitz $\phi$,
\[
0\leq E_{\rm before}\phi(q\cdot X)-E_{\rm after}\phi(q\cdot X)
\leq2\|s-q\|.
\]
For the upper bound replace $q$ by $s$ in both expectations and apply IC for true type $s$.

Linear and capped-linear tests, with $(u-Z)_+=u-Z+(Z-u)_+$ for the reverse sign, therefore imply, uniformly in $p$,
\begin{align*}
 |h\cdot m_t(p)-(c+kh\cdot p)|&\leq2\delta,\\
 |E_t(u-h\cdot X)_+-E_0(u-h\cdot X)_+|&\leq4\delta.
\end{align*}
The corresponding upper partial moments $E(h\cdot X-u)_+$ differ by at most $6\delta$. Hence, for $0<\epsilon<1$, the endpoint atoms imply masses at least $\beta-6\delta/\epsilon$ in $\{h\cdot X>1-\epsilon\}$ and $\gamma-6\delta/\epsilon$ in $\{h\cdot X<-1+\epsilon\}$. Selecting positive-part submeasures gives endpoint-mean errors at most their masses times $\sqrt{2\epsilon}$ and leaves mass at most $k+12\delta/\epsilon$. Hence
\[
 \sup_p\|m_t(p)-ch\|\leq
 k+24\delta/\epsilon+\sqrt{2\epsilon}=k+o(1)
\]
for $\epsilon=\delta^{2/3}$.

The support functions $H_t(y)=\sup_p y\cdot m_t(p)$ have locally uniform subsequential limits $H$. The bound and the $h$-mean estimate at $p=\pm h$ force $m_t(\pm h)\to ch\pm kh$ and
$k|h\cdot y|\leq H(y)-ch\cdot y\leq k\|y\|$.
For $v\perp h$, $v\ne0$, the derivative of $H_t(ah+v)$ is
$h\cdot m_t((ah+v)/\|ah+v\|)$ almost everywhere. Integration over finite intervals gives
$H(ah+v)=ca+k\sqrt{a^2+\|v\|^2}+A(v)$.
The upper bound gives $A(v)\leq0$; the lower bound as $a\to\infty$ gives $A(v)\geq0$. Thus $H(y)=ch\cdot y+k\|y\|$, whose unique subgradient at $p\in S$ is $ch+kp$. Compactness and the subgradient inequalities, also for varying $p$, yield
\begin{equation}\label{eq:mixed-antipodal-trace}
 \sup_{p\in S}\|m_t(p)-(ch+kp)\|\longrightarrow0.
\end{equation}
For $k=0$ the translated-ball bound suffices.

\noindent\textit{Step 3: Identification.}
Let $G_w$ be the integral in \eqref{eq:control-mean}. Dominated convergence off the reports' null great spheres makes it continuous. Monotonicity of $w$ gives finite linear IC directionwise, while
$a_d^{-1}\int|u\cdot v|\,d\sigma_d=1$ bounds its norm by one; it has the same block curls as $m$. At signed-antipodal opponents its sign table is
$c\operatorname{sgn}(h\cdot v)+k\operatorname{sgn}(p\cdot v)$, so it equals $ch+kp$ there and has a uniform continuous trace. By \eqref{eq:mixed-antipodal-trace}, $F=m-G_w$ has uniformly vanishing signed traces in every placement of the reports.

Lemma~\ref{lem:zero-trace} therefore gives $m=G_w$ almost everywhere on $S^n$.

Finally recover every actual profile. Fix an own menu agreeing almost everywhere with continuous $G_w$. Finite IC gives its support $H(p)=p\cdot G_w(p)$ on a dense set and hence everywhere. The comparison support's selected subgradient is $G_w$, and its supporting-plane residual is at most
$\|G_w(\widehat z)-G_w(\widehat y)\|\|z-y\|=o(\|z-y\|)$; it is therefore differentiable away from zero. Every actual selected mean is consequently $G_w$.

Fubini first gives a full-measure tail set $E_1\subset S^{n-1}$ on which the menu argument yields equality on $S\times E_1$. Inductively, take a Borel full-measure subset of $E_j$ and full-measure sections in block $j+1$. Applying the menu argument for every arbitrary prefix gives equality on $S^{j+1}\times E_{j+1}$. Induction to $j=n$ covers every actual profile and proves \eqref{eq:control-mean}. Thus continuity and orthogonal equivariance of the mean are conclusions, not assumptions on the law.
\end{proof}

\subsection{Planar risk is fixed by binary control}\label{app:planar-control}
\begin{lemma}\label{lem:planar-residual-insurance}
Under the general mechanism hypotheses of Theorem~\ref{thm:control-identification}, the planar scalar-law identity \eqref{eq:planar-control-statement} holds at every actual profile.
\end{lemma}
\begin{proof}
Let $P_3$ be the induced three-camp law, $w_i=w_{A_i}$ and $\kappa=1-\sum_iw_i$. For distinct reports and $\kappa>0$, the actual residual $Q$ in \eqref{eq:residual} has mean $g_3$ by Lemma~\ref{lem:three-mean}. When the origin lies in the relative interior of the report triangle, outside null great spheres the three projection signs cannot all agree, so their majority is their sum and $g_3=M:=p_1+p_2+p_3$. Let $L$ be the report plane and $\Pi_L$ its orthogonal projection. Put $B_P=\mathbb E_{P_3}XX^\top$, $B_Q=\mathbb E_QXX^\top$, and define the covariance on $L$ by
\[
 V=\left.\Pi_L(B_Q-MM^\top)\Pi_L\right|_L.
\]
We show that $V$ vanishes using the incentives of $P_3$.

\noindent\textit{Step 1: The envelope identity.}
Fix opponents and move one camp's report along a unit-speed circle path $p(\theta)$ while the origin remains in the triangle's relative interior. Telescoping individual ICV-SP within the camp makes truth attain the following supremum for $\phi(u)=u-u^2/2$:
\[
 U(\theta)=\sup_{s\in S}\left\{
 p(\theta)\cdot m_P(s)-\tfrac12p(\theta)^\top B_P(s)p(\theta)\right\}.
\]
Uniform Lipschitz bounds make the supremum locally absolutely continuous. The actual truthful test touches it, so at differentiability points $U'=p'\cdot m_P-p'^\top B_Pp$. Since $m_P=(w_i+\kappa)p+\text{constant}$, its own derivative is orthogonal to $p$. Hence $h_P=p^\top B_Pp$ is locally absolutely continuous with $h_P'=2p'^\top B_Pp$. Now
\[
 h_Q=\frac{h_P-w_i-\sum_{j\ne i}w_j(p\cdot p_j)^2}{\kappa}.
\]
Opponent derivatives cancel, and the own atom contributes zero since $p'\cdot p=0$. Thus $h_Q'=2p'^\top B_Qp$; subtracting $(p\cdot M)^2$ gives
\begin{equation}\label{eq:planar-variance-envelope}
 a'=2p'^\top Vp\quad\text{a.e.},\qquad a=p^\top Vp.
\end{equation}

\noindent\textit{Step 2: Boundary decay.}
Along the entire interval of such reports, write its angular coordinate as $s\in(0,u)$, $0<u<\pi$. The unit-norm constraint gives
\[
 \operatorname{tr}V\leq1-\|M\|^2
 =8\sin(s/2)\sin((u-s)/2)\cos(u/2)
 \leq2\sin u\min\{s,u-s\}.
\]
In the frame $(p,p')$, positive semidefiniteness and \eqref{eq:planar-variance-envelope} bound $|(\sqrt{a+\eta})'|$ by the square root of this continuous variance bound. Integrate from either endpoint, where the bound forces $a\to0$, and let $\eta\downarrow0$. At every actual point,
\[
 a(s)\leq\tfrac89\sin u\min\{s^3,(u-s)^3\}.
\]
Starting the integral inside the interval and approaching its endpoint requires no lottery continuity.

\noindent\textit{Step 3: Vanishing near the boundary.}
Fix an oriented isometry $\mathcal R:\R^2\to L$ with $p_3=\mathcal R(1,0)$, and write
\[
 p_1=\mathcal R(-\cos x,\sin x),\quad p_2=\mathcal R(-\cos y,-\sin y),
 \qquad x,y>0,\quad u=x+y<\pi.
\]
Set $f_i=p_i^\top Vp_i$ and
$h_1=f_1/(\sin^2x\sin y)$, $h_2=f_2/(\sin x\sin^2y)$.
The coordinates $x,y$ change only camps 1,2, respectively. By \eqref{eq:planar-variance-envelope}, $h_1$ is locally absolutely continuous in $x$ for fixed $y$, and $h_2$ in $y$ for fixed $x$. Solving the three covariance contractions gives
\begin{align*}
 \partial_xh_1&=\frac{h_2}{\sin u}
 -\frac{f_1}{\sin^3x\sin u}
 -\frac{f_3\sin u}{\sin^3x\sin^2y}\leq\frac{h_2}{\sin u},\\
 \partial_yh_2&=\frac{h_1}{\sin u}
 -\frac{f_2}{\sin^3y\sin u}
 -\frac{f_3\sin u}{\sin^2x\sin^3y}\leq\frac{h_1}{\sin u}.
\end{align*}
With $z=\pi-x-y$, the cubic bound gives
\[
 0\leq h_1\leq\tfrac89\frac{\min\{x^3,z^3\}}{\sin^2x},
 \qquad
 0\leq h_2\leq\tfrac89\frac{\min\{y^3,z^3\}}{\sin^2y}.
\]
In particular $h_1(0,y)=h_2(x,0)=0$ as actual boundary limits.

Choose $\epsilon>0$ with $e c_\epsilon^2>1$, where $c_\epsilon=\sin\epsilon/\epsilon$. On $x+y<\epsilon$, the supremum $M_0=\sup h_1/x$ is finite. First integrate the second path inequality using the measurable bound $h_1(s,r)\leq M_0s$, then the first using its resulting explicit integral bound. Since $\sin t\geq c_\epsilon t$ there, for $b=y/x$,
\[
 \frac{h_1(x,y)}x
 \leq\frac{M_0}{c_\epsilon^2}
 \int_0^1\frac{t}{t+b}\log\frac{t+b}{t}\,dt
 \leq\frac{M_0}{e c_\epsilon^2}.
\]
The last inequality uses $(\log v)/v\leq1/e$. Taking suprema gives $M_0=0$, hence $h_1=h_2=0$. Only explicit measurable majorants were integrated.

\noindent\textit{Step 4: Extension to the full planar domain.}
For any $R<\pi$ and $\lambda>1$, put $k_i=h_i/\tan(u/2)^\lambda$ on $u<R$. These functions are bounded by the cubic estimates and vanish near $u=0$. Their common nonnegative supremum $K$ is finite. Since $(\log\tan(u/2))'=1/\sin u$, the path inequalities give
\[
 \partial_xk_1\leq\frac{K-\lambda k_1}{\sin u},\qquad
 \partial_yk_2\leq\frac{K-\lambda k_2}{\sin u}.
\]
The integrating factor and zero incoming boundary limits imply $k_i\leq K/\lambda$. Thus $K\leq K/\lambda$ and $K=0$. Letting $R\uparrow\pi$ gives $f_1=f_2=0$ at every such profile. Positivity gives $Vp_1=Vp_2=0$; these directions span $L$, so $V=0$ on $L$. Thus $\mathbb E_Q\|\Pi_LX-M\|^2=\operatorname{tr}V=0$.

For distinct planar reports whose triangle does not contain the origin in its relative interior, the majority sign equals the angular median's sign almost everywhere. Thus $g_3$ is that unit median, and the residual is its Dirac mass. At collisions, complementarity gives the merged binary mass $1-w_k=w_i+w_j+\kappa$, exactly as in \eqref{eq:planar-control-statement}. If $\kappa=0$, the guaranteed atoms exhaust probability. These are actual-profile arguments, with no limit of the original law. They prove the stated scalar identity in all cases.
\end{proof}

\subsection{Quantitative control and risk}\label{app:fixed-d-risk}
\begin{proof}[Proof of Proposition~\ref{prop:fixed-d-risk}]
The finite label average $\bar P(r)=(n!)^{-1}\sum_\pi P(r_\pi)$ is anonymous and preserves ICV-SP. Since $w_A=\omega_{|A|}$, it also preserves every binary lottery and both tested efficiencies. Steps 1--6 derive weight bounds, writing $P$ for this average. The original rule has the same weights; its peak guarantees and planar scalar law then give \eqref{eq:asymmetric-risk-bound} and \eqref{eq:fixed-d-risk}. Step 7 treats perturbations of the original rule directly.

\noindent\textit{Step 1: Grouped cap geometry.}
In $\R^4$, put
\[
 \begin{gathered}
 p_0=\frac{\sqrt{29}}8,\quad s_0=\frac1{\sqrt{32}},\quad
 u_0=\frac14,\quad \rho=\frac78,\quad a=e_1,\quad b=e_2,\\
 h=(-p_0,-p_0,s_0,u_0),\quad c=(-p_0,-p_0,s_0,-u_0).
 \end{gathered}
\]
These directions are unit, with $h\cdot a=h\cdot b=-p_0$ and $h\cdot c=\rho$.
For $d=4$ use center $h$ and leaves $a,b,c$.
For $d\geq5$, put $k=d-4$ and choose a group count $G\in\{1,\ldots,k\}$.
Partition the $k$ extra coordinates into blocks of positive integer dimensions
$r_j$ satisfying $\sum_jr_j=k$ and $r_j\leq\lceil k/G\rceil\leq2k/G$.
In block $j$, choose unit regular-simplex vertices $w_{ji}$,
$i=1,\ldots,r_j+1$, and set $\epsilon_j=1/(4\sqrt G r_j)$.
Besides $b,c$, use leaves
\[
 \frac{(\epsilon_j a,w_{ji})}{\sqrt{1+\epsilon_j^2}}
 \quad(1\leq j\leq G,\ 1\leq i\leq r_j+1),
 \qquad l_G=k+G+2
\]
with the added coordinates supported in their respective blocks. There are $l_G$ leaves. Zero coordinates of $h,b,c$ are suppressed below.
The caps are
\begin{equation}\label{eq:star-caps}
 \begin{gathered}
 h\cdot x\geq\rho,\quad c\cdot x\geq\rho,\quad b\cdot x\geq-p_0,\\
 \epsilon_j(a\cdot x+p_0)+w_{ji}\cdot y_j\geq0\quad\text{for all }j,i.
 \end{gathered}
\end{equation}
For $d=4$ replace the last constraints by $a\cdot x\geq-p_0$.
Define
\[
 \alpha=2\sqrt{178}-4\sqrt{29},\qquad
 A=\begin{cases}\alpha,&d=4,\\ \sqrt{713}-4\sqrt{29},&d\geq5,\end{cases}
\]
and the affine functional
\begin{equation}\label{eq:star-functional}
 \ell(x,y)=8(h\cdot x-\rho)+8(c\cdot x-\rho)
       +\alpha(b\cdot x+p_0)+A(a\cdot x+p_0).
\end{equation}
Its linear part $L=(8h+8c+\alpha b+Aa,0)$ satisfies $\|L\|<28$ and $\ell(h)=1$.
The cap vertex is $v=(-p_0,-p_0,-s_0,0)$, with $\|v\|^2=15/16$.

The inverse directions for rows $a,b,h,c$ are
\[
 \begin{gathered}
 d_a=(1,0,p_0/s_0,0),\quad d_b=(0,1,p_0/s_0,0),\\
 d_h=(0,0,1/(2s_0),1/(2u_0)),\quad
 d_c=(0,0,1/(2s_0),-1/(2u_0)).
 \end{gathered}
\]
Each has norm less than four. We have $v+d_h/8=h$, $v+d_c/8=c$,
$v\cdot d_a=v\cdot d_b=-2p_0$ and $\|d_a\|^2=\|d_b\|^2=31/2$.
The identities
\[
 \alpha^2+8\sqrt{29}\alpha=248,\qquad
 A^2+8\sqrt{29}A=249\quad(d\geq5)
\]
show that $(v+d_b/\alpha,0)$ and every $(v+d_a/A,z/A)$ with $\|z\|=1/4$ are unit. For $d=4$, $v+d_a/\alpha$ is unit.

To see why these points separate the caps, put
$q_h=h\cdot x-\rho$, $q_c=c\cdot x-\rho$, $q_b=b\cdot x+p_0$ and
$t=a\cdot x+p_0$. In the cap cone these quantities are nonnegative.
Indeed summing any block's constraints gives $t\geq0$.
If $t>0$, the simplex identities give
\[
 \frac{y_j}{t}\in\operatorname{conv}\{\epsilon_j r_jw_{ji}:1\leq i\leq r_j+1\}.
\]
The vertices $z$ of this product of simplices all satisfy
$\|z\|^2=\sum_j(\epsilon_jr_j)^2=1/16$.
Thus $(x,y)-(v,0)$ is a nonnegative combination of
$(d_h,0),(d_c,0),(d_b,0)$ and vectors $(d_a,z)$, with total coefficients
$q_h,q_c,q_b,t$ in these four families. Equivalently it is a combination of the corresponding unit edge points minus $(v,0)$, with total weight
$\Lambda=8q_h+8q_c+\alpha q_b+At=\ell(x,y)$.
If $t=0$, the block constraints force all $y_j=0$ and the same representation holds. In dimension four use $td_a$ directly. Hence
\begin{equation}\label{eq:star-edge-bound}
 \|(x,y)\|\leq\|v\|+(1-\|v\|)\ell(x,y)
 \quad\text{whenever the caps hold and }\ell(x,y)<1.
\end{equation}
In particular $\ell\geq1$ on their intersection with the unit sphere.

\noindent\textit{Step 2: A dimension-explicit separating mean.}
For an integration direction $z$, let $H(z)=\operatorname{sgn}(h\cdot z)$.
Let $F_\star(z)$ equal $H(z)$ unless every leaf sign is $-H(z)$, in which case it equals $-H(z)$. Define the comparison vector
\[
 G_\star=\frac1{a_d}\int_S zF_\star(z)\,d\sigma_d(z).
\]
We use this vector only to compare means. Its separation gap satisfies
\begin{equation}\label{eq:star-separation-gap}
 \begin{aligned}
 g_{k,G}:=1-\ell(G_\star)&\geq e^{-100k}(G/k)^{k/2}&& (d\geq5),\\
 g_4:=1-\ell(G_\star)&\geq e^{-100}&& (d=4).
 \end{aligned}
\end{equation}

We need a conditional Gaussian inequality. Let independent
$U\sim N(0,u_0^2)$ and $V_\mu\sim N(\mu,s_0^2)$, so $u_0^2=2s_0^2$.
Put $m(\mu)=\mathbb E[V_\mu\mid U>|V_\mu|]$ and
\[
 \theta=\arctan\sqrt2,\qquad
 \beta=1-\frac{\sin(2\theta)}{2\theta}.
\]
For $\mu>0$ we claim $m(\mu)\geq\beta\mu$.
Writing $\overline\Phi$ for the standard normal upper tail, direct differentiation and the value at zero verify
\[
 \overline\Phi(z)=\frac1\pi\int_0^\infty
       \frac{e^{-(1+t^2)z^2/2}}{1+t^2}\,dt\qquad(z\geq0).
\]
The conditional density of $V_\mu$ is proportional to
$e^{-(v-\mu)^2/(2s_0^2)}\overline\Phi(|v|/u_0)$.
Completing the square expresses it as a mixture of normal densities with mean
$\lambda_t\mu$ and variance $s_0^2\lambda_t$, where
\[
 \lambda_t=\frac1{1+(s_0^2/u_0^2)(1+t^2)}.
\]
The mixing weights are proportional to
$\sqrt{\lambda_t}(1+t^2)^{-1}e^{\lambda_t\mu^2/(2s_0^2)}$.
Relative to $\mu=0$, this increasing exponential tilt cannot decrease the mean of $\lambda_t$: for increasing $f$ and an independent copy $T'$,
$2\operatorname{Cov}(T,f(T))=\mathbb E[(T-T')(f(T)-f(T'))]\geq0$.
Consequently
\[
 \frac{m(\mu)}\mu\geq
 \frac{\mathbb E[V_0^2\mid U>|V_0|]}{s_0^2}
 =\frac1\theta\int_{-\theta}^\theta\sin^2\varphi\,d\varphi=\beta.
\]
The last identity uses the rotational symmetry of $(V_0/s_0,U/u_0)$:
the conditioning event is the wedge of half-angle $\theta$ about its positive second axis, and the radial second moment is two.

The coefficients in \eqref{eq:star-functional} satisfy
\begin{equation}\label{eq:star-coefficients}
 0<\alpha\leq A<\frac{24\sqrt{29}}{25}<16p_0\beta.
\end{equation}
To verify the last inequality without numerical approximations, let
$z=3-2\sqrt2$. Then $z/(1+z^2)=1/6$ and
$\theta=\pi/4+\arctan z>3/4+1/6=11/12$.
Since $\sin(2\theta)=2\sqrt2/3<20/21$, we obtain
$\beta>37/77>12/25$.
For $d\geq5$, the middle inequality follows by squaring:
$713\cdot625=445625<445904=124^2\cdot29$.
Also $2\sqrt{178}<\sqrt{713}$, so the same bound holds for $d=4$.

These estimates also give $16p_0\beta-A>16p_0/1925>1/200$.
For $d\geq5$, let $(Z,W)$ be standard Gaussian in $\R^4\oplus\R^k$.
The positive-center flip event is
\[
 D=\{h\cdot Z>0,\ c\cdot Z<0,\ b\cdot Z<0,
       \ \epsilon_j a\cdot Z+w_{ji}\cdot W_j<0\text{ for all }j,i\}.
\]
The split inequalities imply $a\cdot Z<0$.
Conditional on $-Z_1=x>0$ and $-Z_2=y>0$, write
$V=p_0(x+y)+s_0Z_3$, $U=u_0Z_4$ and $\mu=p_0(x+y)$.
The first two inequalities are $U>|V|$.
The split event, of probability $q(x)$, is independent of $U,V$.
Writing $F(\mu)=\Pr(U>|V_\mu|)$ gives
\[
 \begin{aligned}
 &\mathbb E[L\cdot(Z,W)\mathbf1_D\mid -Z_1=x,-Z_2=y]\\
 &\qquad=q(x)F(\mu)[16m(\mu)-Ax-\alpha y]
 \geq q(x)F(\mu)(x+y)/200.
 \end{aligned}
\]
This is nonnegative everywhere in the quadrant.

The split event is a product of simplices of block radii $x/(4\sqrt G)$, whose product has radius $x/4$.
A regular $r$-simplex of circumradius $R$ has volume
$\sqrt{r+1}(1+1/r)^{r/2}R^r/r!\geq R^r/r!$.
For $\sqrt k\leq x\leq\sqrt k+1\leq2\sqrt k$, the Gaussian density and $r_j!\leq r_j^{r_j}$ therefore give
\[
 \begin{aligned}
 q(x)&\geq (2\pi)^{-k/2}e^{-x^2/32}
        \frac{(x/(4\sqrt G))^k}{\prod_jr_j!}\\
 &\geq 3^{-k}e^{-k/8}\left(\frac{\sqrt{G/k}}8\right)^k
 \geq e^{-4k}(G/k)^{k/2}.
 \end{aligned}
\]
Also restrict $0\leq y\leq1$. Then $\mu/u_0<9\sqrt k$.
The event $|Z_3|\leq1$ and $\mu/u_0+1<Z_4<\mu/u_0+2$ implies $U>|V|$, since $u_0>s_0$.
Its probability is at least $e^{-65k}$: the $Z_4$ interval has length one and density at least $e^{-121k/2}/3$, while
$\Pr(|Z_3|\leq1)\geq2e^{-1/2}/3$.
The densities of $x$ and $y$ on the two unit intervals are at least $e^{-4k}$ and $e^{-2k}$, respectively. Since $x\geq1$ and $1/200>e^{-6}$, integration gives
\[
 I:=\mathbb E[L\cdot(Z,W)\mathbf1_D]
 \geq e^{-81k}(G/k)^{k/2}.
\]
Gaussian radial normalization and the antipodal flip event imply, for $a_G=\sqrt{2/\pi}$,
\[
 G_\star=h-\frac4{a_G}\mathbb E[(Z,W)\mathbf1_D],\qquad
 1-\ell(G_\star)=4I/a_G.
\]
This proves the first line of \eqref{eq:star-separation-gap}.
For $d=4$, omit $W$, use $q(x)=1$, $A=\alpha$, and impose $a\cdot Z<0$.
Restricting $1\leq x\leq2$, $0\leq y\leq1$ gives the same estimates with $k=1$, proving its second line.

\noindent\textit{Step 3: A penalty outside the caps.}
Let $D(z)$ be the sum of deficits in the normalized cap rows, so each split deficit divides the last row of \eqref{eq:star-caps} by $\sqrt{1+\epsilon_j^2}$.
Write $D=D_{\rm base}+D_{\rm split}$, separating the $h,c,b$ rows from the split rows.
For a unit point $z=(x,y)$, raise negative $q_h,q_c,q_b$ to zero using the inverse directions $d_h,d_c,d_b$.
Independently raise $t=a\cdot x+p_0$ to
\[
 t'=\max\left\{t,\max_{j,i}\frac{-w_{ji}\cdot y_j}{\epsilon_j}\right\}.
\]
The inner maximum is nonnegative and
\[
 0\leq t'-t\leq
 \max_j\frac{\sqrt{1+\epsilon_j^2}}{\epsilon_j}D_{\rm split}
 \leq16kD_{\rm split}.
\]
Correct $x$ by $(t'-t)d_a$ and leave $y$ fixed. The resulting $z'$ satisfies every cap. The bounds $\|d_\nu\|<4$ and $\alpha,A<6$ give
\[
 \|z'-z\|\leq4D_{\rm base}+64kD_{\rm split}\leq64kD,
 \qquad \ell(z')-\ell(z)\leq8D_{\rm base}+96kD_{\rm split}\leq96kD.
\]
Since $1-\|v\|>1/32$, \eqref{eq:star-edge-bound} yields
\begin{equation}\label{eq:star-global-penalty}
 \ell(z)\geq1-K_{k,G}D(z),\qquad K_{k,G}=2200k.
\end{equation}
Indeed $64k/(1-\|v\|)+96k<2144k<2200k$; if $\ell(z')\geq1$ the bound is immediate. In $d=4$, correct the four inverse coordinates directly; $K_4=2200$ remains a valid bound.

For the one-group geometry, a sharper penalty coefficient strengthens the balanced risk bound. For $G=1$, or for $d=4$, the normalized report-row matrix $E$ is square and invertible. Write the threshold vector as $b_*$, so $E^{-1}b_*=(v,0)$, and express
$\ell(z)=\sum_j a_j[(Ez)_j-(b_*)_j]$.
The coefficients are $a_h=a_c=8$, $a_b=\alpha$, and
$a_j=A\sqrt{1+\epsilon^2}/[\epsilon(k+1)]$ on split rows, with $\epsilon=1/(4k)$; for $d=4$ use $a_a=\alpha$.
Correcting by $E^{-1}(b_*-Ez)_+$ as in \eqref{eq:star-edge-bound} gives another valid penalty coefficient
\[
 K_d^{\rm geo}=\frac{\|E^{-1}\|_{2\to2}}{1-\|v\|}+\max_ja_j.
\]
Let $g_d$ denote the actual separating gap for this one-group geometry (and $g_4$ in dimension four).

\noindent\textit{Step 4: The control recurrences.}
Consider any population $n$ and write $\beta_t=1-\omega_t=\omega_{n-t}$.
At the actual profile $(h^t,r_1^q,\ldots,r_l^q,(-h)^s)$, use the $l=l_G$ leaves above, with $t,q\geq1$ and $s=n-t-lq\geq0$.
If $\beta_t>0$, put
\[
 Q=\frac{P-\omega_t\delta_h}{\beta_t},\qquad
 \gamma=\frac{\beta_{t+q}}{\beta_t}.
\]
Lemma~\ref{lem:peak-mass} makes $Q$ an actual probability. We impose no incentives on $Q$.
If exactly $j$ leaf directions agree in sign with $h$, complementarity and \eqref{eq:control-mean} give its residual mean integrand
$H(1-2\beta_{t+jq}/\beta_t)$.
For $j=0$ this is $F_\star=-H$; for $j\geq1$ it differs from $F_\star=H$ by at most $2\gamma$.
Testing against a unit direction and using $a_d^{-1}\int|u\cdot z|\,d\sigma_d(z)=1$ gives
\begin{equation}\label{eq:star-mean-approximation}
 \|\mathbb E_QX-G_\star\|\leq2\gamma.
\end{equation}

For leaf $r_j$, put $c_j=r_j\cdot h$ and move its $q$ reporters to $h$ one at a time, using true direction $r_j$ and evaluation $\min(z,c_j)$.
The terminal $h$-atom is at least $\omega_{t+q}$, so its expected capped value is at least $c_j-(1+c_j)\beta_{t+q}$.
Telescoping these individual incentive inequalities and subtracting the guaranteed $h$-atom gives
\[
 \mathbb E_Q(c_j-r_j\cdot X)_+\leq(1+c_j)\gamma.
\]
Moving the $t$ central reporters to leaf $c$, with true direction $h$ and cap $\rho$, similarly gives
$\mathbb E_Q(\rho-h\cdot X)_+\leq(1+\rho)\gamma$.
Thus $\mathbb E_QD\leq2(l+1)\gamma$.
Separation, the cap penalty and \eqref{eq:star-mean-approximation} imply
\[
 g_{k,G}\leq[2(l_G+1)2200k+56]\gamma\leq22056k^2\gamma.
\]
As $22056k^2\leq e^{20k}$, complementarity yields
\begin{equation}\label{eq:grouped-control-recurrence}
 \begin{gathered}
 \omega_{r-q}\geq\eta_{k,G}\omega_r,\qquad
 \eta_{k,G}=e^{-120k}(G/k)^{k/2},\\
 1\leq r\leq n-1,\qquad q\geq1,\qquad (k+G+2)q\leq r.
 \end{gathered}
\end{equation}
If $\beta_t=0$ the underlying inequality is automatic. The argument permits an empty antipodal remainder.
In $d=4$, four deficits and Step 2 give the factor $e^{-120}$ whenever $3q\leq r$.

For the one-group geometry, the same argument also retains
\begin{equation}\label{eq:capacity-recurrence}
 \eta_d=\min\left\{\frac12,\frac{g_d}{2dK_d^{\rm geo}+2\|L\|}\right\}>0,
 \qquad\omega_{r-q}\geq\eta_d\omega_r\quad((d-1)q\leq r\leq n-1).
\end{equation}

\noindent\textit{Step 5: Descent to balanced camps.}
Let $n=3m$, $m\geq d-2$ and $r_0=\lceil3m/2\rceil\leq2m$. Monotonicity and complementarity give $\omega_{r_0}\geq1/2$.
First, with $l=d-1$, repeatedly choose
$q=\min\{\lfloor r/l\rfloor,r-m\}$ in \eqref{eq:capacity-recurrence}.
Every unfinished state has $r\geq m+1\geq l$.
Each nonfinal step contracts $r$ by at least the factor $1-1/(2l)$.
Since $(1-1/(2l))^{2l}<e^{-1}<1/2$, at most $2l$ steps reach $m$.
Therefore $3\omega_m/2\geq c_d$ as stated.

For the dimension-explicit bound, let $d\geq5$ and $k=d-4$.
While $r>m$ and $r\geq2k+2$, use $G=k$ and
$q=\min\{\lfloor r/(2k+2)\rfloor,r-m\}$ in \eqref{eq:grouped-control-recurrence}.
Every nonfinal step sends $r$ to at most $[1-1/(4k+4)]r$. Since $r_0\leq2m$ and $[1-1/(4k+4)]^{4k+4}<1/2$, at most $4k+4$ such steps occur.
If $m<r<2k+2$ remains, take $q=1$ and $G=r-k-2$.
The condition $m\geq k+2$ ensures $1\leq G\leq k-1$, and $l_G=r$.
The subsequent group counts are distinct. Since
$\sum_{G=1}^k\log(k/G)\leq k$, their total negative logarithmic factor is at most $120.5k^2$.
The first portion costs at most $480k^2+480k$, so
\[
 \frac32\omega_m\geq\frac34 e^{-600.5k^2-480k}
 \geq e^{-1000(k+4)^2}.
\]
The low-$r$ portion is empty for $k=1$.
In $d=4$, the six-step descent with factor $e^{-120}$ gives
$3\omega_m/2\geq(3/4)e^{-720}>e^{-16000}$ for $m\geq2$.
The planar law \eqref{eq:planar-control-statement} identifies variance with $3\omega_m/2$, proving \eqref{eq:fixed-d-risk}.

\noindent\textit{Step 6: Arbitrary camp shares.}
For \eqref{eq:control-share-bound}, fix $d-2\leq s\leq n$ and start at $r_0=\lceil n/2\rceil$.
If $s\geq r_0$, $\omega_s\geq1/2$ suffices.
Otherwise $r_0\leq n-1$ and, for $d\geq5$, use the preceding descent toward $s$ with $L_0=2k+2$.
The number of high-$r$ steps is at most $2L_0\log(n/s)+2$: all but possibly the last contract by $1-1/(2L_0)$ and remain above $s$.
The low-$r$ group counts are again distinct because $s\geq k+2$.
Consequently
\[
 \omega_s\geq\tfrac12\exp\{-(480k^2+480k)\log(n/s)-120.5k^2-240k\}
 \geq b_d(n,s).
\]
For $d=4$, every unfinished state has $r\geq s+1\geq3$.
Using $q=\min\{\lfloor r/3\rfloor,r-s\}$ gives at most $6\log(n/s)+2$ steps, whose cost is at most $720\log(n/s)+240$.
This also implies the stated bound.

\noindent\textit{Step 7: Individual report perturbations.}
Return to the original rule $P$. For a nonempty proper $A$, a unit direction $v$, $0<\eta<2$ and $D=\sum_{i\in A}\|r_i-v\|$, we first prove
\begin{equation}\label{eq:coalition-concentration}
 \mathbb E_{P(r)}(v\cdot X-1+\eta)_+
 \geq w_A(\eta-2D^2).
\end{equation}
If $w_A=0$ the claim is immediate. Otherwise fix outsiders at $-v$. Their guaranteed atom makes
$Q=[P(r_A,(-v)_{A^c})-(1-w_A)\delta_{-v}]/w_A$
an actual probability. The fixed subtracted atom cancels in each incentive comparison, so $Q$ inherits linear IC for $A$. Initially all members report $v$ and $Q=\delta_v$. Restore their reports one at a time. If $a,b$ are the residual means before and after restoring $r_i$, true-$r_i$ IC gives $r_i\cdot(b-a)\geq0$. Put $y=1-v\cdot a$, $z=1-v\cdot b$. Ball-valued means give
\[
 z-y\leq(r_i-v)\cdot(b-a)
 \leq\sqrt2\|r_i-v\|(\sqrt z+\sqrt y).
\]
Dividing, with the zero-denominator case immediate, and telescoping gives $1-v\cdot\mathbb E_QX\leq2D^2$. The positive-part inequality proves \eqref{eq:coalition-concentration} at this intermediate profile. Restoring outsiders from $-v$ cannot decrease its left side: for $\phi(u)=\min\{u,-1+\eta\}$,
$\phi(-v\cdot X)=-1+\eta-(v\cdot X-1+\eta)_+$,
so true-$(-v)$ IC has precisely this implication.

Now put $t_d=e^{-1000d^2}$, $\gamma=1-c/t_d$, $\rho=\gamma/32$ and $\delta=\gamma^{3/2}/(128m)$. Write the reference profile as $(v_1^{A_1},v_2^{A_2},v_3^{A_3})$, where $|A_k|=m$ and $\sum_kv_k=0$. The caps
$C_k=\{x:v_k\cdot x\geq1-\rho^2/2\}$
have radius $\rho$ in Euclidean distance and are disjoint. Since each camp's total displacement is at most $m\delta$, \eqref{eq:coalition-concentration} and the tent bound $(\rho^2/2)\mathbf1_{C_k}$ give
$P(r)(C_k)\geq\omega_m(1-\gamma/4)\geq(2t_d/3)(1-\gamma/4)$.
For $l\in A_i$, let $q=r_l$, $M=\mathbb E_{P(r)}q\cdot X$ and $y_k=v_i\cdot v_k$. On $C_k$,
$|q\cdot X-y_k|\leq\delta+\rho=:e$ and
$(q\cdot X-M)^2\geq\max\{0,(y_k-M)^2-4e\}$.
Integrate over the disjoint caps, use their common lower mass, and then drop the nonnegative clipping. Since $\sum_k(y_k-M)^2=3/2+3M^2$ and $12e\leq3\gamma/4$,
\[
 \operatorname{Var}_{P(r)}(q\cdot X)
 \geq t_d(1-\gamma/4)(1-\gamma/2)
 >t_d(1-\gamma)=c.\qedhere
\]
\end{proof}

\subsection{Two diagnostics of individual control}\label{app:pair-control-proof}\label{subsec:control-diagnostics}
The preceding control representation and recurrence give two tests for random dictatorship: one based on pairs of participants, the other on the local response of alignment risk to reports.

In dimension four, pairwise additivity already characterizes the whole rule. Under full-domain ICV-SP and binary efficiency, for every $n\geq2$,
\begin{equation}\label{eq:pair-control-diagnostic}
 w_{\{i,j\}}=w_{\{i\}}+w_{\{j\}}\quad(i\ne j)
 \quad\Longleftrightarrow\quad
 P\text{ is fixed-weight random dictatorship}.
\end{equation}
Thus any other rule in four dimensions gives some pair strictly more guaranteed control than its members separately. This conclusion requires neither symmetry nor report regularity. The control recurrence gives a short proof.

\begin{proof}[Proof of \eqref{eq:pair-control-diagnostic}]
The converse is immediate. Suppose all pair weights are additive. For $n=2$ the pair is $N$, so singleton weights sum to one and the binary law proves the claim. Let $n\geq3$. Form the finite label average $P^\circ(r)=(n!)^{-1}\sum_\pi P(r_\pi)$, which is anonymous, ICV-SP and binary efficient. Write its weights as $\omega_j$, and put
\[
 u=\omega_1=\frac1n\sum_iw_{\{i\}},\qquad \rho=1-nu\geq0.
\]
Pair additivity of the original controls gives $\omega_2=2u$. Suppose $\rho>0$. At every profile, each maximal common-report camp $A$ has a guaranteed atom of mass at least $\omega_{|A|}\geq |A|u$. Thus
\[
 \widetilde P(r)=\frac{P^\circ(r)-u\sum_i\delta_{r_i}}{\rho},\qquad
 \theta_j=\frac{\omega_j-ju}{\rho}
\]
define actual sphere probabilities and normalized, complementary, monotone, superadditive controls. In particular $\theta_1=\theta_2=0$. By linearity of \eqref{eq:control-mean}, subtracting the singleton terms gives the represented mean $G_\theta$ for $\widetilde P$; this does not apply the incentive theorem to a residual rule.

The cap comparisons used in Steps 3--4 of Appendix~\ref{app:fixed-d-risk} survive this subtraction. For a true report $p$, alternative $s$, and cap $\phi_t(z)=\min(z,t)$ with $t\leq p\cdot s$, the changing singleton has the same value $\phi_t(1)=\phi_t(p\cdot s)=t$ at both reports. Other subtracted singleton atoms cancel. Every leaf-to-center and center-to-leaf comparison in that argument uses precisely such a cap, at the corresponding report inner product. Peak guarantees likewise give residual camp atoms at least $\theta_{|A|}$, after subtraction of all the singleton masses in the camp.

The four-dimensional cap separation uses only these comparisons, the represented mean and actual nonnegative residual probabilities. It therefore gives the same positive recurrence for $\theta$:
\[
 \theta_{r-q}\geq e^{-120}\theta_r
 \qquad(1\leq r\leq n-1,\ q\geq1,\ 3q\leq r).
\]
No full ICV property of $\widetilde P$ is asserted. If $n\geq4$, start with $\theta_{n-1}=1$ and iterate $q=1$ down to $r=3$. This implies $\theta_2>0$, a contradiction. For $n=3$, complementarity directly contradicts $\theta_1=\theta_2=0$. Therefore $\rho=0$.

The original singleton weights consequently sum to $nu=1$. For each maximal common-report group $A$, its guaranteed mass is at least $w_A\geq\sum_{i\in A}w_{\{i\}}$. These masses exhaust probability, proving fixed-weight random dictatorship at every original profile.
\end{proof}

The second diagnostic applies in every $d\geq4$. Fix perpendicular unit vectors $a,c$, and order the $n\geq3$ labels once. For $k=1,\ldots,n-2$, put $A_k=\{1,\ldots,k\}$, $B_k=\{k+1\}$, and $C_k=\{k+2,\ldots,n\}$. Along the report path
\[
 r^k(t)=\bigl(a^{A_k},[-\cos(t)a+\sin(t)c]^{B_k},c^{C_k}\bigr),
 \qquad F_k(t)=\mathbb E_{P(r^k(t))}(c\cdot X)^2,
\]
$F_k$ is the actual alignment second moment of camp $C_k$. Its one-sided derivatives at zero exist, and their difference $J_k=F_k'(0-)-F_k'(0+)$ satisfies
\begin{equation}\label{eq:nested-kink-sum}
 J_k\geq0,\qquad
 \sum_{k=1}^{n-2}J_k=2\left[1-\sum_{i=1}^n w_{\{i\}}\right].
\end{equation}
Consequently the rule is fixed-weight random dictatorship if and only if these $n-2$ second-moment paths are differentiable at zero. Every other rule has a positive derivative jump on at least one path. This does not require or imply discontinuity of lottery laws.

\begin{proof}
For any three-camp partition $(A,B,C)$ along the same path, write $v_1=w_A$, $v_2=w_B$, $v_3=w_C$ and $\kappa=1-v_1-v_2-v_3$. For small $t\geq0$, the angular median of the three reports is $c$, so $c\cdot g_3=1$. For small $t<0$, the three directions surround the origin and $g_3=a+b(t)+c$, giving $c\cdot g_3=1+\sin t$. These formulas also agree at the actual antipodal profile $t=0$. Theorem~\ref{thm:control-identification} therefore gives
\[
 F(t)=v_2\sin^2t+v_3+
 \begin{cases}
 \kappa,&t\geq0,\\
 \kappa(1+\sin t)^2,&t<0.
 \end{cases}
\]
Hence $F'(0-)-F'(0+)=2\kappa$. With $A_{n-1}=\{1,\ldots,n-1\}$, complementarity gives
$\kappa_k=w_{A_{k+1}}-w_{A_k}-w_{\{k+1\}}$ for the nested partitions. Summing telescopes to \eqref{eq:nested-kink-sum}. If every jump vanishes, singleton guarantees sum to one and exhaust the lottery at every profile, including coincident reports. Conversely random dictatorship gives smooth second moments along each path. No regularity of the original rule has been assumed.
\end{proof}

In particular, everywhere differentiability of the raw second-moment map $r\mapsto\mathbb E_{P(r)}XX^\top$ forces random dictatorship. Quantitatively, some tested path has a derivative jump at least $2[1-\sum_iw_{\{i\}}]/(n-2)$. These statements concern the specified paths; they do not assert that $n-2$ is the minimum number of differentiability tests.

\section[Approximate incentives and mean-preserving insurance]{\raggedright Approximate incentives and mean-preserving \mbox{insurance}}\label{app:insurance-error-proof}
\subsection{A population-uniform lower bound}\label{app:insurance-error-lower}
\begin{proof}[Proof of the lower bound in Proposition~\ref{prop:insurance-error}]

Write $\eta=(4d)^{-(d-1)}/[1+2d(d+1)]$ and $C_d=3\eta^{2d}/4$. We first show that some $\tau_d>0$, independent of $m\geq d-1$, satisfies
\begin{equation}\label{eq:robust-fixed-d-risk}
 nI(P)+\Delta(P)\leq\tau_d
 \quad\Longrightarrow\quad \mathcal L(P,r_m)\geq C_d/2.
\end{equation}

Fix a unit $h$ and regular-simplex vertices $u_1,\ldots,u_d$ in $h^\perp$,
with $\sum_j u_j=0$ and $u_i\cdot u_j=-1/(d-1)$ for $i\ne j$.
Set $a=1/(4d)$, $c=-a$, and $r_j=ch+\sqrt{1-a^2}u_j$.
Every $x\in S$ satisfies
\begin{equation}\label{eq:simplex-penalty}
 1-h\cdot x\leq4dD(x),\qquad
 D(x)=(c-h\cdot x)_++\sum_{j=1}^d(c-r_j\cdot x)_+.
\end{equation}
Indeed, write $x=zh+y$, $y\perp h$, and $f=1-z\in[0,2]$.
The simplex identity $\sum_j u_ju_j^\top=dI/(d-1)$ implies
$\min_j u_j\cdot y\leq-\|y\|/(d-1)$: otherwise, for $y\ne0$,
the positive weights $[1+(d-1)u_j\cdot y/\|y\|]/d$ would express
the unit vector $y/\|y\|$ as a strict convex combination of distinct unit vectors.
Consequently one leaf deficit is at least
$[\sqrt{1-a^2}\sqrt{f(2-f)}/(d-1)-af]_+$.
For $0<f\leq3/2$ this is at least
$f[\sqrt5/(4(d-1))-1/(4d)]\geq f/(4d)$.
For $3/2\leq f\leq2$, the central deficit is
$f-1-a\geq f-5/4\geq f/6\geq f/(4d)$.
The case $f=0$ is immediate, proving the global bound.

For an integration direction $v$, let $H=\operatorname{sgn}(h\cdot v)$
and $R_j=\operatorname{sgn}(r_j\cdot v)$.
Define $F_\star=-H$ when every $R_j=-H$, and $F_\star=H$ otherwise;
put $G_\star=a_d^{-1}\int_S vF_\star\,d\sigma_d(v)$.
This vector will supply a lower bound on the mean loss forced by the cap constraints.
On the polar caps $|h\cdot v|>\sqrt{1-a^2}$ every leaf sign is $-H$:
for $z=h\cdot v>\sqrt{1-a^2}$,
$r_j\cdot v\leq-az+\sqrt{1-a^2}\sqrt{1-z^2}<0$, and antipodality gives the other cap.
The density of $z$ is proportional to $(1-z^2)^{(d-3)/2}$, so direct integration gives
\begin{equation}\label{eq:star-mean-loss}
 1-h\cdot G_\star
 =\frac2{a_d}\int_{\{R_j=-H\ \forall j\}}|h\cdot v|\,d\sigma_d(v)
 \geq2a^{d-1}.
\end{equation}
Here the ratio of the $|z|$ integral over the two polar caps to its full integral is $a^{d-1}$.

The associated integer descent is also independent of the sharper configuration in Appendix~\ref{app:fixed-d-risk}. For $m\geq d-1$, start at $r_0=\lceil3m/2\rceil$. While $r>m$, subtract $q=\min\{\lfloor r/d\rfloor,r-m\}$. Every unfinished state has $r\geq m+1\geq d$, so $q\geq1$. Unless the step finishes, the next state is at most $(1-1/(2d))r$. Since $r_0\leq2m$ and $(1-1/(2d))^{2d}<e^{-1}<1/2$, at most $2d$ steps reach $m$.

\noindent\textit{Step 1: A fixed-input control restriction.}
Consider an exact ICV-SP rule efficient on $\mathcal E_2^\perp$, with one central input $0$, $d$ leaf inputs and optionally one further input. Its control weights need not be anonymous. We claim
\begin{equation}\label{eq:weighted-star}
 \max_{1\leq j\leq d}(1-w_{\{0,j\}})\geq\eta(1-w_{\{0\}}).
\end{equation}
Use $h,r_1,\ldots,r_d$ and the optional report $-h$ defined above. Put $b_0=1-w_{\{0\}}$ and $b_1=\max_j(1-w_{\{0,j\}})$. If $b_0=0$ the claim is immediate. Otherwise the peak guarantee makes $Q=(P-w_{\{0\}}\delta_h)/b_0$ an actual probability at this profile. In the sign coalition agreeing with $h$, the optional input is absent. If no leaf agrees, the residual integrand in the mean representation is exactly $-\operatorname{sgn}(h\cdot v)$; if a leaf $j$ agrees, monotonicity bounds its missing control weight by $b_1$. Hence the comparison mean $G_\star$ defined above satisfies
\[
 \|\mathbb E_QX-G_\star\|\leq2b_1/b_0.
\]
Changing leaf $j$ to $h$, the actual final $h$-atom is at least $w_{\{0,j\}}$. With $c=-1/(4d)$, the original capped-utility comparison gives
$\mathbb E_Q(c-r_j\cdot X)_+\leq(1+c)b_1/b_0$.
Changing the central input to one leaf gives the same bound for $(c-h\cdot X)_+$. In both comparisons the subtracted $h$-atom has capped value $c$. Apply the whole-sphere penalty and polar-cap gap, equations~\eqref{eq:simplex-penalty} and~\eqref{eq:star-mean-loss}, to obtain
\[
 2(4d)^{-(d-1)}-2b_1/b_0
 \leq4d(d+1)(1-1/(4d))b_1/b_0.
\]
This proves \eqref{eq:weighted-star}. Only comparisons of the original rule were used; no incentive property of $Q$ was assumed.

\noindent\textit{Step 2: Stability of finitely many features.}
The rules remain defined on the full report domain; only the input count and the collection of outcome moments are finite. Fix an input count $\ell$ and orthogonal unit vectors $e,f$. For every $J\subseteq\{1,\ldots,\ell\}$, write
\[
 x_J(Q)=e\cdot\int x\,dQ(e^J,f^{J^c})(x).
\]
For an exact rule these are its binary weights, by Lemma~\ref{lem:binary}. For approximate rules they are only mean coordinates. There is a nondecreasing modulus $\delta_{\ell,d}(t)\to0$ as $t\downarrow0$ such that $I(Q),\Delta(Q)\leq t$ implies that $x(Q)$ lies within $\delta_{\ell,d}(t)$, in sup norm, of the weight vector of an exact $\ell$-input rule.

To prove this, place all assignments in the compact product
$\Prob_B(S)^{S^\ell}$, with the weak topology in every probability coordinate. Exact incentive inequalities for continuous evaluations and exact efficiency on the tested profiles are closed: each uses fixed report coordinates and continuous outcome functions.

Continuous evaluations suffice. Indeed, for any $\phi\in\mathcal U$, its Bernstein polynomials
\[
 B_k\phi(u)=\sum_{j=0}^k\binom{k}{j}
 \phi(-1+2j/k)t^j(1-t)^{k-j},\qquad t=(u+1)/2,
\]
remain normalized, nondecreasing and concave, by their first and second finite differences. They are bounded in $[0,1]$, converge at every interior point, and equal $\phi$ at both endpoints. Bounded convergence recovers the full incentive class, including its endpoint jump, after taking an exact limit.

Thus the exact class is compact and nonempty, and its finite feature image is compact. If the asserted modulus failed, a sequence of approximate rules with errors tending to zero would have a cofinal product-weak convergent subnet. Its limit would be exact, and continuity of the finite features would contradict their positive distance from the exact image. The compactness argument uses finitely many report coordinates at a time. The argument also applies to an anonymous subclass, which is closed, and to any additional finite list of continuous outcome moments at fixed reports.

Choose one modulus $\delta(t)$ dominating the finitely many cases $\ell=2,3,d+1,d+2$. Include the anonymous three-input case with the equilateral moment $R$ as an additional feature. Its exact image obeys $R=3w_{\{1\}}/2$, by Theorem~\ref{thm:control-identification}. Consequently, for an approximate anonymous three-input rule,
\begin{equation}\label{eq:three-input-feature}
 |R(Q,(p_1,p_2,p_3))-(3/2)x_{\{1\}}(Q)|
 \leq(5/2)\delta(t).
\end{equation}

\noindent\textit{Step 3: Control coordinates for the original agents.}
For any original $P$, average its actual laws over the finitely many label permutations to obtain $\bar P$. This rule is anonymous, with $I(\bar P)\leq I(P)$, $\Delta(\bar P)\leq\Delta(P)$ and $R(\bar P,r_m)=\mathcal L(P,r_m)$. Let $T=nI(P)+\Delta(P)$ and
\[
 b_s=e\cdot\int x\,d\bar P(e^s,f^{n-s})(x),\qquad0\leq s\leq n.
\]
Group the original agents into any fixed number of nonempty blocks. A block deviation telescopes at most $n$ individual incentive comparisons, so its incentive error is at most $nI(P)$ and its efficiency distance is at most $\Delta(P)$. Applying the preceding finite-feature bounds to two and three blocks gives
\begin{align}
 |b_s+b_{n-s}-1|&\leq2\delta(T),&&1\leq s<n,\label{eq:approx-complement}\\
 b_u&\leq b_v+2\delta(T),&&1\leq u<v<n.\label{eq:approx-monotone}
\end{align}
For example, the three blocks of sizes $u,v-u,n-v$ compare the exact nearby coordinates $w_{\{1\}}\leq w_{\{1,2\}}$. All blockings refer to the same original numbers $b_s$.

Next take a central block of size $t$, $d$ leaf blocks of size $q$, and the nonempty remainder if there is one. For $t,q\geq1$ and $t+dq\leq n$, equation~\eqref{eq:weighted-star} for the nearby exact control vector gives
\[
 1-b_{t+q}\geq\eta(1-b_t)-(1+\eta)\delta(T).
\]
Using \eqref{eq:approx-complement} at $t=n-r$ and $t+q$ yields
\begin{equation}\label{eq:approx-recurrence}
 b_{r-q}\geq\eta b_r-3(1+\eta)\delta(T)
          \geq\eta b_r-6\delta(T),
 \qquad dq\leq r\leq n-1.
\end{equation}
At $r_0=\lceil3m/2\rceil$, equations~\eqref{eq:approx-complement}--\eqref{eq:approx-monotone} imply $b_{r_0}\geq1/2-2\delta(T)$.

The integer descent proved above, with decrements $\min\{\lfloor r/d\rfloor,r-m\}$, reaches $m$ in at most $2d$ steps. Iterating \eqref{eq:approx-recurrence} therefore gives
\[
 b_m\geq\tfrac12\eta^{2d}-(2+12d)\delta(T).
\]
The equal-block three-input contraction is anonymous. Apply \eqref{eq:three-input-feature} to it; its binary coordinate is exactly $b_m$ and its equilateral moment is $R(\bar P,r_m)$. Hence
\[
 \mathcal L(P,r_m)\geq C_d-(11/2+18d)\delta(T).
\]
The finite feature lists and their moduli depend only on $d$, not $m$. Since $\delta(t)\to0$, a positive $\tau_d$ gives \eqref{eq:robust-fixed-d-risk}. Rotating a rule before this argument proves the same threshold at every orientation, without a neutrality assumption or limits at moving report coordinates.

Finally take $b_d=\bar\delta_d=\tau_d/2$, decreasing $\tau_d$ to at most one if needed. An eligible rule with $\mathcal L(P,r_m)=0$ and $\Delta(P)\leq\bar\delta_d$ cannot satisfy \eqref{eq:robust-fixed-d-risk}; hence $I(P)>b_d/n$. Taking the infimum proves the lower bound in \eqref{eq:insurance-error-order}.
\end{proof}

\subsection{A matching full-domain construction}\label{app:insurance-error-upper}
\begin{proof}[Proof of the upper bound in Proposition~\ref{prop:insurance-error}]

Fix $d\geq4$ and $n\geq1$.

\noindent\textit{Step 1: Construction.}
For a report profile $r$, put
\[
 \mu=\frac1n\sum_i r_i,\qquad M=\frac1n\sum_i r_ir_i^{\mathsf T},
 \qquad \lambda_1\geq\lambda_2\geq\lambda_3\geq\cdots\geq\lambda_d,
\]
where the $\lambda_j$ are the eigenvalues of $M$. Define
\[
 \chi(t)=\min\{1,\max\{0,8(t-1/4)\}\},\qquad
 \alpha=\chi(\lambda_2-\|\mu\|^2)\chi(\lambda_2-\lambda_3).
\]
If $\alpha>0$, let $\Pi$ be the orthogonal projector onto the lower spectral subspace of $M$, of dimension $d-2$ and let $K_\Pi$ be uniform probability on $S^{d-1}\cap\operatorname{ran}\Pi$. Set
\begin{equation}\label{eq:spectral-insurance-rule}
 P_n(r)=(1-\alpha)\frac1n\sum_i\delta_{r_i}+\alpha K_\Pi.
\end{equation}
At $\alpha=0$ use random dictatorship. On the active set the spectral gap exceeds $1/4$, so the projector is unique, including ties within either block. Projecting a standard Gaussian and normalizing realizes $K_\Pi$ as an actual probability. The active projector is continuous by gap separation. Coupling the projected Gaussian realizations gives weak continuity of $K_\Pi$ there. At $\alpha=0$, the difference of the mixture from random dictatorship in any continuous test $f$ is bounded by $2\|f\|_\infty\alpha$. Thus the full rule is weakly continuous and hence jointly Borel. Label invariance and Gaussian orthogonal invariance give anonymity and neutrality, including reflections.

For a two-position profile with shares $t,1-t$ at $a,b$,
\[
 M=\mu\mu^{\mathsf T}+t(1-t)(a-b)(a-b)^{\mathsf T}.
\]
On the $(d-1)$-dimensional space $(a-b)^\perp$ its Rayleigh quotient is at most $\|\mu\|^2$, so min--max gives $\lambda_2\leq\|\mu\|^2$ and $\alpha=0$. The coincident case is immediate. Hence the rule is random dictatorship at every such profile, including antipodes, and satisfies both efficiencies there. At balanced equilateral disagreement with report plane $L$, $\mu=0$ and $M=\frac12\operatorname{Proj}_L$. Thus $\alpha=1$, $\Pi=\operatorname{Proj}_{L^\perp}$, proving the required zero alignments.

\medskip
\noindent\textit{Step 2: The effect of one report change.}
Primes denote the profile after a single replacement. Directly,
\[
 \|\mu-\mu'\|\leq2/n,\quad
 \|M-M'\|_{\rm op}\leq2/n,\quad
 \bigl|\|\mu\|^2-\|\mu'\|^2\bigr|\leq4/n.
\]
Min--max bounds each eigenvalue change by $2/n$. Since $\chi$ is $8$-Lipschitz and valued in $[0,1]$,
\begin{equation}\label{eq:spectral-weight-change}
 |\alpha-\alpha'|\leq80/n.
\end{equation}
Suppose $n\geq32$ and both coefficients are positive. Put $\delta=\|M-M'\|_{\rm op}\leq1/16$. An upper-block eigenvalue of $M$ and a lower-block eigenvalue of $M'$ differ by more than $3/16$. For corresponding orthonormal eigenvectors $u_i,v_j'$, $i\leq2<j\leq d$,
\[
 (\lambda_i-\lambda_j')u_i\cdot v_j'=u_i\cdot(M-M')v_j'.
\]
The matrix of numerator inner products has two rows and operator norm at most $\delta$, so its Frobenius norm is at most $\sqrt2\delta$. Squaring and summing gives $\|(I-\Pi)\Pi'\|_F\leq\sqrt2\delta/(3/16)$. The trace identity for equal-rank projectors then gives
\begin{equation}\label{eq:spectral-projector-change}
 D:=\|\Pi-\Pi'\|_{\rm op}
 \leq\|\Pi-\Pi'\|_F
 =\sqrt2\|(I-\Pi)\Pi'\|_F
 \leq\frac{32}{3}\delta<\frac{32}{n}.
\end{equation}
The rank of $M$ may change; on the active set, $\Pi$ always projects onto a $(d-2)$-dimensional spectral subspace.

\medskip
\noindent\textit{Step 3: A uniform bound on utility changes.}
If $X$ is uniform on the unit circle of a two-dimensional plane $E$, its alignment with a fixed unit $p$ has law $A\cos\Theta$, where $A=\|\operatorname{Proj}_E p\|$ and $\Theta$ is uniform on $[0,2\pi]$. For $-1<t\leq1$, let
\[
 f_t(u)=\min\{(u+1)/(1+t),1\}.
\]
For $A>0$ and $|t|<A$, elementary integration gives
\[
 \frac{d}{dA}\mathbb Ef_t(A\cos\Theta)
 =-\frac{\sqrt{A^2-t^2}}{\pi A(1+t)}.
\]
The expectation is constant on $A<|t|$. Write $A=\sin\beta$. The identity
\[
 A^2(1+t)^2-(1-A^2)(A^2-t^2)=(A^2+t)^2\geq0
\]
bounds the absolute $\beta$ derivative by $1/\pi$. Continuity at the transition and endpoints, and the formula $\mathbb Ef_0(A\cos\Theta)=1-A/\pi$, therefore yield
\begin{equation}\label{eq:cap-angular-modulus}
 \left|\mathbb Ef_t(A\cos\Theta)-\mathbb Ef_t(B\cos\Theta)\right|
 \leq\frac{|\arcsin A-\arcsin B|}{\pi}.
\end{equation}

Nonnegative combinations of capped-linear utilities extend this bound to every nondecreasing concave $\phi:[-1,1]\to[0,1]$. To see this without imposing endpoint continuity, replace only $\phi(-1)$ by its right limit. The resulting $\phi_c$ is continuous and concave on the closed interval; monotonicity and concavity also force continuity at $1$. The circle alignment laws have no atom at $-1$, even for $A=0$ or $1$, so this replacement preserves their evaluations. Let $\phi_k$ be a piecewise linear interpolant on knots $-1=t_0<\cdots<t_k=1$, with nonnegative decreasing slopes $l_j$. Exactly,
\[
 \phi_k(u)=\phi_c(-1)+2l_k f_1(u)
 +\sum_{1\leq j<k}(l_j-l_{j+1})(1+t_j)f_{t_j}(u).
\]
The coefficients are nonnegative and sum to $\phi_c(1)-\phi_c(-1)\leq1$. Uniform interpolation convergence proves \eqref{eq:cap-angular-modulus} for $\phi$.

For a uniform unit vector $W$ in $\R^k$, $k=d-2\geq2$, write $W_1=R\cos\Theta$, where $R\in[0,1]$ and $\Theta$ is uniform and independent of $R$ by planar rotational invariance. Conditional on $R$, the circle bound applies with amplitudes $AR,BR$. Since $\beta\mapsto\arcsin(R\sin\beta)$ is one-Lipschitz, averaging yields the same bound with $A,B$. Thus \eqref{eq:cap-angular-modulus} also bounds every normalized evaluation of the subspace-sphere alignment laws.

For two orthogonal projectors of the same rank, $D=\|\Pi-\Pi'\|_{\rm op}\leq1$. The minimum angular distances of $p$ to their subspace unit spheres are $\arccos\|\Pi p\|$ and $\arccos\|\Pi'p\|$; they differ by at most $\arcsin D$. Indeed, for $D<1$, project and normalize a closest unit vector in one subspace onto the other; its angular displacement is at most $\arcsin D$. Apply the spherical triangle inequality in both directions. For $D=1$ the assertion is immediate. Taking complementary angles gives
\[
 |\arcsin\|\Pi p\|-\arcsin\|\Pi'p\||\leq\arcsin D.
\]
Using \eqref{eq:cap-angular-modulus} and $\arcsin D\leq\pi D/2$, for every allowed evaluation,
\begin{equation}\label{eq:circle-evaluation-modulus}
 \left|\int\phi(p\cdot x)\,dK_\Pi(x)
 -\int\phi(p\cdot x)\,dK_{\Pi'}(x)\right|\leq D/2.
\end{equation}

\medskip
\noindent\textit{Step 4: The incentive error.}
For any fixed true direction and evaluation, the random-dictatorship evaluation changes by at most $1/n$. The absolute change under \eqref{eq:spectral-insurance-rule} is bounded by this amount, plus $|\alpha-\alpha'|$, plus $\min\{\alpha,\alpha'\}$ times the subspace-evaluation difference. If either coefficient is zero the last term vanishes. Otherwise, for $n\geq32$, \eqref{eq:spectral-weight-change}--\eqref{eq:circle-evaluation-modulus} bound the change by $97/n<100/n$. For $n<32$ the range bound one suffices; it also allows truncation by one for every $n$. This proves the upper bound in \eqref{eq:insurance-error-order} for every unilateral deviation and completes the construction.
\end{proof}

\section{The boundary of three-camp risk identification}\label{app:rank-boundary}

The planar restriction in Theorem~\ref{thm:control-identification} matters even when binary control is fixed. For every $d\geq4$ and $n\geq3$, we construct two anonymous, orthogonally neutral, jointly Borel, full-domain ICV-SP rules that satisfy both efficiencies on $\mathcal E_2$. They have the same mean at every profile and the same binary controls
\begin{equation}\label{eq:three-committee-control}
 \omega_s=\frac{\binom{s}{3}+\binom{s}{2}(n-s)}{\binom n3},
 \qquad 0\leq s\leq n,
\end{equation}
where binomial coefficients vanish outside their integer ranges. At every three-camp profile whose directions are linearly independent, every participant has different alignment variance under the two rules. At rank at most two, the theorem identifies all reported alignment laws for every admissible rule with these controls. Thus rank at most two is the exact pointwise domain of three-camp risk identification for this control vector. The construction uses the control vector in \eqref{eq:three-committee-control}.

We perturb the three-person committee rule without changing its mean, control or incentives. Uniform feasibility and incentive bounds allow one perturbation size to work on the full report domain. Averaging over committees then gives the two population rules.

\subsection{A three-person kernel with variable reported risk}

For a triple $t=(t_1,t_2,t_3)$, let $W=\operatorname{span}\{t_1,t_2,t_3\}$ and define
\[
 g(t)=\frac1{a_d}\int_S vM_3(\operatorname{sgn}(t_1\cdot v),
       \operatorname{sgn}(t_2\cdot v),\operatorname{sgn}(t_3\cdot v))\,d\sigma_d(v),
 \qquad B=1-\|g\|^2.
\]
Here $M_3$ is majority of three signs. The Gaussian representation
$g=\mathbb E[Z M_3]/\mathbb E|Z_1|$ shows that its value is independent of ambient dimension and lies in $W$. Testing against any unit vector gives $\|g\|\leq1$. For a fixed pair of opponents, changing one's sign changes the majority only when the opponents disagree; reporting truth maximizes each directional contribution to one's mean alignment. Hence $g$ satisfies finite linear incentives. With two reports equal, $g$ is their common direction.

At rank three, let $T=[t_1\ t_2\ t_3]$ and write
\[
 \begin{gathered}
 K=T^\top T=\begin{pmatrix}1&x&y\\x&1&z\\y&z&1\end{pmatrix},\qquad D=\det K,\\
 s_x=\sqrt{1-x^2},\qquad s_y=\sqrt{1-y^2},\qquad s_z=\sqrt{1-z^2}.
 \end{gathered}
\]
Define
\begin{equation}\label{eq:noisy-gram}
 \begin{gathered}
 S_0=\operatorname{diag}(s_z,s_y,s_x),\quad
 \eta=\frac{D}{s_xs_ys_z},\quad N=\sqrt D\,TK^{-1}S_0^{-1},\\
 Q=N^\top N=S_0^{-1}\operatorname{adj}(K)S_0^{-1},\qquad
 A=\eta^2NQN^\top.
 \end{gathered}
\end{equation}
The columns of $N$ are unit dual normals, $Q$ is positive definite with diagonal one and $\det Q=\eta^2$. Thus $A$ is positive definite on $W$, zero on $W^\perp$, and $\operatorname{tr}A=\eta^2\operatorname{tr}(Q^2)\leq9\eta^2$. At lower ranks set $A=0$.

Put
\begin{equation}\label{eq:noisy-epsilon}
 C=3072\pi e^{9/2},\qquad L=39\pi/2,\qquad
 \varepsilon_* =\min\{(18C)^{-1},(256C^2)^{-1},(12L^2)^{-1}\}.
\end{equation}
Fix $0<\varepsilon\leq\varepsilon_*$ and $k_\varepsilon=\sqrt{3\varepsilon}$.
At rank three draw $U$ uniformly on the unit sphere in $W$ and, independently, $Z$ uniformly on the unit sphere in $W^\perp$. Set
\begin{equation}\label{eq:noisy-kernel}
 Y=g+k_\varepsilon A^{1/2}U,\qquad
 X=Y+\sqrt{1-\|Y\|^2}\,Z.
\end{equation}
At rank at most two use $X=g+\sqrt B Z$ with $Z$ uniform on the normal sphere. Denote this rule by $H_\varepsilon$ and denote the normal lift at every rank by $H_0$. The next two subsections prove that \eqref{eq:noisy-kernel} is feasible and ICV-SP on the full report domain.

\subsection{Uniform slack for sphere feasibility}

We prove the two estimates
\begin{equation}\label{eq:noisy-radial-bounds}
 \operatorname{tr}A\leq9CB,\qquad
 g^\top Ag\leq\frac{4C^2}{3}B^2.
\end{equation}
These bounds hold uniformly over every positive-definite Gram matrix, including arbitrary approaches to rank loss.

Work in the three-dimensional $W$. If $g\ne0$, put $h=g/\|g\|$ and let $M$ denote the majority of Gaussian report signs. For a standard Gaussian $V$ in $W$, radial normalization gives
\[
 B\geq1-\|g\|=\frac4{\mu_3}
 \mathbb E\bigl[|h\cdot V|\mathbf1_{\{M\ne\operatorname{sgn}(h\cdot V)\}}\bigr],
 \qquad \mu_3=2\sqrt{2/\pi}.
\]
Normalize report coordinates by their conditional standard deviations:
$t_i\cdot V=(\sqrt D/(S_0)_{ii})\xi_i$.
Their exact density is
\[
 f_\xi(y)=(2\pi)^{-3/2}\eta e^{-y^\top Qy/2}
 \geq c_0\eta\quad(y\in[-1,1]^3),\qquad c_0=(2\pi)^{-3/2}e^{-9/2},
\]
because $\operatorname{tr}Q=3$ and $\|y\|^2\leq3$.
Also $V=N\xi$; set $b=N^\top h$.
For any $b\in\R^3$, the cube majority satisfies
\[
 \int_{[-1,1]^3}|b\cdot y|
 \mathbf1_{\{M_3(\operatorname{sgn}y)\ne\operatorname{sgn}(b\cdot y)\}}\,dy
 \geq\|b\|/1024.
\]
To verify this, normalize $\|b\|=1$ and choose $j$ with $|b_j|\geq1/\sqrt3$.
Take $y_j$ with the sign of $b_j$ and magnitude in $[3/4,1]$, and both other coordinates in $[-1/8,0]$.
On this box majority is negative while
$b\cdot y\geq3/(4\sqrt3)-\sqrt2/8>1/4$.
Its volume is $1/256$, so the integral is at least $(1/256)(1/4)=1/1024$; homogeneity covers general $b$.
It follows that
\begin{equation}\label{eq:noisy-strong-slack}
 B\geq\frac{4c_0}{1024\mu_3}\eta\|N^\top h\|
   =\frac3{2C}\eta\|N^\top h\|.
\end{equation}
The singular values of $N$ have squared sum three and product $\eta$; the product of the two largest is at most $3/2$. Hence
$\|N^\top h\|\geq2\eta/3$ and $\eta^2\leq CB$.
Since $\|Q\|\leq3$,
\[
 g^\top Ag\leq3\eta^2\|N^\top h\|^2\leq4C^2B^2/3,
\]
which proves \eqref{eq:noisy-radial-bounds}.
If $g=0$, the second estimate is immediate and the first follows from
$\eta^2=\det Q\leq1$, $B=1$ and $C>1$.

For every realization of $U$ in \eqref{eq:noisy-kernel},
\[
 \begin{aligned}
 \|Y\|^2
 &\leq\|g\|^2+2k_\varepsilon\sqrt{g^\top Ag}
                       +k_\varepsilon^2\|A\|\\
 &\leq\|g\|^2+B(4C\sqrt\varepsilon+27C\varepsilon)\leq1.
 \end{aligned}
\]
Indeed the chosen interval makes the first bracketed term at most $1/4$ and the second at most $27/(256C)<3/4$.
The normal component is orthogonal to $Y$, so $X$ is on the sphere.
Conditional normal means vanish, and $\mathbb EU=0$, giving $\mathbb EX=g$.
All rank strata and Gram formulas are Borel. With $K=T^\top T$ at every rank, the span projection is $P_W=TK^\dagger T^\top$, where $K^\dagger$ is the Moore--Penrose inverse, Borel on the rank strata. We use the principal positive-semidefinite square root $A^{1/2}$. Uniform subspace-sphere laws can be realized by projecting independent ambient Gaussians and normalizing; the projections are nonzero almost surely on each positive-dimensional subspace used here. Consequently $H_\varepsilon$ is jointly Borel. The formulas commute with permutations and orthogonal maps, giving anonymity and neutrality. At binary reports the majority peak has $B=0$, so both rules select it with certainty.

\subsection{Finite incentives at every report pair}

Fix opponents $a,b$ and define, on the ambient space,
\[
 f_{ab}(x)=\frac1{a_d}\int_{\{v:(a\cdot v)(b\cdot v)<0\}}
                   |x\cdot v|\,d\sigma_d(v).
\]
The majority identity gives
$g(x/\|x\|,a,b)=(a+b)/2+\nabla f_{ab}(x)$ for $x\ne0$.
Thus, for unit truth $p$ and alternative $s$,
\begin{equation}\label{eq:noisy-mean-slack}
 \delta=p\cdot(g_p-g_s)
 =f_{ab}(p)-f_{ab}(s)-\nabla f_{ab}(s)\cdot(p-s)\geq0.
\end{equation}
If $a\ne\pm b$, put $E=\operatorname{span}\{a,b\}$, $q=1-(a\cdot b)^2$,
$H(x)=\|\operatorname{Proj}_{E^\perp}x\|^2$ and
$X_a(x)=\|x\|^2-(a\cdot x)^2$, with $X_b$ defined similarly. Define
\[
 r_{ab}(x)=\frac{\sqrt q\,H(x)^{3/2}}{\sqrt{X_a(x)X_b(x)}},
\]
with value zero on $E$. If $a=\pm b$, set $r_{ab}=0$.
For a unit rank-three report $t$, direct substitution in \eqref{eq:noisy-gram} gives
\begin{equation}\label{eq:noisy-gradient}
 r_{ab}(t)^2=t^\top A_t t,
 \qquad \nabla r_{ab}(t)=A_t t/r_{ab}(t).
\end{equation}
Here is a direct check of the gradient identity. Write $x=t\cdot a$, $y=t\cdot b$, $z=a\cdot b$; then
$r_{ab}(t)=\eta\sqrt D/\sqrt q$.
Differentiation of the displayed formula gives gradient inner products
$r_{ab}(t)$, $r_{ab}(t)(yz-x)/(1-y^2)$ and
$r_{ab}(t)(xz-y)/(1-x^2)$ against $t,a,b$, respectively.
The identity $T^\top N=\sqrt D S_0^{-1}$ gives the same three products for $A_t t/r_{ab}(t)$. Both vectors lie in $\operatorname{span}\{t,a,b\}$, so they agree.

We now prove that $Lf_{ab}-r_{ab}$ is convex, with the uniform constant $L$ in \eqref{eq:noisy-epsilon}.
It is enough first to work in four dimensions: every comparison $(p,s,a,b)$ spans at most four dimensions, and all these functions are invariant under embedding that span. In particular, for a standard ambient Gaussian $V$,
\[
 f_{ab}(x)=\frac{\mathbb E[|x\cdot V|\mathbf1_{\{(a\cdot V)(b\cdot V)<0\}}]}
                   {\mathbb E|V_1|}.
\]
The Gaussian coordinates orthogonal to $\operatorname{span}\{x,a,b\}$ integrate out.
At a rank-three unit $t$, put
\[
 \begin{gathered}
 x=t\cdot a,\quad y=t\cdot b,\quad X=\sqrt{1-x^2},\quad Y=\sqrt{1-y^2},\\
 \widehat a=(a-xt)/X,\qquad\widehat b=(b-yt)/Y.
 \end{gathered}
\]
Write $c=\cos\theta=\widehat a\cdot\widehat b$ and $s_\theta=\sin\theta>0$.
An orthonormal basis of $t^\perp$ is
\[
 e_+=\frac{\widehat a+\widehat b}{\sqrt{2(1+c)}},\qquad
 e_-=\frac{\widehat a-\widehat b}{\sqrt{2(1-c)}},\qquad e_0\perp e_+,e_-.
\]
Differentiating the integral for $f_{ab}$ gives its tangent Hessian as three times the second-moment matrix on the two disagreement lunes of the unit sphere in $t^\perp$. The factor is
$2(2/\pi)/a_4=3$, since $a_4=4/(3\pi)$.
In polar coordinates in the $(e_+,e_-)$ plane, integrating over the two angle intervals of length $\theta$ gives
\begin{equation}\label{eq:noisy-lune-hessian}
 \nabla^2f_{ab}(t)|_{t^\perp}
 =\pi^{-1}\operatorname{diag}(\theta-s_\theta,\theta+s_\theta,\theta).
\end{equation}
For the other Hessian put
\[
 u=\frac{xY-yXc}{\sqrt q},\qquad v=\frac{yX-xYc}{\sqrt q}.
\]
The identities $q=(xY-yXc)^2+X^2s_\theta^2
=(yX-xYc)^2+Y^2s_\theta^2$ give $|u|,|v|\leq1$.
Differentiating $r_{ab}$ yields the following tangent matrix:
\begin{equation}\label{eq:noisy-radius-hessian}
 \begin{pmatrix}
 \frac{s_\theta(1-c)}2(3u^2+2uv+3v^2-2)&\frac32s_\theta^2(u^2-v^2)&0\\
 \frac32s_\theta^2(u^2-v^2)&\frac{s_\theta(1+c)}2(3u^2-2uv+3v^2-2)&0\\
 0&0&s_\theta(1+u^2+v^2)
 \end{pmatrix}.
\end{equation}
For an explicit differentiation formula, let $P=\operatorname{Proj}_{E^\perp}$, $P_a=I-aa^\top$, $P_b=I-bb^\top$. For any tangent $w$ the second derivative divided by $r=r_{ab}(t)$ is
\[
 \begin{aligned}
 \frac{\nabla^2r(t)[w,w]}{r(t)}
 &=\left(\frac{3w^\top Pt}{t^\top Pt}
 -\frac{w^\top P_at}{t^\top P_at}-\frac{w^\top P_bt}{t^\top P_bt}\right)^2\\
 &\quad+3\left(\frac{w^\top Pw}{t^\top Pt}
 -\frac{2(w^\top Pt)^2}{(t^\top Pt)^2}\right)
 -\sum_{j\in\{a,b\}}\left(\frac{w^\top P_jw}{t^\top P_jt}
 -\frac{2(w^\top P_jt)^2}{(t^\top P_jt)^2}\right).
 \end{aligned}
\]
Substitution of $e_+,e_-,e_0$ gives \eqref{eq:noisy-radius-hessian}.
For $w=w_+e_++w_-e_-+w_0e_0$, its quadratic form is bounded above by
\[
 5s_\theta(1-c)w_+^2+3s_\theta^2|w_+w_-|
       +5s_\theta(1+c)w_-^2+3s_\theta w_0^2.
\]
Young's inequality bounds the middle term by
$\tfrac32s_\theta(1-c)w_+^2+\tfrac32s_\theta(1+c)w_-^2$.
Finally
\[
 s_\theta(1-c)\leq3(\theta-s_\theta),\qquad
 s_\theta(1+c)\leq\theta+s_\theta,\qquad s_\theta\leq\theta.
\]
The first inequality follows by differentiating the difference, whose derivative is $2(1-c)^2$ and whose value at zero is zero. These bounds give
$\nabla^2r_{ab}\preceq L\nabla^2f_{ab}$ on the tangent space. Degree-one homogeneity makes both radial and mixed radial--tangent Hessian components zero and scales the unit-point inequality to every nonzero point. Thus the inequality holds in the full space at rank three.

The extension across rank loss requires no differentiability at the origin.
Differentiation under the integral makes $f_{ab}$ continuously differentiable away from zero.
The same is true of $r_{ab}$, with gradient zero on nonzero points of $E$.
Away from the lines through $a,b$, this follows from its cubic numerator and positive denominator. Near a nonzero multiple of $a$, let $z=\operatorname{Proj}_{a^\perp}x$. Then $H(x)\leq\|z\|^2$, $X_a(x)=\|z\|^2$, while $X_b(x)$ stays positive, so $r_{ab}(x)=O(\|z\|^2)$.
Its differentiated formula
\[
 \sqrt q\left[
 \frac{3\sqrt H\operatorname{Proj}_{E^\perp}x}{\sqrt{X_aX_b}}
 -\frac{H^{3/2}P_ax}{X_a^{3/2}\sqrt{X_b}}
 -\frac{H^{3/2}P_bx}{\sqrt{X_a}X_b^{3/2}}\right]
\]
has respective norms $O(\|z\|)$, $O(\|z\|)$ and $O(\|z\|^2)$. The argument at multiples of $b$ is identical.
Let $h=Lf_{ab}-r_{ab}$. Its restriction to a line avoiding zero is convex: such a line either lies in $E$, where $r_{ab}=0$, or crosses $E$ at most once, and the continuous derivative joins the convex pieces.
Both functions are even, homogeneous of degree one and continuous at zero.
For $x\ne0$ choose $y$ not collinear with $x$. The midpoint inequality on the offset line through $x+\tau y$ and $-x+\tau y$ gives, as $\tau\downarrow0$, $h(x)\geq0$.
On a line through zero, $h(tx)=|t|h(x)$ is therefore convex as well.
This proves global convexity, including antipodal comparisons.

Its supporting inequality and \eqref{eq:noisy-mean-slack} now imply
\begin{equation}\label{eq:noisy-bregman}
 r_{ab}(p)-p\cdot\nabla r_{ab}(s)\leq L\delta.
\end{equation}
At nonzero rank drops both $r$ and its gradient are zero.
If $a=\pm b$, every truthful alignment under the normal lift is certain; finite linear incentives and Jensen already prove ICV-SP.
Otherwise suppose the true triple has rank three.
Its alignment under $H_\varepsilon$ is uniform with mean $m_p=p\cdot g_p$ and half-width $k_\varepsilon r_{ab}(p)$, because a projection of uniform probability on the two-sphere is uniform on an interval.
If the alternative triple has rank three, $p\cdot Y_s$ is uniform with mean $m_s=p\cdot g_s$ and half-width $k_\varepsilon v_s$, where $v_s=\sqrt{p^\top A_s p}$.
Positive-semidefinite Cauchy--Schwarz and \eqref{eq:noisy-gradient} give
$v_s\geq p\cdot\nabla r_{ab}(s)$.
At lower-rank alternatives put $Y_s=g_s$ and $v_s=0$; the same bound holds.
By \eqref{eq:noisy-bregman} and $k_\varepsilon L\leq1/2$,
\[
 m_p\geq m_s,\qquad
 m_p-k_\varepsilon r_{ab}(p)\geq m_s-k_\varepsilon v_s.
\]

For completeness, two possibly degenerate uniform laws are ordered by every increasing-concave evaluation whenever their means and lower endpoints are so ordered.
Write them as $m_j+w_jV$ with $V$ uniform on $[-1,1]$.
If $w_1>w_2$, then
$(m_1+w_1V)-(m_2+w_2V)=(m_1-w_1)-(m_2-w_2)+(1+V)(w_1-w_2)\geq0$ pointwise.
If $w_1\leq w_2$, monotonicity first lowers $m_1$ to $m_2$; contraction of the centered interval and concavity then give
$\mathbb E\phi(m_2+w_1V)\geq\mathbb E\phi(m_2+w_2V)$.
Indeed, for $w_2>0$, set $\lambda=w_1/w_2$ and apply concavity to
$m_2+w_1V=\lambda(m_2+w_2V)+(1-\lambda)m_2$, followed by Jensen.
All intermediate alignments stay in $[-1,1]$.
Conditional on $Y_s$, normal noise has mean zero. Therefore
\[
 \mathbb E\phi(p\cdot X_p)=\mathbb E\phi(p\cdot Y_p)
 \geq\mathbb E\phi(p\cdot Y_s)\geq\mathbb E\phi(p\cdot X_s).
\]
If the true triple has rank at most two, its alignment is certain and linear IC followed by Jensen proves the same conclusion for every alternative rank.
These arguments include bounded concave evaluations with a downward jump at $-1$. Jensen follows from an interior supporting line when the mean is interior; an endpoint mean forces a degenerate law. Pointwise monotonicity and finite concavity remain valid at endpoints, and positive-width uniform laws have no endpoint atoms.
The four-report span reduction extends the proof to every $d\geq4$. Thus one fixed $\varepsilon$ gives a ICV-SP rule on the full report domain.

\subsection{Identical control with different risk at every nonplanar triple}

For any $n\geq3$, sample a three-element subset $J$ of participants uniformly, independently of reports, and apply $H_0$ or $H_\varepsilon$ to their reports. Denote the resulting full-domain rules by $P_n^0,P_n^\varepsilon$.
Every unilateral incentive comparison holds component by component: a committee excluding the deviator is unchanged, while one including it obeys the three-person ICV-SP inequality. Averaging preserves incentives and the stated symmetries and Borel property.
Their means coincide at every profile. At binary reports each committee selects its majority peak, which gives \eqref{eq:three-committee-control}. Such endpoint lotteries satisfy both binary efficiencies. At distinct reports $a,b$, both endpoints are efficient. The SD cutoff at one would require any dominating law to put mass at least $\omega_s$ at $a$ and $1-\omega_s$ at $b$, exhausting probability. At unanimity the common peak is uniquely efficient.

At three nonempty camps of sizes $n_1,n_2,n_3$ with distinct directions $p_1,p_2,p_3$, put
\[
 \tau=\frac{n_1n_2n_3}{\binom n3},\qquad
 \mu_i=p_i\cdot g(p_1,p_2,p_3),\qquad a_i=p_i^\top A(p_1,p_2,p_3)p_i.
\]
A committee either has a strict-majority camp $j$, with probability $\omega_{n_j}$, or contains one member of each camp, with probability $\tau$.
Thus $\sum_j\omega_{n_j}+\tau=1$.
At rank three, every camp-$i$ participant has the alignment laws
\begin{equation}\label{eq:rank-three-score-laws}
 \begin{aligned}
 P_n^0:&\quad \sum_j\omega_{n_j}\delta_{p_i\cdot p_j}+\tau\delta_{\mu_i},\\
 P_n^\varepsilon:&\quad \sum_j\omega_{n_j}\delta_{p_i\cdot p_j}
       +\tau\operatorname{Unif}[\mu_i-\sqrt{3\varepsilon a_i},
                               \mu_i+\sqrt{3\varepsilon a_i}].
 \end{aligned}
\end{equation}
This includes unselected participants: every reported direction is present in a one-from-each committee, so its normal noise is orthogonal to their reports too.
Since means coincide, the exact variance difference is
\begin{equation}\label{eq:rank-three-variance-gap}
 \operatorname{Var}_{P_n^\varepsilon}(p_i\cdot X)
 -\operatorname{Var}_{P_n^0}(p_i\cdot X)=\tau\varepsilon a_i>0.
\end{equation}
The same pair of global rules works for every rank-three camp profile, size vector and labeling.
At rank at most two the two rules themselves coincide on three-camp profiles, and the planar identification theorem fixes reported alignment laws for the full class with these controls.

For orthonormal directions, $K=I$, $A=\operatorname{Proj}_W$ and $a_i=1$.
At equal camps of size $m$,
\[
 \tau=\frac{2m^2}{(3m-1)(3m-2)}\longrightarrow\frac29.
\]
Thus the variance gap need not vanish as population grows.
Varying $\varepsilon$ in $[0,\varepsilon_*]$ gives the interval of increments
$[0,\tau\varepsilon_*a_i]$, with all participants' increments linked by the same $A$.
At every three-camp profile, each participant weakly prefers $P_n^0$ for every permitted evaluation, strictly at rank three for strictly concave evaluations and $\varepsilon>0$.
This three-camp welfare comparison follows from the mean-preserving spread in \eqref{eq:rank-three-score-laws}.

\phantomsection
\addcontentsline{toc}{section}{References}
\begingroup
\normalsize
\setlength{\bibsep}{0pt plus 6pt}
\setlength{\parskip}{0pt}

\interlinepenalty=0
\bibliographystyle{plainnat}
\bibliography{references}
\endgroup
\end{document}